\documentclass[sigconf,edbt]{acmart-edbt2022}

\def\BibTeX{{\rm B\kern-.05em{\sc i\kern-.025em b}\kern-.08em
    T\kern-.1667em\lower.7ex\hbox{E}\kern-.125emX}}

\usepackage{booktabs} 

\setcopyright{rightsretained}

\acmDOI{}

\acmISBN{978-3-89318-092-9}

\acmConference[EDBT 2023]{26th International Conference on Extending Database Technology (EDBT)}{28th March-31st March, 2023}{Ioannina, Greece} 
\acmYear{2023}

\usepackage{tabu,booktabs}
\usepackage{multirow}
\usepackage[caption=false]{subfig}
\usepackage{graphicx}
\usepackage{epstopdf}
\usepackage{svg}
\usepackage{balance}
\usepackage{todonotes}
\usepackage{color, colortbl}

\usepackage{xspace}
\newcommand{\PaperAcronym}{REIN\xspace}
\newcommand{\quotes}[1]{``#1''}

\begin{document}
\title{\PaperAcronym: A Comprehensive Benchmark \\ Framework for Data Cleaning Methods in ML Pipelines}
\titlenote{This work was supported (in part) by the Federal Ministry of Education and Research through grants 02L19C155, 01IS21021A (ITEA project number 20219).}

\author{Mohamed Abdelaal, Christian Hammacher, Harald Schöning}
\affiliation{%
	\institution{Software AG, Darmstadt, Germany}
	\streetaddress{Uhlandstrasse 12}
	\postcode{D-64297}
}
\email{first.last@softwareag.com}

\renewcommand{\shortauthors}{M. Abdelaal et al.}

\begin{abstract}
Nowadays, machine learning (ML) plays a vital role in many aspects of our daily life. In essence, building well-performing ML applications requires the provision of high-quality data throughout the entire life-cycle of such applications.  Nevertheless, most of the real-world tabular data suffer from different types of discrepancies, such as missing values, outliers, duplicates, pattern violation, and inconsistencies. Such discrepancies typically emerge while collecting, transferring, storing, and/or integrating the data. To deal with these discrepancies, numerous data cleaning methods have been introduced. However, the majority of such methods broadly overlook the requirements imposed by downstream ML models. As a result, the potential of utilizing these data cleaning methods in ML pipelines is predominantly unrevealed. In this work, we introduce a comprehensive benchmark, called \PaperAcronym\footnote{The source code, data, and other artifacts have been made available at \url{https://github.com/mohamedyd/rein-benchmark}}, to thoroughly investigate the impact of data cleaning methods on various ML models. Through the benchmark, we provide answers to important research questions, e.g., where and  whether data cleaning is a necessary step in ML pipelines. To this end, the benchmark examines 38 simple and advanced error detection and repair methods. To evaluate these methods, we utilized a wide collection of ML models trained on 14 publicly-available datasets covering different domains and encompassing realistic as well as synthetic error profiles. 
\end{abstract}

%
%



\maketitle


\section{Introduction}

With the advent of modern computing technologies, many industries nowadays are developing robust ML models capable of analyzing big and complex data while delivering fast and accurate results on vast scales. Such results are typically harnessed by organizations and businesses to make better decisions without or with minimal human intervention. However, the correctness of such decisions broadly depends on the quality of the available data. According to a recent Gartner research \cite{moore18}, organizations believe poor data quality to be responsible for an average of \$15 million per year in losses. Another study by IBM in 2016 \cite{redman16} revealed that poor data quality costs the US economy \$3.1 trillion per year. These studies illustrate that data quality problems are predominantly expensive and pervasive. 

For decades, data quality has been an active research area. In this context, the data management community tackled the data quality problems as a part of the ETL workflows. Accordingly, numerous proposals have been introduced to automatically detect and/or repair data discrepancies~\cite{fahes18,holodetect19,holoclean17,nadeef13,baran20,katara15}. In fact, only a small fraction of these proposals considered the heterogeneity profiles of data errors while discovering and repairing the erroneous instances. In other words, most proposed techniques are dedicated to serve only one error type. Moreover, most of such methods have been developed in isolation from the downstream ML applications. Thus, the consequences of adopting such cleaning methods for predictive tasks are broadly concealed. Accordingly, a challenge of selecting the most well-suited cleaning strategies (i.e., combinations of error detection and repair methods) in ML pipelines arises. 

In this paper, we tackle this challenge through introducing a benchmark framework, referred to as \PaperAcronym. The main goal of \PaperAcronym is to thoroughly investigate the interplay between data cleaning and ML modeling. Through extensive experiments, \PaperAcronym examines plenty of cleaning strategies in combination with various ML models, covering classification, regression, clustering, and AutoML models. In \PaperAcronym, we evaluate the error detection and repair methods while being adopted as stand-alone methods and as components in ML pipelines. To this end, it is necessary to possess the ground truth of the available dirty datasets. Nevertheless, it is not usually straightforward to find such datasets which are also well-suited for ML tasks. Another challenge of conducting such a comprehensive study is the scale of the intended experiments. The number of models to be trained are exploded due to involving plenty of error detection and repair methods (cf. Section~\ref{sec:overview}). For such detection and repair methods, it is also crucial to provide the necessary configurations and signals, i.e., patterns, rules, and helping functions. Finally, ML models are inherently probabilistic, where resampling may change the results. Hence, we need to validate the conclusions obtained from the ML experiments.

In detail, the paper provides the following contributions: (1) We define an architectural framework to systematically evaluate error detection and repair tools dedicated to tabular data. In addition to the traditional evaluation measures relative to the ground truth, \PaperAcronym enables data scientists and practitioners to properly judge their detection and repair methods using the performance of several predictive models. Moreover, \PaperAcronym utilizes the intersection over union (IoU) metric to quantify the similarities between data cleaning methods. 
(2) We design a benchmark controller that efficiently manages the other components in the framework. Such a controller leverages the design-time knowledge, e.g., the error types and the ML tasks, to broadly sidestep unnecessary experiments, thus reducing the complexity of running the benchmark. (3) We provide a classification of the most prominent error detection and repair methods according to their methodology and the required configurations. (4) We examine the performance of the involved ML models in different scenarios which are characterized by the data version, i.e., ground truth, dirty, or repaired data. (5) We evaluate scalability of the error data cleaning methods through using small, medium, and large datasets as inputs to these methods. Moreover, we evaluate the robustness of such methods through repeating the experiments while monotonically increasing the error rate. (6) We adopt the Wilcoxon signed-rank test with continuity correction to compensate for the randomness inherited in the training process. To the best of our knowledge, \PaperAcronym is the first large-scale benchmark framework which evaluates the data cleaning methods from different perspectives, including detection and repair performance, predictive accuracy, robustness, and scalability. 


%
\vspace{-3mm}\section{Benchmark Overview}\label{sec:overview}
In this section, we introduce the architecture of \PaperAcronym together with our assumptions. \PaperAcronym comprises several data processing and evaluation steps. Specifically, several dirty datasets $\Phi^{-} = \phi_{1}^{-},\cdots, \phi_{n}^{-}, \phi_i^{-} \in \mathbb{R}^{u \times v}$  are used as inputs to different error detectors $\alpha_{1},\cdots, \alpha_{m}$, where the superscript `--' denotes a dirty dataset and $u$, $v$ denote the number of records and attributes in $\phi_{i}^{-}$. Afterward, the erroneous instances, identified by each detection method, are replaced with repair candidates using a number of data repair methods $\beta_{1},\cdots, \beta_{k}$. The result of this step is a new set of repaired datasets $\Phi^{+} = \phi_{i, 1}^{+},\cdots, \phi_{i, \epsilon}^{+}$, where $\epsilon = m \times k$ represents the number of generated repair versions for each dirty dataset $\alpha_i$ and the superscript `+' denotes a repaired dataset. Finally, each repaired dataset $\phi_{i,j}^{+}$ is sampled to train several ML models $\gamma_{1},\cdots, \gamma_{h}$, where $h$ is the number of involved ML models. Thus, the number of ML experiments for each dirty dataset $\phi_{i}^{-}$ is $(\epsilon + 1) \times h \times s$, where each experiment is repeated $s$ times to estimate the means and standard deviations, and the dirty version is added to the number of generated repaired versions.

\begin{figure}[htbp]
	\centering
	\includegraphics[width=0.45\textwidth]{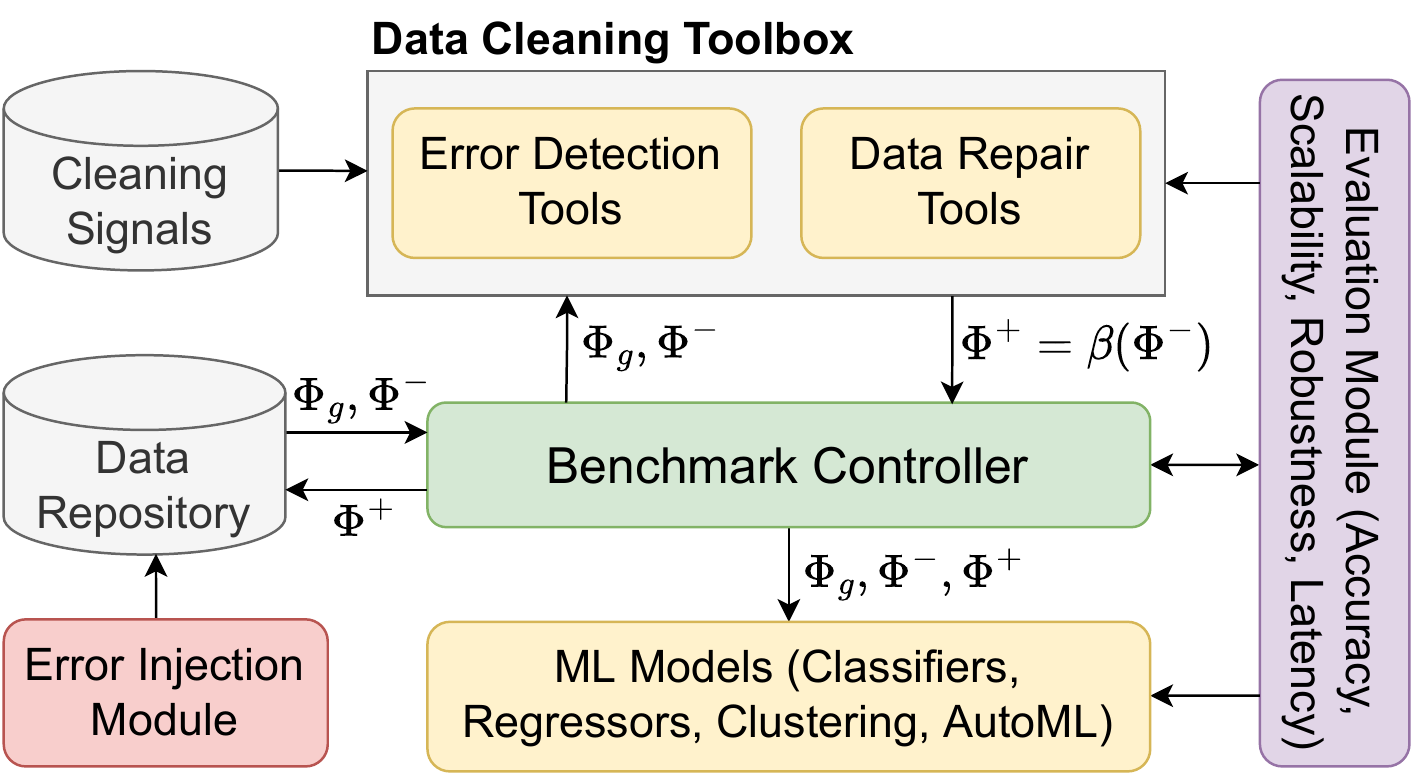}
	\caption{Benchmark architecture}
	\label{fig:architecture}\vspace{-5mm}
\end{figure}

To realize such a large-scale benchmark, we implemented the architecture depicted in Figure~\ref{fig:architecture}. A \textit{data repository}, i.e., PostgreSQL database, is utilized to store the ground truth $\Phi_{g}$, the dirty data $\Phi^{-}$, and the set of generated repaired versions $\Phi^{+}$. To properly control the experiments, an \textit{error injection} module generates different types of errors with various error rates. Practically speaking, the task of error injection is carried out in an offline phase before running the experiments (cf. Section~\ref{sec:data} for more details). Another component is the \textit{data cleaning toolbox}, which is a pool containing all available error detection and repair tools. Some of these tools, such as NADEEF, HoloClean, and OpenRefine, cannot be utilized without providing them with a set of \textit{cleaning signals}. Examples of such signals include functional dependency constraints, integrity constraints, knowledge bases, patterns, and pre-estimated configurations. 

The main component in \PaperAcronym is the \textit{benchmark controller}, which connects all other components in the benchmark. The purpose of such a controller is three-fold: First, it smoothly exchanges the ground truth $\Phi_g$, the dirty $\Phi^{-}$, and the repaired data $\Phi^{+}$ among the different modules. Second, it avoids unnecessary error detection and repair operations exploiting prior knowledge about the dirty datasets. For example, if a dataset is known to have duplicates (e.g., the \textit{Citation} dataset), it is meaningless to run rule violation or outlier detection methods. Third, it exploits the prior knowledge to adapt the data preparation steps in accordance with the associated ML tasks. The last component in the architecture is the evaluation module, which serves the error detection and repair methods as well as the ML models. For instance, the evaluation module leverages several quality metrics to estimate the predictive performance of ML models trained on the ground truth, the dirty and the repaired data. 

Another component is a pool of \textit{ML models} which comprises a wide collection of classification, regression, and clustering methods. Moreover, \PaperAcronym also examines two AutoML algorithms to check the performance of fully-automated pipelines consisting of data cleaning and modeling modules. Finally, an \textit{evaluation module} examines the performance of data cleaning and modeling methods in terms of four metrics, including accuracy, latency, scalability, and robustness (cf. Section~\ref{sec:evaluation}). Due to space constraints, we define in the README file of the source code: (1) how to run the benchmark with/without the ground truth of dirty datasets, and (2) how to readily extend the REIN framework by adding new datasets, ML models, and data cleaning tools. 

%
\section{Data Cleaning Methods}\label{sec:cleaning}
%
%
In this section, we provide an overview of the examined error detection and repair methods. 

\subsection{Error Detection Methods}

In \PaperAcronym, we selected 19 publicly-available error detection methods, which deal with the most common attribute and class errors in tabular data\footnote{Attribute errors occur in the training features, while class errors occur in the labels}. Table~\ref{tab:detectors_cleaners} lists the error detection methods and their targeted error types. Moreover, the table comprises the configurations and/or signals, i.e., patterns, constraints, helping functions, and knowledge bases, necessary for running each detection method. In \PaperAcronym, we classify the error detection methods according to their methodology into two main categories, including (I) Non-learning methods and (II) ML-supported methods. As its name implies, the former category includes the methods and tools which detect errors using either a set of user-provided knowledge base, business rules, integrity constraints, or using a set of statistical measures. Each of these methods and tools typically tackle specific error types, e.g., duplicates, outliers, or missing values. The second category comprises the methods, e.g., Picket, ED2, and RAHA, which formulate the error detection task as a classification problem. These methods initially extract a set of features for each attribute. Such auto-generated features enable a classifier to differentiate between clean and dirty data samples. To train such a classifier, some training samples are selected to be labeled by an oracle. Below, we introduce the error detectors in each category. 

\begin{table*}[tbh]
	\caption{Examined error detection and repair methods (The index (\textit{Idx}) and abbreviation (\textit{Abbr}) are used to refer to the detection and repair methods in the figures of Section~\ref{sec:evaluation})}\vspace{-3mm}
	\label{tab:detectors_cleaners}
	\resizebox{\textwidth}{!}{%
		\begin{tabular}{llcclc||llcclc}
			\toprule
			\textbf{Idx.} & \textbf{Detector}                       & \textbf{Abbr.} & \textbf{Cat.} & \textbf{Tackled Errors}       & \textbf{Configs.}  & \textbf{Idx.} & \textbf{Repair Method}                                & \textbf{Abbr.} & \textbf{Cat.} & \textbf{Tackled Errors}       & \textbf{Configs.}  \\
			\midrule
			K              & KATARA \cite{katara15}                  & ---            & I                 & Pattern violations            & Knowledge Base     & 1              & Ground Truth                                          & GT             & I                 & ---                           & ---      \\
			N              & NADEEF \cite{nadeef13}                  & ---            & I                 & Rule violations & FD Rules, Patterns & 2              & Delete                                                & ---            & I                               & ---                           & ---      \\
			F              & FAHES \cite{fahes18}                    & ---            & I                 & Missing Values                & ---                & 3              & Imputation: Mean-Mode                                 & Impute         & I                 & MV/Outliers       & ---                  \\
			H              & HoloClean \cite{holoclean17}            & Holo           & I                 & Rule violations         & Denial Constraints & 4              & Imputation: Median-Mode                               & Impute         & I                       & MV/Outliers       & ---                  \\
			B              & dBoost \cite{dboost16}                  & ---            & I                 & Outliers                      & Hyperparams        & 5              & Imputation: Mode-Mode                                 & Impute         & I                 & MV/Outliers       & ---                  \\
			O              & OpenRefine \cite{openrefine13}          & OpnR           & I                 & Inconsistencies            & Clusters    & 6              & Imputation: missForest \cite{missforest12}            & MISS-Mix       & I                 & MV/Outliers       & ---                            \\
			I              & Outlier Detector: IF \cite{if2012}      & IF             & I                 & Outliers                      & Hyperparams        & 7              & Imputation: DataWig \cite{datawig19}                  & DataWig-Mix    & I                 & MV/Outliers       & ---                  \\
			S              & Outlier Detector: SD \cite{outliers13}  & SD             & I                 & Outliers                      & Hyperparams        & 8              & Imputation: missForest-missForest \cite{missforest12} & MISS-Sep       & I                 & MV/Outliers       & ---                  \\
			Q              & Outlier Detector: IQR \cite{outliers13} & IQR            & I                 & Outliers                      & Hyperparams        & 9              & Imputation: missForest-DataWig                        & MISS-Datawig   & I                 & MV/Outliers       & ---                  \\
			V              & MV Detector \cite{mvd10}                & MVD            & I                 & Missing Values                & ---                & 10             & Imputation: Decision Tree-missForest                  & DT-MISS        & I                 & MV/Outliers       & Hyperparams          \\
			D              & Key Collision \cite{cleanml19}          & DuplD          & I                 & Duplicates                    & Key Columns        & 11             & Imputation: Bayesian Ridge-missForest                 & Bayes-MISS     & I                 & MV/Outliers       & Hyperparams          \\
			Z              & ZeroER \cite{zeroer20}                  & ---            & I                 & Duplicates                    & Blocking Functions & 12             & Imputation: KNN-missForest                            & KNN-MISS       & I                 & MV/Outliers       & Hyperparams          \\
			C              & CleanLab \cite{cleanlab21}              & ---            & I                 & Mislabels                     & Hyperparams        & 13             & HoloClean \cite{holoclean17}                          & Holo           & I                 & MV/Rule Violation & Denial Constraints   \\
			M              & Min-K \cite{max_min16}                  & Min            & I                 & Holistic                      & Hyperparams        & 14             & OpenRefine \cite{openrefine13}                        & OpenR          & I                 & Inconsistencies   & Clusters             \\
			X              & Max Entropy \cite{max_min16}            & Max            & I                 & Holistic                      & Hyperparams        & 15             & BARAN \cite{baran20}                                  & ---            & I                 & Holistic          & Labels               \\
			T              & Metadata-Driven \cite{metadata18}       & Meta           & II                & Holistic                      & Labels             & 16             & CleanLab \cite{cleanlab21}                            & ---            & II                & Mislabels         & ---                  \\
			R              & RAHA \cite{raha19}                      & ---            & II                & Holistic                      & Labels             & 17             & ActiveClean \cite{activeclean16}                      & ---            & II                & ---               & Repairs, Labels      \\
			E              & ED2 \cite{ed219}                        & ---            & II                & Holistic                      & Labels             & 18             &  BoostClean \cite{boostclean17}                       & ---            & II                & ---               & Repair, Labels       \\
			P              & Picket \cite{picket20}                  & ---            & II                & Holistic                      & ---                & 19             & CPClean \cite{cpclean20}                              & ---            & II                & ---               & Hyperparams, Repairs       
			\\ \bottomrule             
		\end{tabular} \vspace{-4mm}
	}
\end{table*}

\paragraph{Non-Learning Detectors:} The first method in Table~\ref{tab:detectors_cleaners} is KATARA \cite{katara15} which aligns the input dirty dataset with crowdsourced knowledge bases to identify and correct data samples that violate semantic patterns. To detect rule and pattern violations, NADEEF \cite{nadeef13} treats data quality rules holistically via providing an interface for implementing denial constraints and other user-defined functions. Another relevant work is HoloClean \cite{holoclean17} which combines qualitative and quantitative signals, e.g., denial constraints and correlations, in a statistical model that enables detecting and repairing missing values and rule/constraint violations. To identify inconsistencies and pattern violations, the OpenRefine tool \cite{openrefine13} enables users to visually explore the dirty datasets through faceting and filtering operations. FAHES \cite{fahes18} is another tool which detects disguised missing values, e.g., "999999" for a phone number. To this end, FAHES employs a syntactic pattern detection module for categorical data and a density-based outlier detection module for numerical data. To detect explicit missing values, \PaperAcronym implements a method to find empty or \textit{NAN} entries. 

dBoost \cite{dboost16} is an outlier detection method which integrates several of the most widely applied outlier detection algorithms, including histograms, Gaussian, and multivariate Gaussian mixtures. To find the optimal hyperparameters for such algorithms, dBoost employs random search, where the search space is all the possible configurations. Other outlier detection methods involve Standard Deviation (SD), Interquartile Range (IQR) \cite{outliers13}, and Isolation Forest (IF) \cite{if2012}. The former method annotates a cell $x \in A$, where $A$ denotes an attribute, as an outlier if it is $n$ numbers of standard deviations away from the mean of entries in $A$. A more resistant statistical measure is IQR, defined as the difference between the 25th and 75th percentiles of an attribute A, i.e., $IQR_{A} = Q_3 - Q_1$. In this case, an outlier is any value laying outside the range of $[Q_1 - k \times IQR_{A}, Q_3 + k \times IQR_{A}]$, where $k$ and $n$ are hyperparameters. The latter method targets identifying outliers without profiling all data samples. Specifically, the IF method builds an ensemble of isolation binary trees for the dirty dataset, and outliers are the data samples that have shorter average path lengths on the binary trees.

To detect duplicates, \PaperAcronym examines two methods, namely Key Collision \cite{cleanml19} and ZeroER \cite{zeroer20}. The former method requires user-provided information about the key attributes assumed to be unique. In this case, two records can be detected as duplicates whenever they share the same value on the key attributes. The latter method relies on Magellan \cite{magellan16} to generate a set of similarity features. However, ZeroER requires zero labeled examples where it implements a Gaussian Mixture Model to learn the distributions that govern the feature vectors of matches and unmatches. Away from duplicates, CleanLab \cite{cleanlab21} detects noisy labels via exploiting the principles of confident learning to estimate the joint distribution of noisy and true labels. To tackle the heterogeneity of data errors, Min-K and Max Entropy \cite{max_min16} implement an ensemble of other non-learning methods to identify most of the existing erroneous samples in a dataset. Specifically, Min-K considers as errors those samples detected by at least $k$-methods. Alternatively, Max Entropy introduces an entropy-based sampling method to systematically select the order in which the non-learning methods should be executed. 

\paragraph{ML-supported Detectors:} The ML-supported methods, examined in \PaperAcronym, differ in how the features are generated and how the required labeling budget is reduced. For example, the metadata-driven error detection method \cite{metadata18} implements a metadata profiler and a suite of non-learning error detection methods to extract the features. In this case, each non-learning method is represented by a binary feature, where the feature value is one, if the non-learning method recognized this cell to be dirty. To reduce the labeling budget, RAHA \cite{raha19} adopts a semi-supervised algorithm which clusters the samples by similarity and acquires labels on a per-cluster basis, before propagating the acquired labels in each cluster. Similarly, ED2 \cite{ed219} extracts a set of attribute-level, tuple-level, and dataset-level features which define the distribution governing the dataset. Moreover, ED2 exploits active learning to acquire labels for clean/erroneous samples that the classifier is uncertain about. Finally, Picket \cite{picket20} employs self-supervision to train an error detection model without requiring user labels.
 
\vspace{-2mm}\subsection{Data Repair Methods}
In \PaperAcronym, we examine 19 data repair methods which can be classified into two main categories according to their intervention type, namely (I) generic methods and (II) ML-oriented methods. The former category comprises the methods which directly modify the dirty dataset to generate a repaired version. Such modifications can be either removing the dirty cells or replacing them with a set of generated repairs. They are generic in the sense that they seek to improve the data quality, regardless of the downstream application, e.g., ML modeling, data visualization, or data enrichment. Alternatively, the second category comprises methods which jointly optimize the data quality and the performance of downstream ML models. In \PaperAcronym, we also exploit the ground truth of the dirty data to show the performance upper-bound. Below, we introduce the various methods in each category.

\paragraph{Generic Repair Methods} To generate repair values, \PaperAcronym examines several standard and ML-driven imputation methods. The standard imputation methods utilize simple statistical measures, such as mean, median, or mode to generate repairs for the numerical values. For categorical values, we simply leverage the mode, i.e., the most frequent value in the corresponding attribute, as the repair value. Advanced imputation methods are those which build ML models to generate accurate repairs based on information in the entire dataset. For numerical values, \PaperAcronym examines 5 ML-based imputation methods including K-nearest neighbors (KNN), Decision Tree (DT), Bayesian Ridge \cite{scikitlearn11}, missForest based on random forest (RF) \cite{missforest12}, and DataWig based on deep neural networks \cite{datawig19}. For categorical values, we examine both of missForest and DataWig. In particular, missForest iteratively trains an RF model on a set of clean samples (i.e., complete with no outliers) in a first step, before predicting the missing values. Similarly, DataWig implements deep learning modules combined with neural architecture search and end-to-end optimization of the imputation pipeline. 

For mixed-type datasets, missForest and DataWig have two modes of operation, namely \textit{separate} mode and \textit{mixed} mode. In the former mode, each method is executed separately for each data type, referred to as MISS-Sep. The latter mode involves executing each method holistically on all data types, referred to as MISS-Mix and DataWig-Mix, taking into account possible relations between different data types. Another generic method is HoloClean \cite{holoclean17} which precisely infers the repair values via holistically employing multiple cleaning signals to build a probabilistic graph model. To repair pattern violations and inconsistencies, OpenRefine \cite{openrefine13} utilizes Google Refine Expression Language (GREL) as its native language to transform existing data or to create repair values. The last method in this category is BARAN \cite{baran20} which is a holistic configuration-free method for repairing all error types. To this end, BARAN trains incrementally updatable models which leverage the value, the vicinity, and the domain contexts of data errors to propose correction candidates. To further increase the training data, BARAN exploits external sources, such as Wikipedia page revision history.

\paragraph{ML-oriented Repair Methods:} The second category comprises the methods designed to jointly optimize the cleaning and modeling tasks. In other words, these methods focus on selecting the optimal repair candidates with the objective of improving the performance of specific predictive models. Accordingly, these methods assume the availability of repair candidates from other generic methods. For instance, BoostClean \cite{boostclean17} treats the error correction task as a statistical boosting problem where a set of weak learners are composed into a strong learner. To generate the weak learners, BoostClean iteratively selects a pair of detection and repair methods, before applying them to the training set to derive a new model. ActiveClean \cite{activeclean16} is another ML-oriented method, principally employed for models with convex loss functions. It formulates the data cleaning task as a stochastic gradient descent problem. Initially, it trains a model on a dirty training set, where such a model is to be iteratively updated until reaching global minima. In each iteration, ActiveClean samples a set of records and then asks an oracle to clean them to shift the model along the steepest gradient. A similar work is CPClean \cite{cpclean20} which incrementally cleans a training set until it is certain that no more repairs can possibly change the model predictions. 

\section{Data Modeling}\label{sec:models}
In this section, we present a representative set of common ML models utilized for assessing the performance of error detection and repair methods.  
%
%
Table~\ref{tab:ml_models} summarizes the algorithms and whether they are used for classification (C), regression (R), or unsupervised clustering (UC) tasks. As listed in the table, \PaperAcronym examines 12 classifiers, 11 regression models, six clustering algorithms, and two AutoML algorithms. Such vital algorithms are broadly applicable in various real-world application domains, e.g., cybersecurity systems, smart cities, healthcare, e-commerce, agriculture, and many more \cite{machine21}. The rationale behind involving two AutoML algorithms is to evaluate the performance of fully automated ML pipelines, consisting of data cleaning and model building. We are interested in checking whether such algorithms are able to find the best possible combination of model architectures and hyperparameters based on dirty or automatically-repaired datasets. For most of these models, \PaperAcronym exploits the Python implementation of Scikit-learn \cite{scikitlearn11} library for training and testing. For hyperparameter tuning, \PaperAcronym leverages a Bayesian-based informed search method, referred to as Optuna \cite{optuna2019}. However, we did not use Optuna with the AutoML algorithms, since they can automatically select the best hyperparameters. Moreover, we did not use the internal processing pipelines of these algorithms, since we mainly focus on the examined cleaners (listed in Table~\ref{tab:detectors_cleaners}).\vspace{-2mm}
\vspace{-3mm}\begin{table}[htb]
	\caption{Examined ML models}\vspace{-2mm}
	\label{tab:ml_models}
	\resizebox{\columnwidth}{!}{%
		\begin{tabular}{lll||llll}
			\toprule
			\textbf{Algorithm}           & \textbf{C} & \textbf{R} & \textbf{Algorithm}                & \textbf{C} & \textbf{R} & \textbf{UC} \\
			\midrule
			Logistic Regression (Logit)  & \checkmark &            & Linear Regression                 &            & \checkmark &             \\
			Decision Tree (DT)           & \checkmark & \checkmark & Bayes Ridge Regressor (BRidge)    &            & \checkmark &             \\
			Random Forest (RF)           & \checkmark & \checkmark & RANSAC                            &            & \checkmark &             \\
			Linear SVC                   & \checkmark & \checkmark & Gaussian Mixture (GMM)            &            &            & \checkmark  \\
			SGD Classifier               & \checkmark &            & K-Means                           &            &            & \checkmark  \\
			KNN                          & \checkmark & \checkmark & Affinity Propagation (AP)         &            &            & \checkmark  \\
			AdaBoost (AdaB)              & \checkmark & \checkmark & Hierarchical Clustering (HC)      &            &            & \checkmark  \\
			Gaussian Naïve Bayes (GNB)   & \checkmark &            & OPTICS                            &            &            & \checkmark  \\
			Multinomial NB               & \checkmark &            & BIRCH                             &            &            & \checkmark  \\
			XgBoost (XGB) \cite{xgboost16}     & \checkmark & \checkmark & Auto-Sklearn \cite{autosklearn15} & \checkmark & \checkmark &             \\
			Ridge                        & \checkmark & \checkmark & TPOT \cite{tpot20}                & \checkmark & \checkmark &             \\
			Multi-Layer Perception (MLP) & \checkmark & \checkmark &                                   &            &            &  		   \\
			\bottomrule          
		\end{tabular}%
	} \vspace{-3mm}
\end{table}
%

In \PaperAcronym, we evaluate the various error detection and repair methods in five scenarios. Table~\ref{tab:scenarios} summarizes the different scenarios defined in terms of the data version used for training and testing. In addition to the dirty and repaired versions of the data, we utilize the ground truth version to estimate the performance upper-bound. For instance, S1 involves training and testing the ML models on either the dirty or the repaired versions of the data. Conversely, S4 represents the optimal setting in which the ground truth version of the data is employed for training and testing the models. To capture the performance if optimal data cleaning can be achieved in only one phase, \PaperAcronym also considers S3 and S4 in which the ground truth (which simulates optimal data cleaning) is used for training and testing, respectively. Finally, S5 is mainly used with ML-oriented repair methods, which generate ML models as their output.\vspace{-2mm}
\begin{table}[htb]
	\caption{Evaluation scenarios}\vspace{-2mm}
	\label{tab:scenarios}
	\resizebox{\columnwidth}{!}{%
		\begin{tabular}{cccc||ccc}
			\toprule
			\multicolumn{1}{l}{}                  & \multicolumn{3}{c}{\textbf{Train}}                                                                                     & \multicolumn{3}{c}{\textbf{Test}}                                                                                      \\
			\multicolumn{1}{l}{\textbf{Scenario}} & \multicolumn{1}{l}{\textbf{Dirty}} & \multicolumn{1}{l}{\textbf{Repaired}} & \multicolumn{1}{l}{\textbf{Ground Truth}} & \multicolumn{1}{l}{\textbf{Dirty}} & \multicolumn{1}{l}{\textbf{Repaired}} & \multicolumn{1}{l}{\textbf{Ground Truth}} \\
			\midrule
			S1                                    & \cellcolor[rgb]{.58,.8,.85}\checkmark                         & \cellcolor[rgb]{.58,.8,.85}\checkmark                            &                                           & \cellcolor[rgb]{.58,.8,.85}\checkmark                         & \cellcolor[rgb]{.58,.8,.85}\checkmark                            &                                           \\
			S2                                    & \cellcolor[rgb]{.58,.8,.85}\checkmark                         & \cellcolor[rgb]{.58,.8,.85}\checkmark                            &                                           &                                    &                                       & \cellcolor[rgb]{.58,.8,.85}\checkmark                                \\
			S3                                    &                                    &                                       & \cellcolor[rgb]{.58,.8,.85}\checkmark                                & \cellcolor[rgb]{.58,.8,.85}\checkmark                         & \cellcolor[rgb]{.58,.8,.85}\checkmark                            &                                           \\
			S4                                    &                                    &                                       & \cellcolor[rgb]{.58,.8,.85}\checkmark                                &                                    &                                       & \cellcolor[rgb]{.58,.8,.85}\checkmark                                \\
			S5                                    &                                    & \cellcolor[rgb]{.58,.8,.85}\checkmark                            &                                           & \cellcolor[rgb]{.58,.8,.85}\checkmark                         &                                       &   
			\\ \bottomrule                                       
		\end{tabular} \vspace{-3mm}
	}
\end{table}

In general, the obtained results in each scenario may vary owing to ML randomness. Therefore, it is crucial to scrutinize the results obtained in each scenario before drawing conclusions. In this regard, \PaperAcronym leverages A/B hypothesis testing to improve our confidence in the interpretation of the obtained results. Generally, an A/B hypothesis test can be exploited to quantify how likely it is to observe two data samples given the assumption that the samples have the same distribution \cite{janez06}. In \PaperAcronym, the A/B hypothesis tests can statistically predict whether an ML model behaves similarly in different scenarios. An initial step in the test procedure is to clearly define the null hypothesis $H_{0}$ and the alternative hypothesis $H_a$. In \PaperAcronym, the null hypothesis $H_{0}$ states that an ML model has circa the same performance in two different scenarios, e.g., S1 and S4, regardless of the data version. Conversely, the alternative hypothesis $H_a$ states that the ML model behaves differently in S1 and S4. The statistical significance is estimated in terms of the \textit{p-value}, i.e., the probability that an observed difference between S1 and S4 could have occurred by random chance. To estimate the p-value, we utilize the non-parametric Wilcoxon signed-rank test \cite{janez06}. The main advantage of such a test lies in making no assumptions about the sampling distributions, e.g., being Gaussian. Specifically, we opted for the \textit{two-tailed} version of the test, since it is not a priori known whether the discrepancy between the results of S1 and S4 will be in favor of S1 or S4. After computing the p-value, we can compare it with the significance level $\alpha$ to estimate whether to reject the null hypothesis $H_0$. In particular, we can reject the null hypothesis $H_0$ if \textit{p-value} $ < \alpha$. Otherwise, we conclude that the obtained results in the compared scenarios support the alternative hypothesis $H_{a}$.
%
\vspace{-3mm}\section{Benchmark Data}\label{sec:data}
\begin{table*}[t]
	\caption{Dataset characteristics\vspace{-2mm}}
	\label{tab:datasets}
	\resizebox{\textwidth}{!}{%
		\begin{tabular}{lcccccccc}
			\toprule
			\multicolumn{1}{c}{\textbf{Dataset}} & \textbf{\# Rows} & \textbf{\# Columns} & \textbf{\# Numerical} & \textbf{\# Categorical} & \textbf{Error Rate} & \textbf{Errors}             & \textbf{Domain} & \textbf{ML Tasks} \\
			\midrule
				Beers \cite{beers17}                 & 2410            & 11                 & 6                    & 5                      & 0.16                    & MVs, Rule Violations, Typos & Business        & C                 \\
				Citation \cite{magellandata}         & 5005            & 2                  & 1                    & 1                      & 0.2                     & Duplicates, Mislabels      & Research        & C                 \\
				Adult \cite{adult96}                 & 45223           & 15                 & 7                    & 8                      & 0.58                    & Rule Violations, Outliers   & Social          & C                 \\
				Breast Cancer \cite{ucidata19}       & 700             & 12                 & 12                   & 0                      & 0.08                    & MVs, Typos, Outliers        & Healthcare      & C                 \\
				Smart Factory \cite{smartfactory22}  & 23645           & 19                 & 19                   & 0                      & 0.153                   & MVs, Outliers               & Manufacturing   & C                 \\
				Nasa \cite{nasa22}                   & 1504            & 6                  & 6                    & 0                      & 0.08                   & MVs, Outliers               & Manufacturing   & R                 \\
				Bikes \cite{bike13}                  & 17378           & 16                 & 16                   & 0                      & 0.1                    & Rule Violations,  outliers   & Business        & R                 \\
				Soil Moisture \cite{soilmoisture18}  & 679             & 129                & 129                  & 0                      & 0.01                   & MVs, Outliers                & Agriculture     & R                 \\
				3D Printer \cite{print3d}            & 50              & 12                 & 10                   & 2                      & 0.05                    & Duplicates, MVs, Implicit MVs     & Manufacturing   & R                 \\
				Mercedes \cite{mercedes17}           & 4210            & 378                & 370                  & 8                      & 0.05                    & Outliers, MVs, Implicit MVs     & Manufacturing   & R                 \\
				Water \cite{water93}                 & 527             & 38                 & 38                   & 0                      & 0.14           & Outliers, Implicit MVs             & Manufacturing   & UC                \\
				HAR \cite{har13}                     & 70000          & 4                  & 3                    & 1                      &  0.13                       &  Outliers, MVs         & Wearables       & UC                \\
				Power \cite{power12}                 & 1456            & 24                 & 24                   & 0                      & 0.037                     &  Typos, MVs, Implicit MVs  & Energy          & UC                \\
				Soccer \cite{soccer16}               & 180228          & 44                 & 40                   & 4                      & 0.27                      & Rule violations, outliers, MVs, Implicit MVS                          & Business        & --                \\
				\bottomrule              
			\end{tabular}%
		} \vspace{-3mm}
	\end{table*}
In this section, we elaborate on the real-world datasets and how to inject errors into them. To systematically select appropriate datasets for running the benchmark, it is necessary to define a set of requirements in light of the objectives of \PaperAcronym. Such objectives revolve around estimating the performance of each detector/repair method separately without considering the subsequent stages of the ML pipeline and examining the impact of these methods on the performance of the downstream predictive models in different scenarios. Accordingly, the datasets, involved in \PaperAcronym, have to fulfill the following conditions: (1) the existence of a complete and clean ground truth version; (2) the existence of associated predictive tasks, e.g., classification, regression, or clustering; (3) the existence of different data types, e.g., categorical, numerical, and/or text; and (4) the existence of different realistic error profiles. In fact, we collected two datasets, i.e., \textit{Beers} and \textit{Citation}, that satisfy these conditions. However, it is not straightforward to find other datasets satisfying our requirements. 

To overcome such a challenge, we opted for injecting different types of errors into a set of real-world datasets. Consequently, we can predominantly control the experiments through obtaining several versions of each dataset along with the ground truth. In addition to the aforementioned requirements, we are also eager to select datasets covering multiple application domains, e.g., business, medical, and industrial, where the data originated in different domains usually have different characteristics. Moreover, we selected datasets of different sizes, ranging from a couple of hundred samples to a couple of hundred thousands, to precisely test the efficiency of the various data cleaning methods. Table~\ref{tab:datasets} summarizes the examined datasets and the characteristics of their ground truth. 

To inject errors into the real-world datasets, \PaperAcronym leverages the BART tool \cite{bart15} which provides a systematic control over the amount of errors and how hard these errors are to be repaired. To inject errors using BART, we use a set of denial constraints to generate different attribute and class errors, such as rule violation, outliers, nulls, duplicates, and mislabels. Furthermore, we also employ a Python library, referred to as \textit{error generator}, to generate highly realistic errors \cite{generator2018}. Examples of such error are typos based on keyboards, implicit missing values, Gaussian noise, and value swapping. To automatically generate FD rules, \PaperAcronym leans on the FDX profiler \cite{profiler20} which formulates the task of learning functional dependencies as a sparse regression problem. After generating the FD rules, we manually convert them into denial constraints to be used with BART and the rule-based error detection and repair methods, e.g., HoloClean and NADEEF. 
\vspace{-3mm}\section{Performance Evaluation}\label{sec:evaluation}
%
%
In this section, we assess the effectiveness and efficiency of various error detection and repair methods. We first describe the setup of our evaluations, before discussing the results and the lessons learned throughout this study.
%
\vspace{-3mm}\subsection{Experimental Setup}
%
In \PaperAcronym, we utilize several metrics to assess the quality of results at different stages of a typical ML pipeline. In the error detection phase, we leverage precision, recall, F1 score, IoU, and runtime to evaluate the effectiveness and efficiency. In this context, the precision $P$ denotes the fraction of relevant instances, e.g., actual erroneous cells, among the detected instances, i.e. $P = \tfrac{t_{p}}{t_{p} + f_{p}}$ where $t_{p}$ and $f_{p}$ are true positives and false positives, receptively. The recall $R$ is defined as the fraction of erroneous instances that are detected, i.e. $R = \tfrac{t_{p}}{t_{p} + f_{n}}$ where $f_{n}$ denotes false negatives. The F1 score denotes the harmonic mean of precision and recall where $F1 = 2 . \tfrac{P . R}{P + R}$. Such metrics define the quality of detection relative to the ground truth. Nevertheless, it is also significant to identify the similarities between the detected erroneous cells obtained by different detection methods. Hence, we adopt the \textit{Intersection over Union} (IoU) metric. Assume that $N_{a}, N_{b}$ are the detected erroneous cells by detectors $a$ and $b$. Accordingly, the IoU metric between detectors $a$ and $b$ is computed as $\tfrac{|N_a \cap N_b|}{|N_a| + |N_b| - |N_a \cap N_b|}$. For these computations, we consider only the true positives, since the false positives may lead to misleading results. Finally, the runtime is the time elapsed while traversing an entire dataset to identify the erroneous cells.

In the error repair phase, we differentiate between the numerical and the categorical attributes. For the former type, we employ the root mean square error (RMSE) as a distance measure between the repaired values and their ground truth. In fact, some error types, e.g., typos and outliers, turn the numerical instances into categorical ones. To properly compute the RMSE metric, we filtered out the transformed instances which have not been detected and repaired. For the latter data type, we employ precision, recall, and F1 measures. In this context, the precision is defined as the fraction of correctly repaired data errors relative to the number of repaired data errors. The recall is defined as the fraction of correctly repaired data errors relative to the number of data errors. We also report the runtime to quantify the time elapsed while generating the repairs. In the ML modeling phase, we utilize precision, recall, and F1 measures for the classification models. For clustering methods which require the number of clusters $k$ as an input, we utilize the Silhouette index to find a well estimate for the value of $k$. For the A/B statistical test, we set the Type I error rate $\alpha$ to 0.05. All experiments have been repeated ten times with different random seeds that control the train-test split, and the means of the ten runs are reported. 
We run all the experiments on an Ubuntu 16.04 LTS machine with 32 2.60 GHz cores and 264 GB memory. Due to space constraints, the results of many experiments have been omitted.

\vspace{-3mm}\subsection{Error Detection}
\begin{figure*}[htbp] 
	\centering
	\subfloat[Beers-Accuracy]{\label{fig:beers_detect_acc}\includegraphics[width=0.21\textwidth]{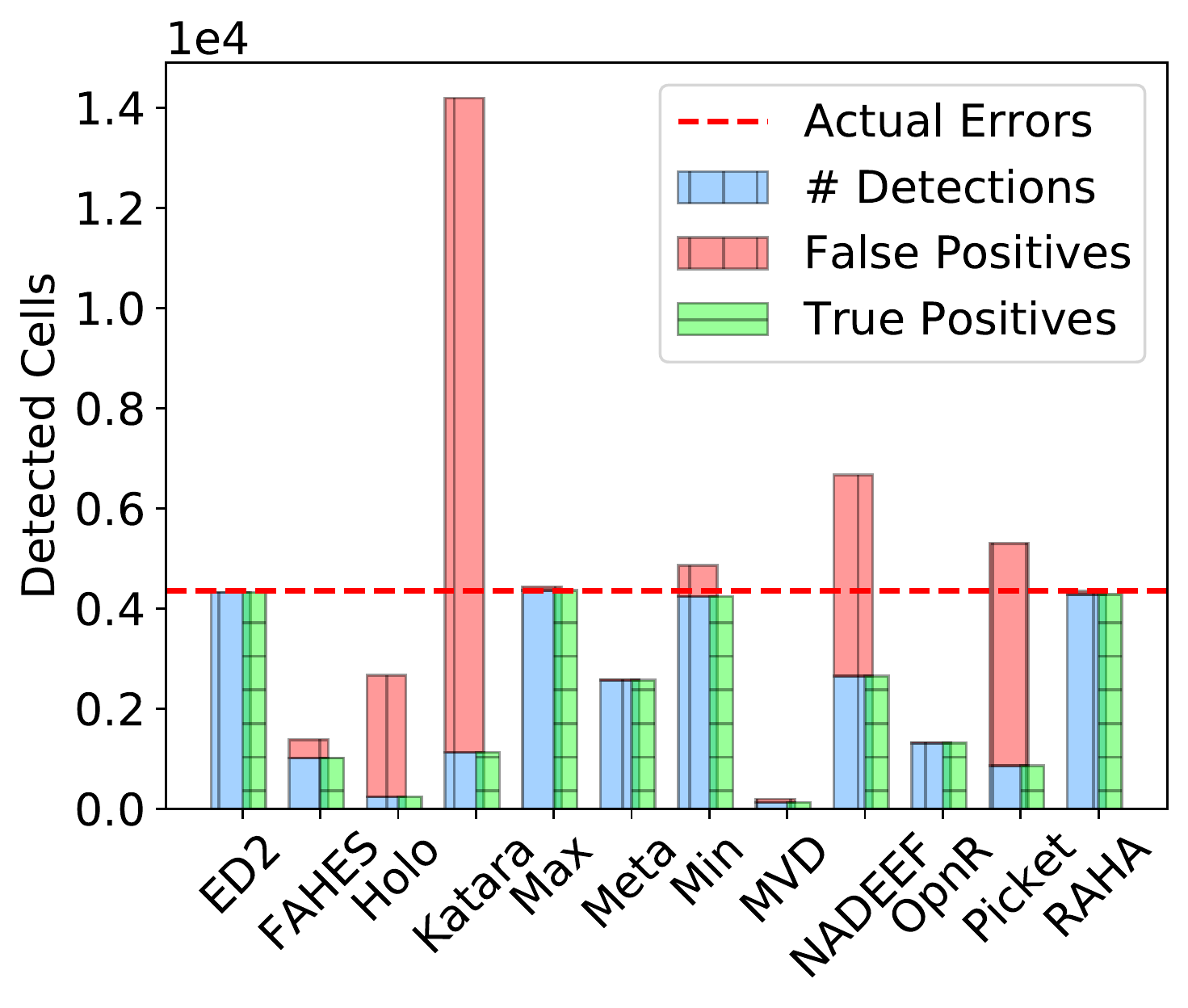}}
	\subfloat[Beers-IoU]{\label{fig:beers_iou}\includegraphics[width=0.19\textwidth]{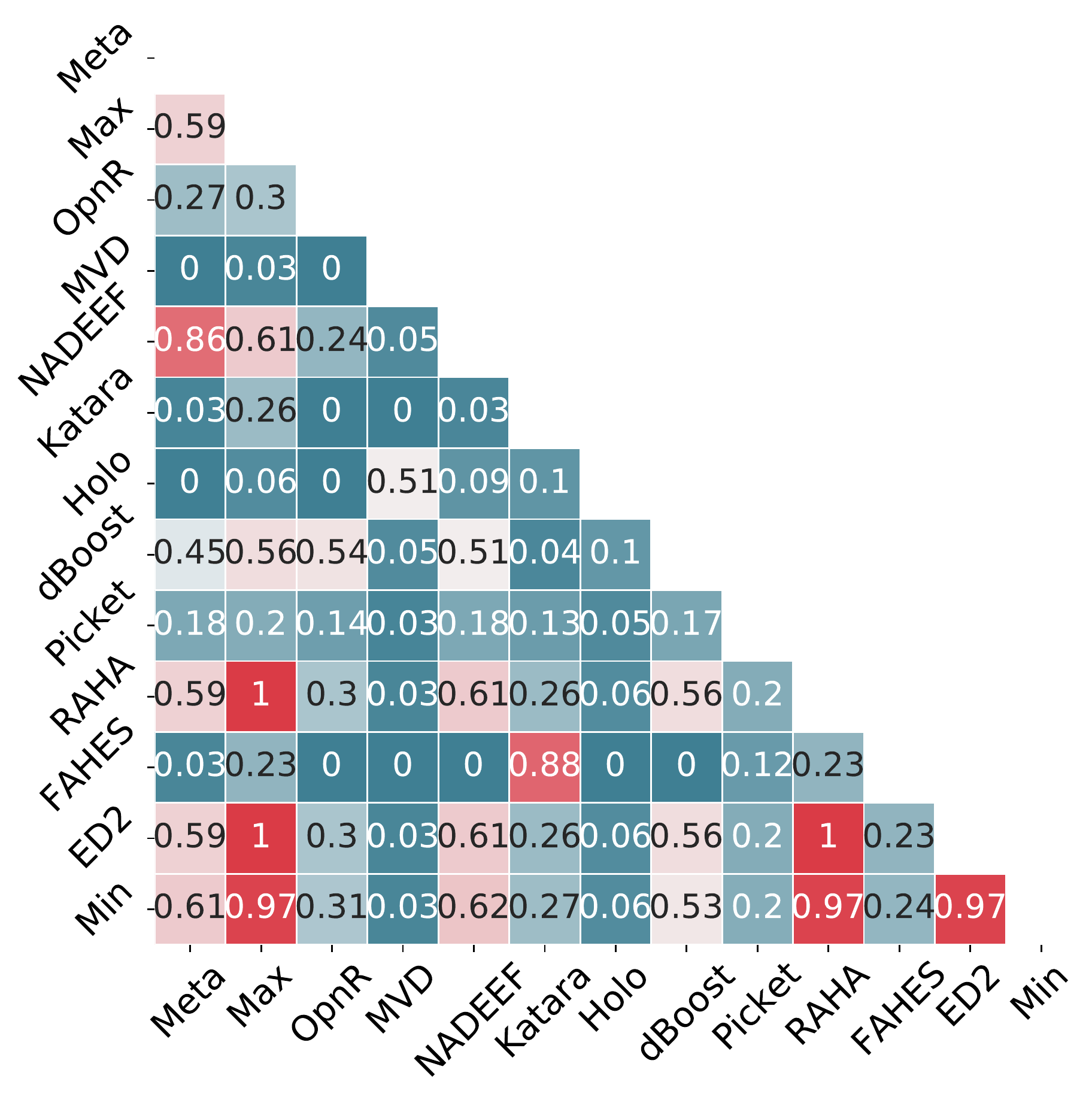}} 
	\subfloat[Beers-Runtime]{\label{fig:beers_detect_time}\includegraphics[width=0.21\textwidth]{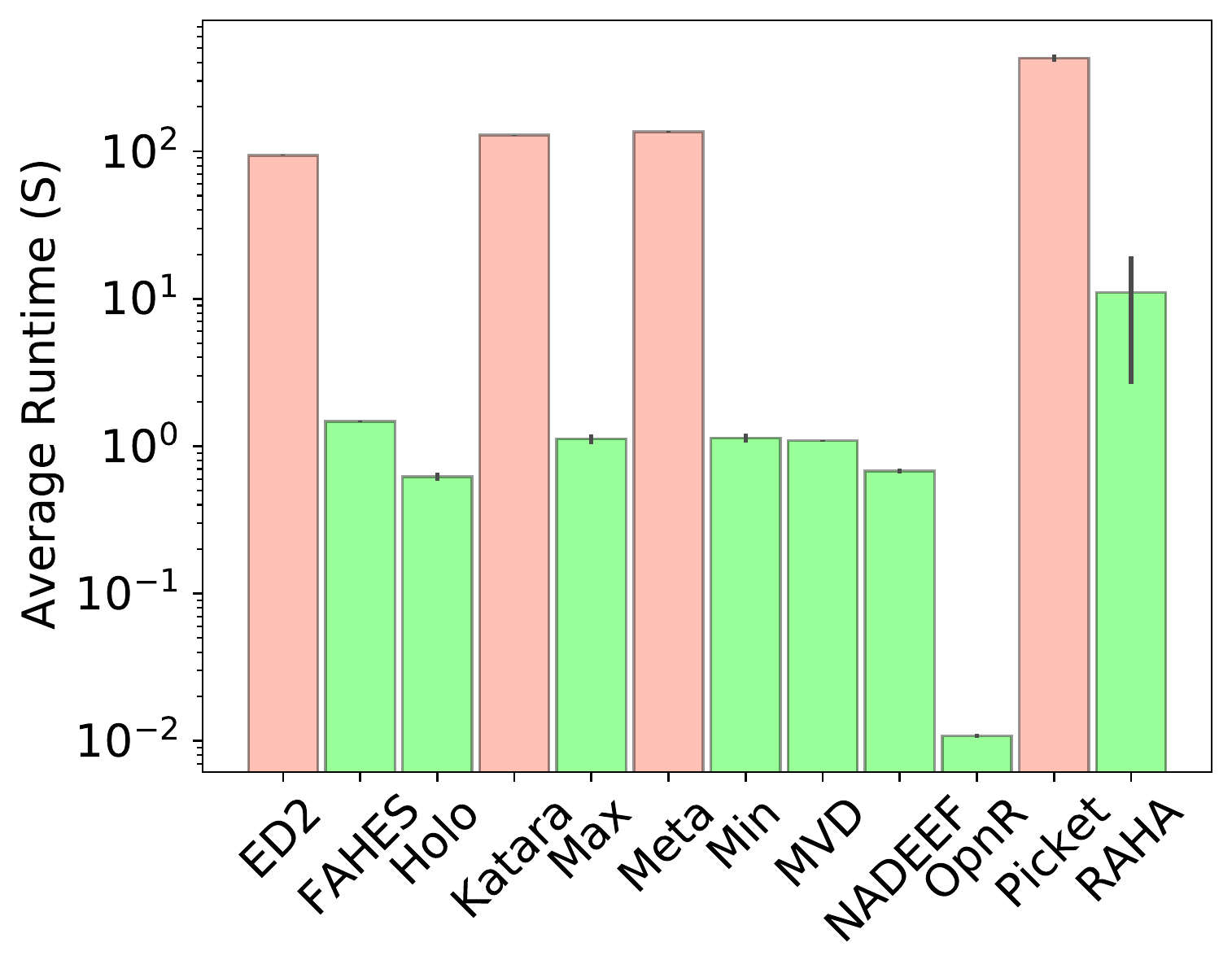}}
	\subfloat[Citation-Accuracy]{\label{fig:citation_detect_acc}\includegraphics[width=0.22\textwidth]{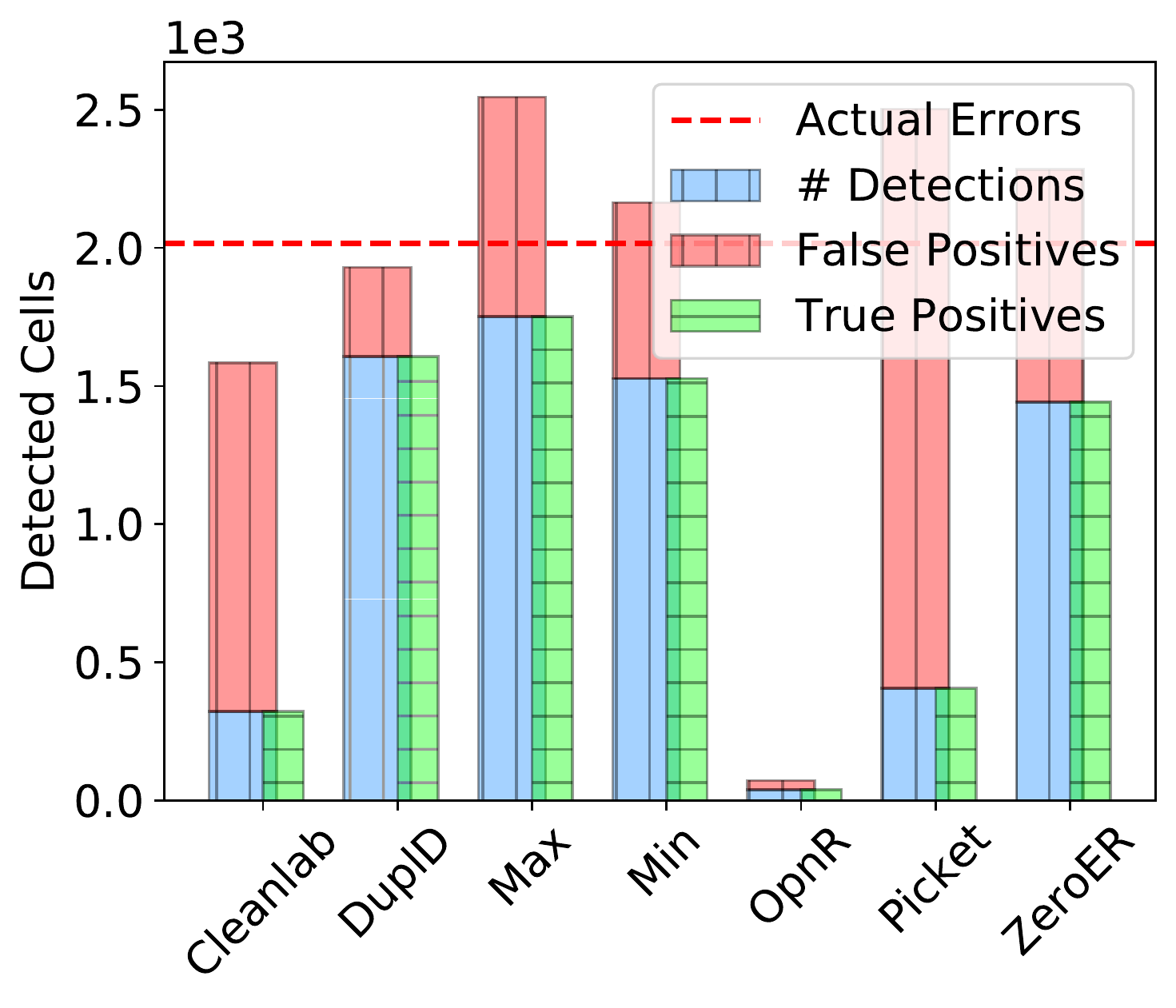}} 
	\subfloat[Citation-IoU]{\label{fig:citation_iou}\includegraphics[width=0.19\textwidth]{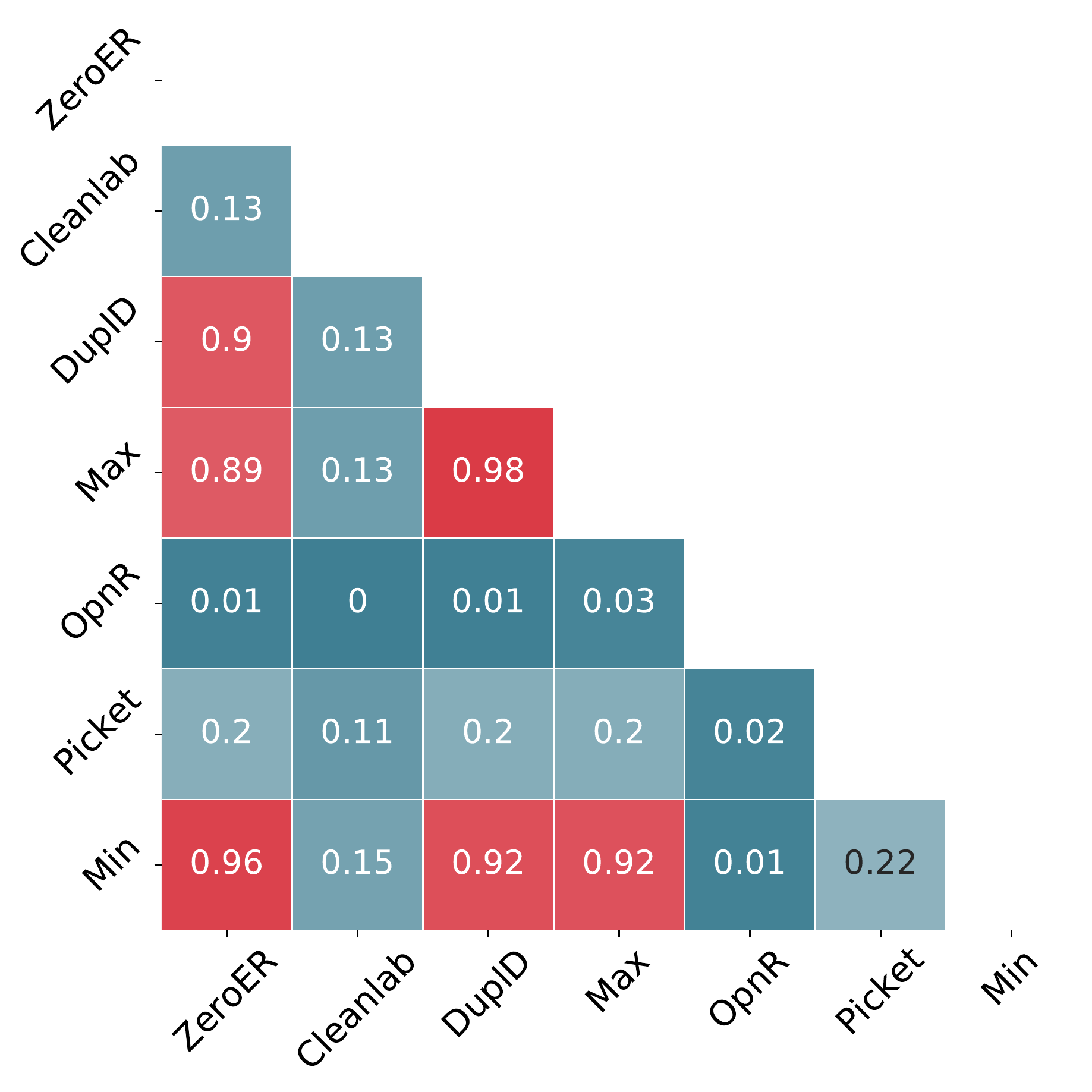}} \hfill
	%
	\subfloat[Adult-Accuracy]{\label{fig:adult_detect_acc}\includegraphics[width=0.21\textwidth]{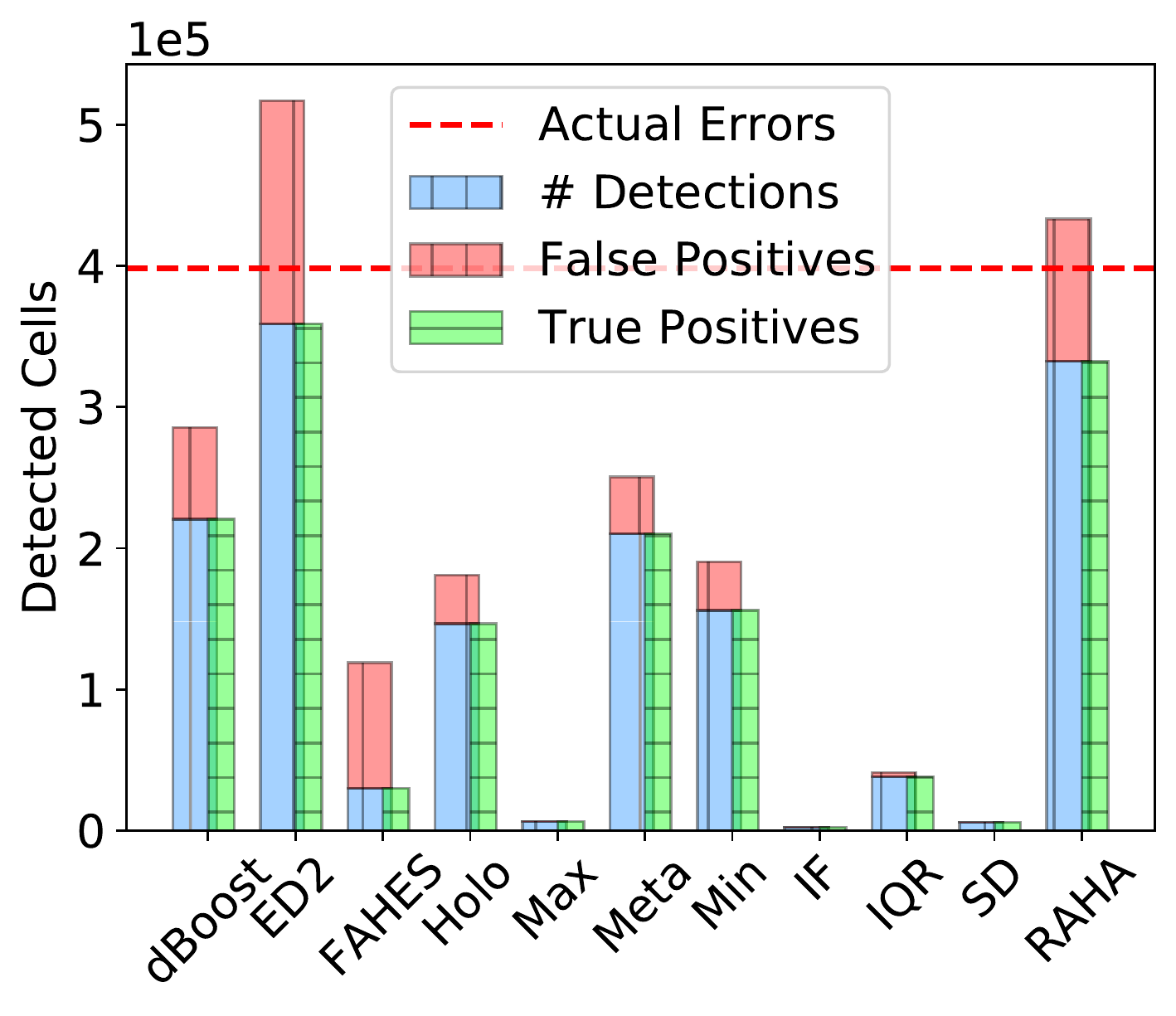}}
	%
	%
	\subfloat[Adult-Runtime]{\label{fig:adult_detect_time}\includegraphics[width=0.21\textwidth]{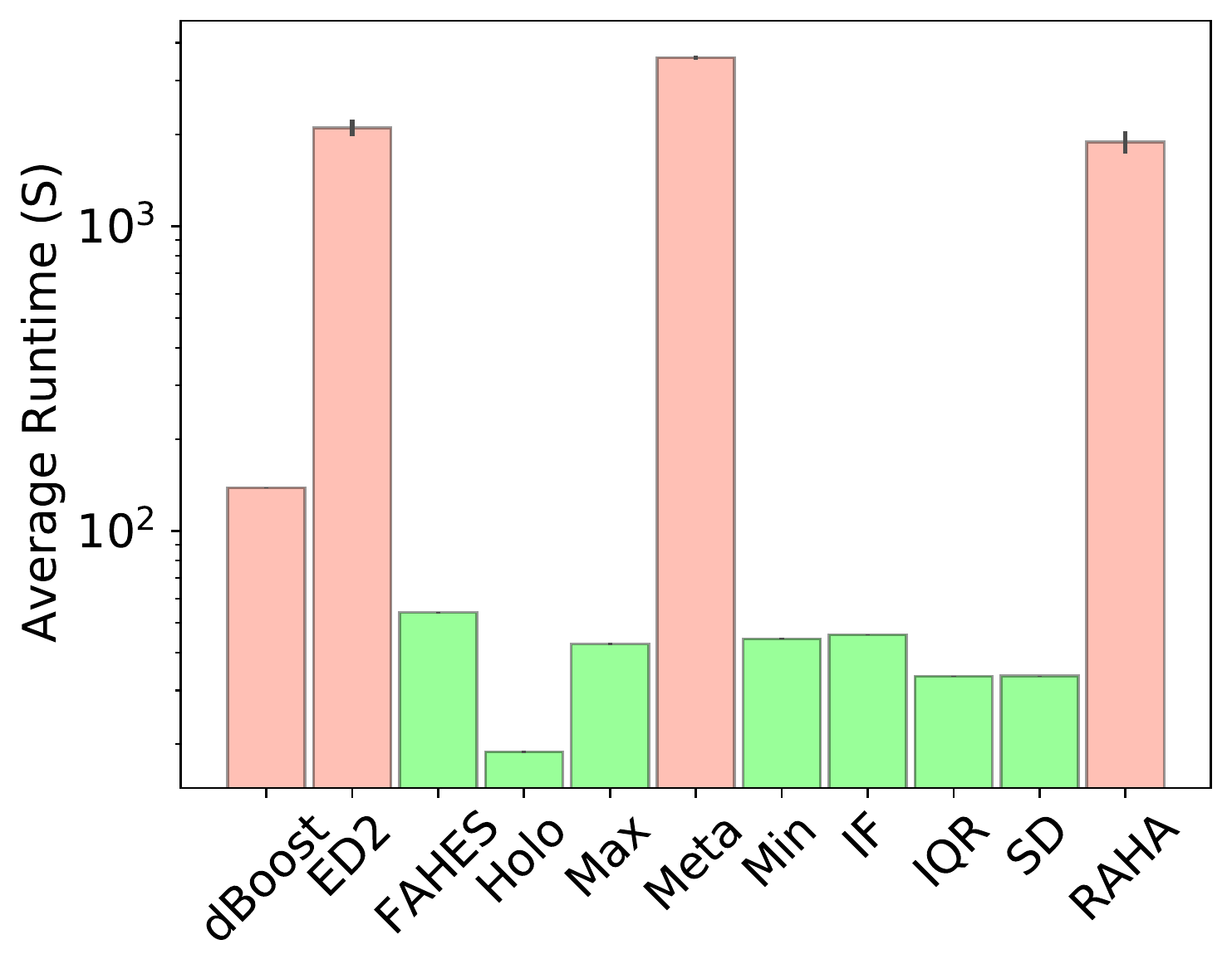}} 
	%
	%
	\subfloat[SmartFactory-Accuracy]{\label{fig:sf_detect_acc}\includegraphics[width=0.21\textwidth]{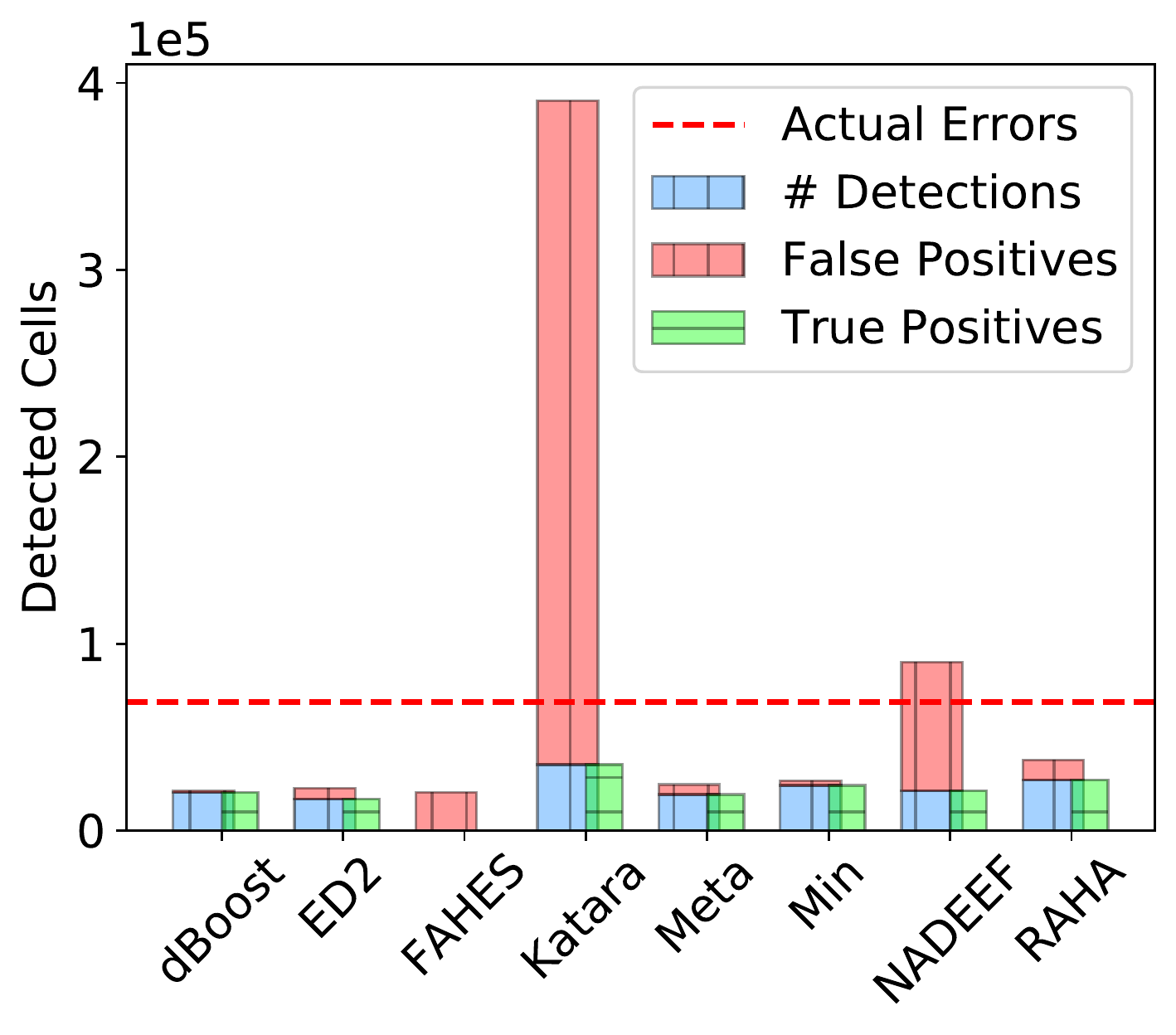}}
	\subfloat[SmartFactory-IoU]{\label{fig:sf_iou}\includegraphics[width=0.18\textwidth]{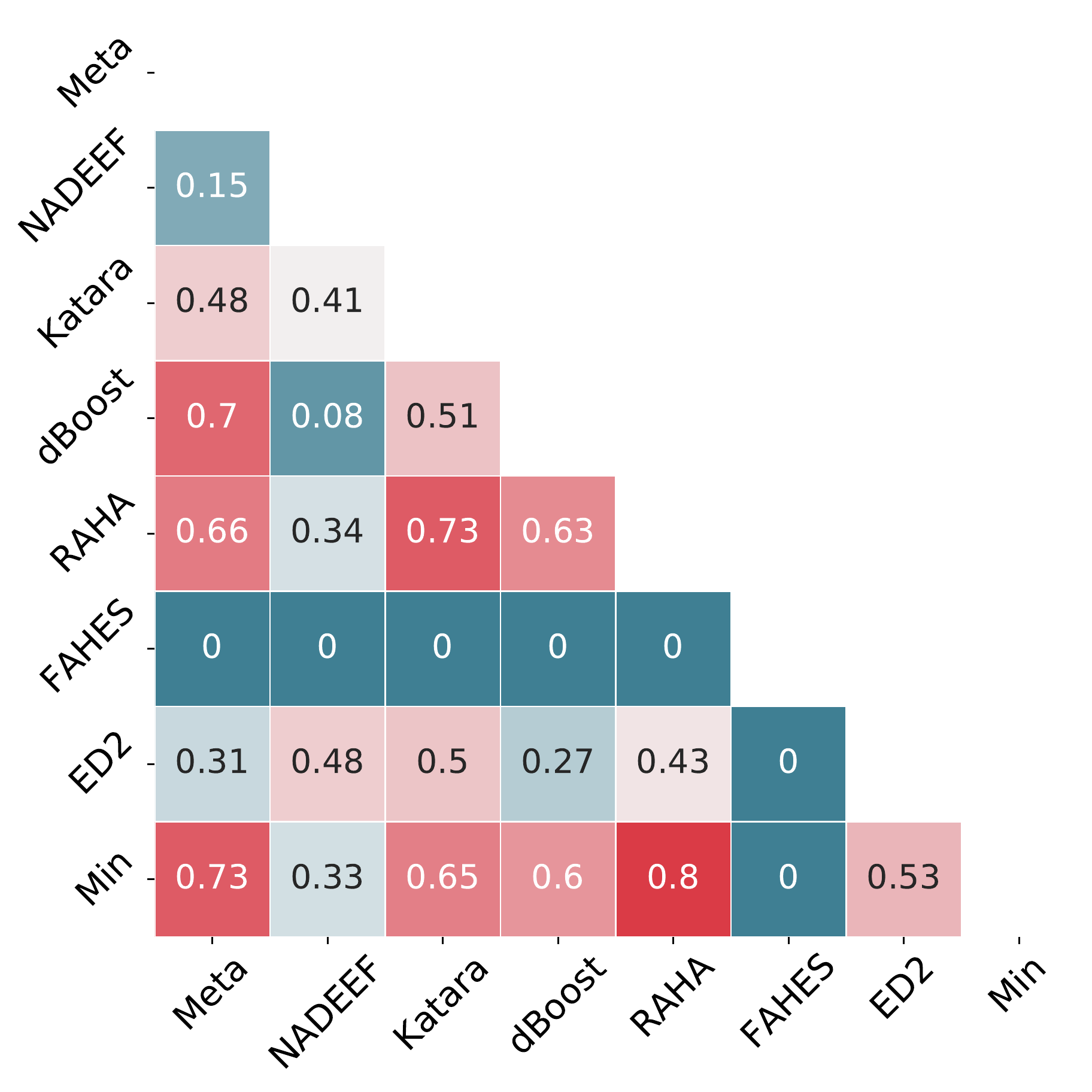}}
	\subfloat[SmartFactory-Runtime]{\label{fig:sf_detect_time}\includegraphics[width=0.21\textwidth]{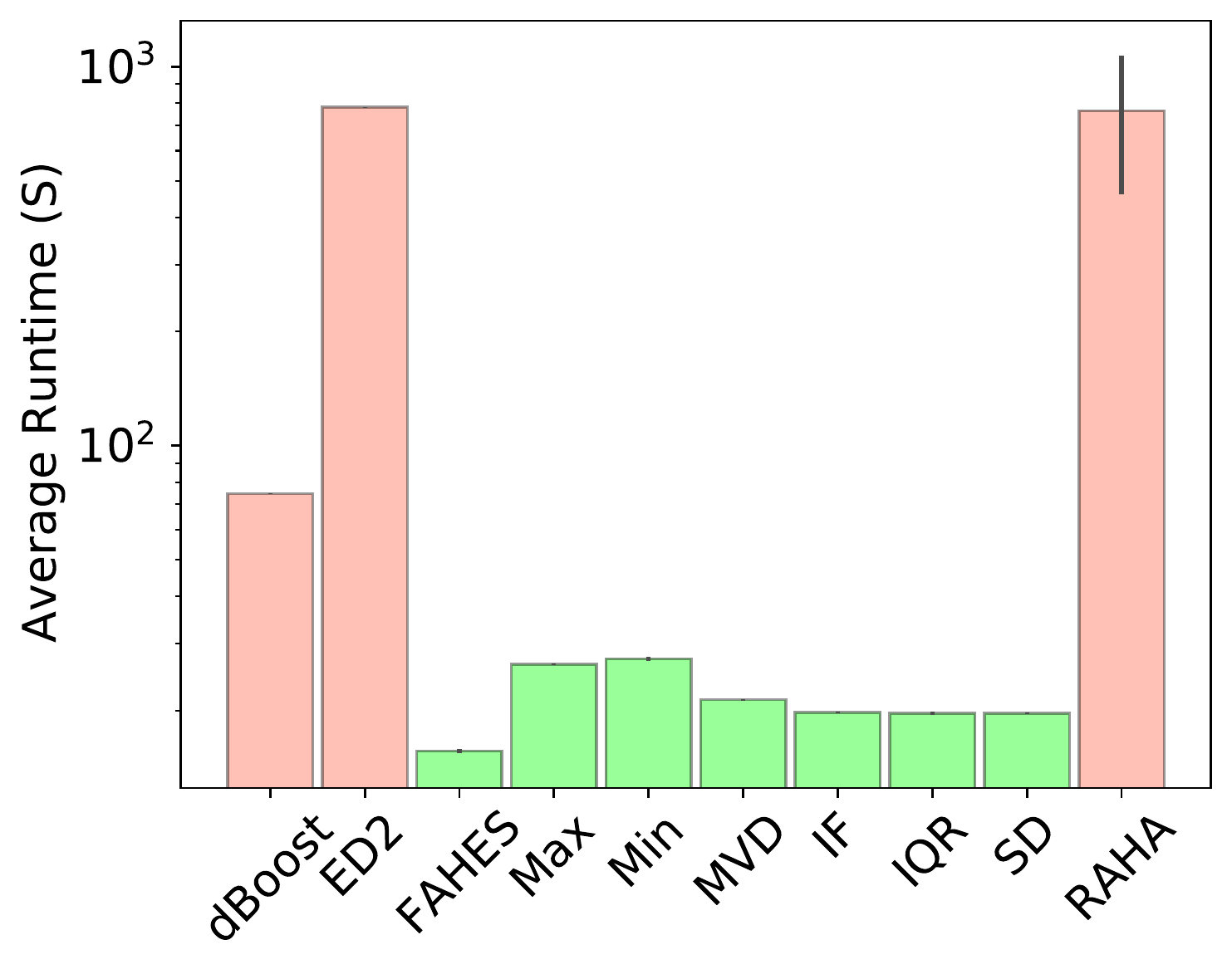}} \hfill	
	\subfloat[Nasa-Accuracy]{\label{fig:nasa_detect_acc}\includegraphics[width=0.21\textwidth]{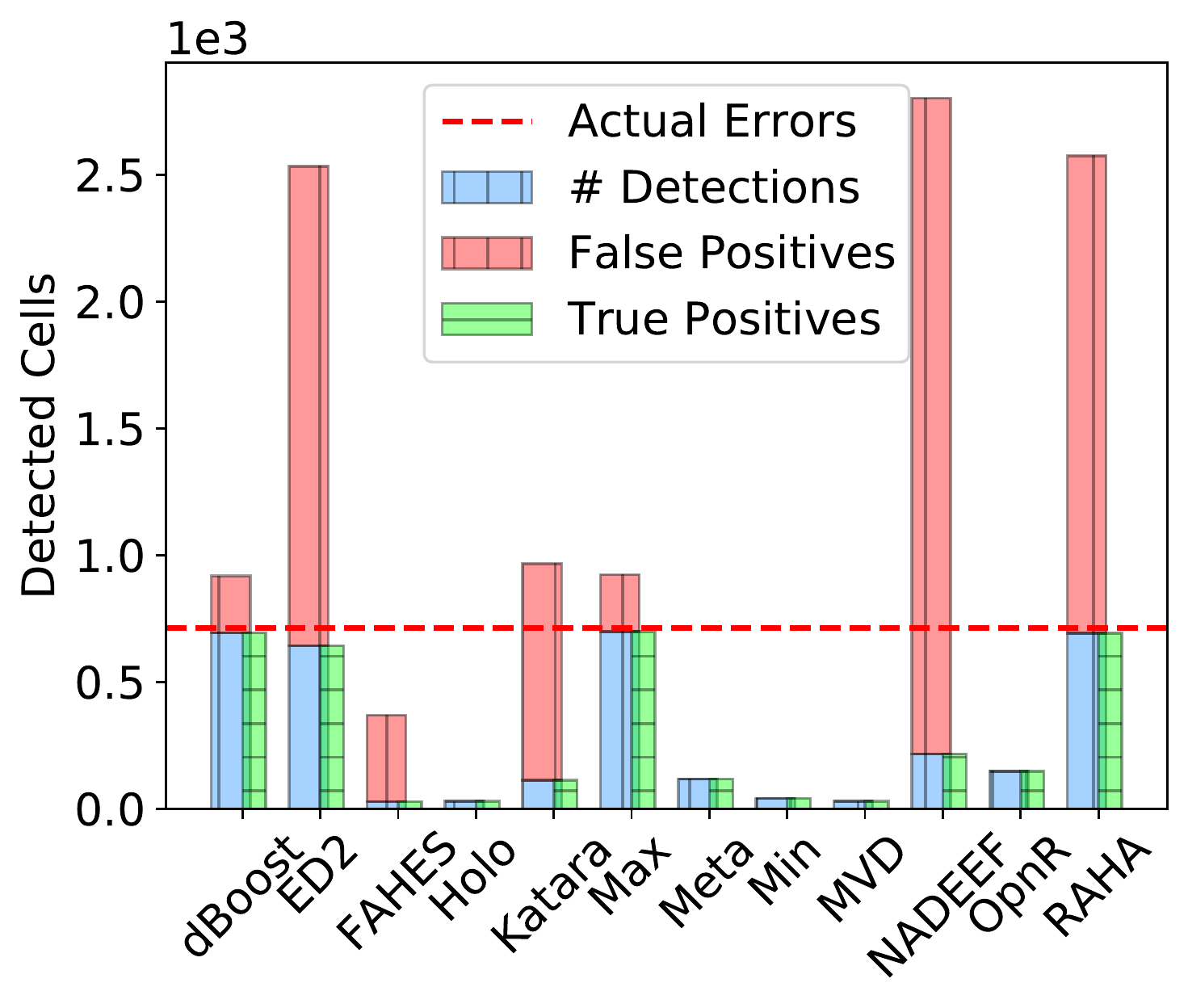}} 
	\subfloat[Nasa-IoU]{\label{fig:nasa_iou}\includegraphics[width=0.18\textwidth]{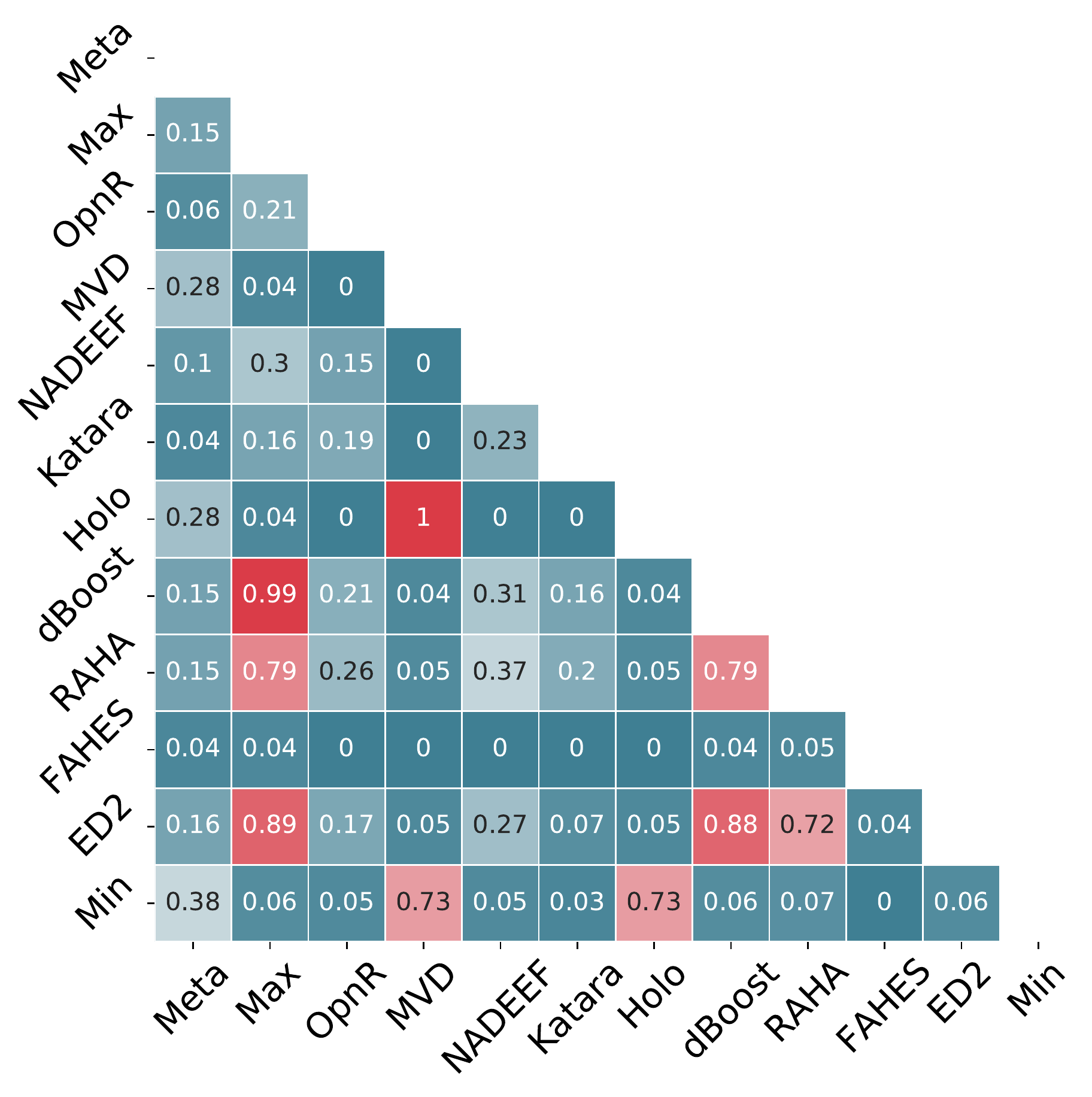}}
	\subfloat[Nasa-Runtime]{\label{fig:nasa_detect_time}\includegraphics[width=0.21\textwidth]{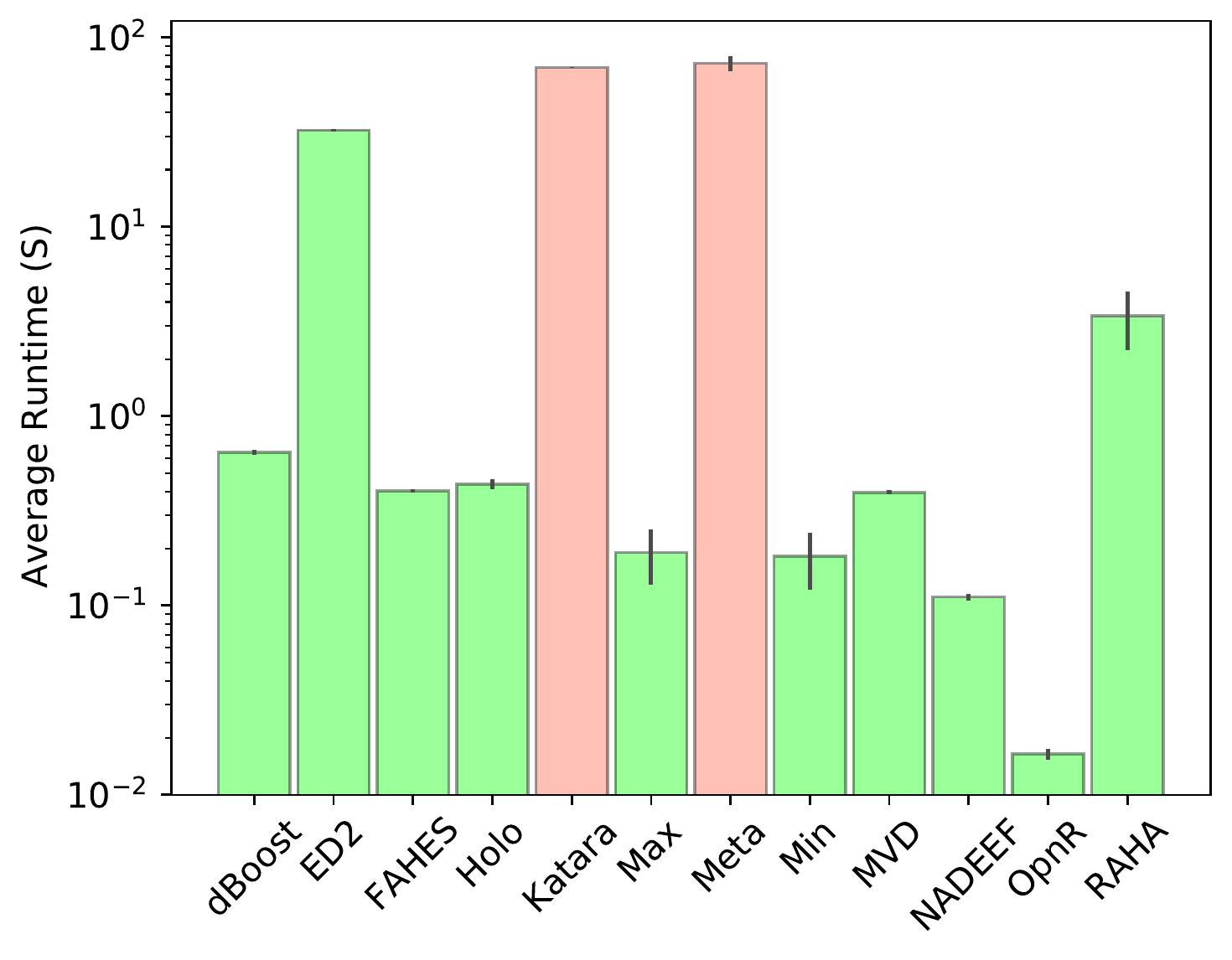}}
	%
	\subfloat[Bikes-Accuracy]{\label{fig:bike_detect_acc}\includegraphics[width=0.21\textwidth]{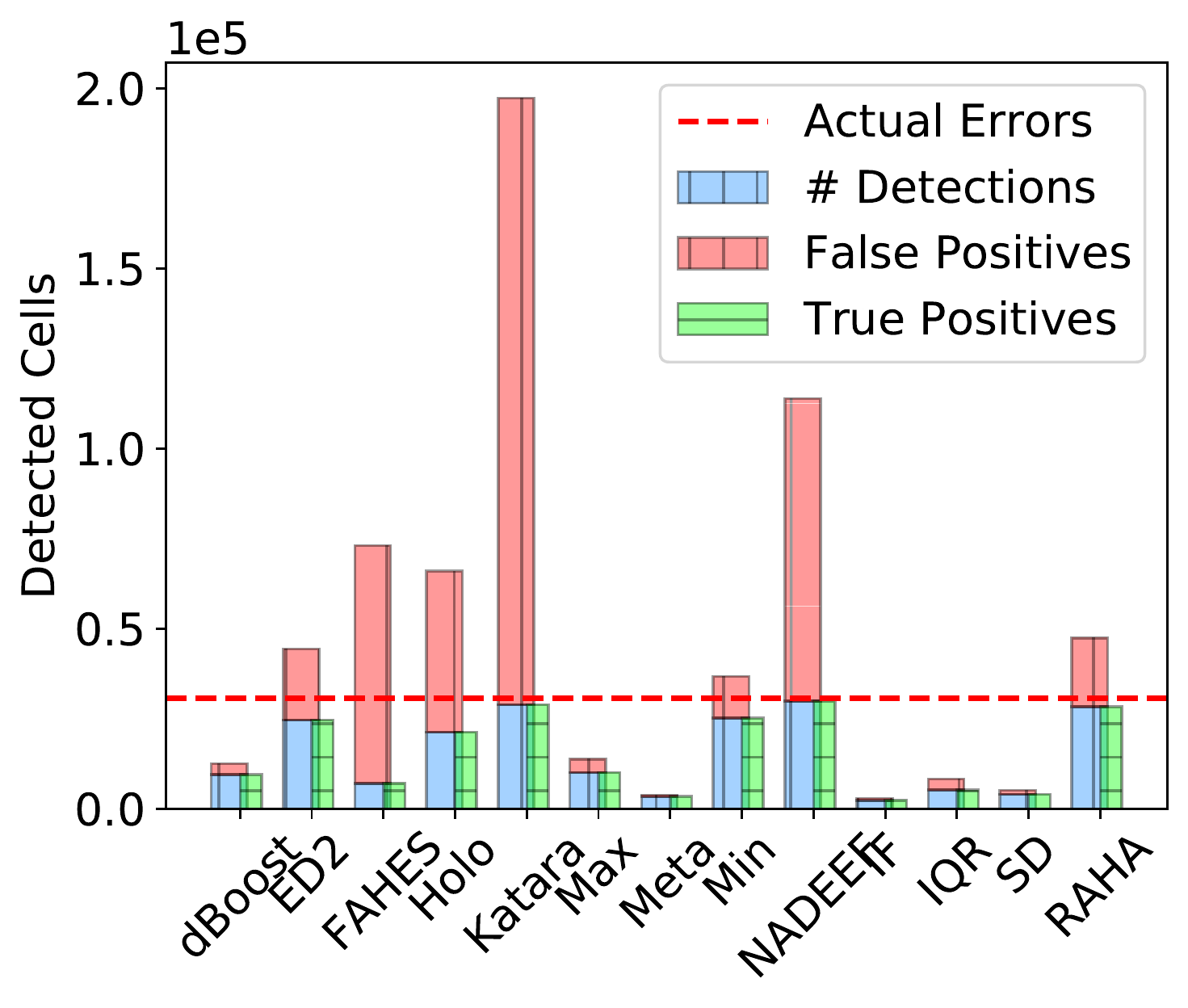}} 
	%
	%
	\subfloat[Bikes-Runtime]{\label{fig:bike_detect_time}\includegraphics[width=0.21\textwidth]{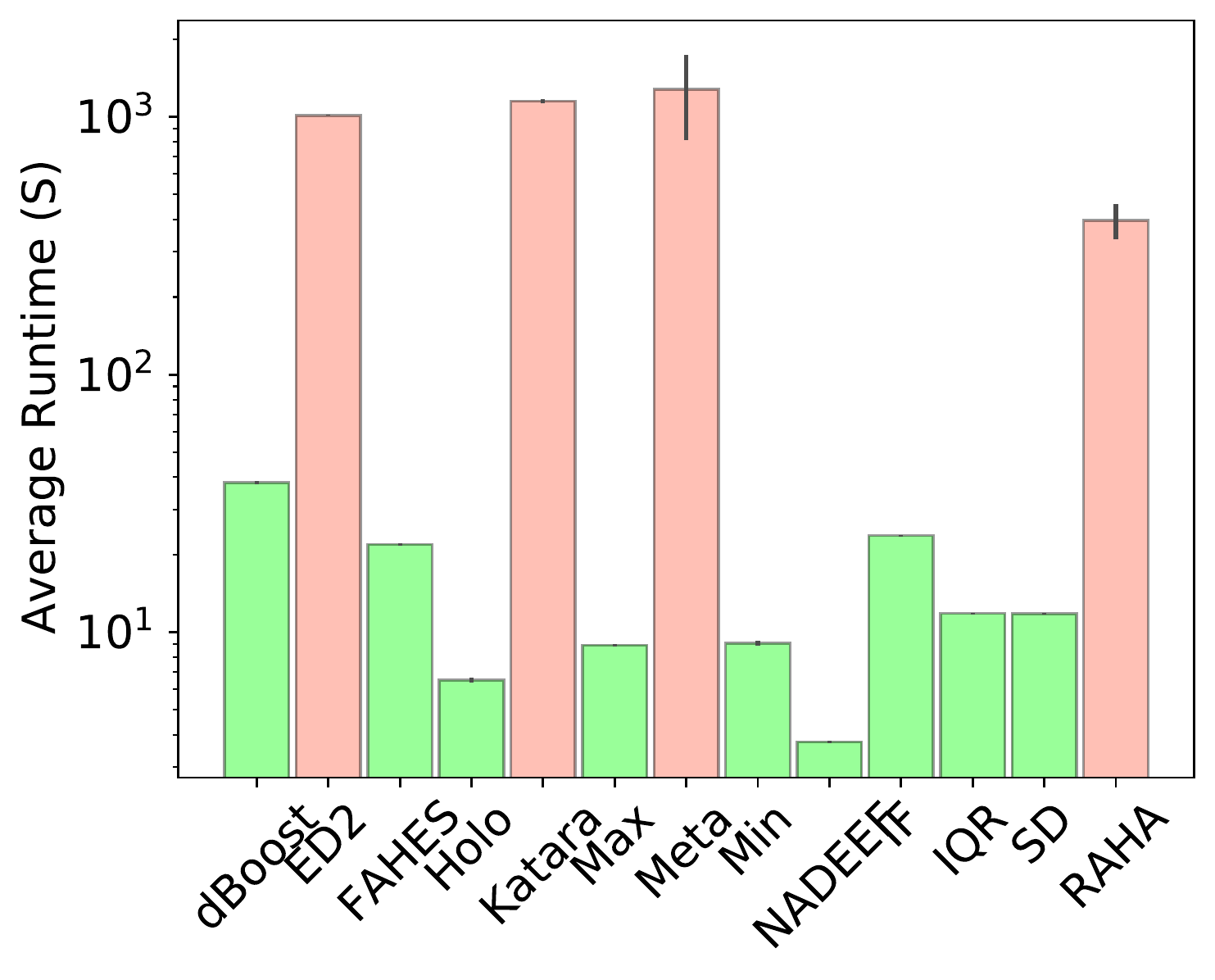}} \hfill
	\subfloat[Water-Accuracy]{\label{fig:water_detect_acc}\includegraphics[width=0.21\textwidth]{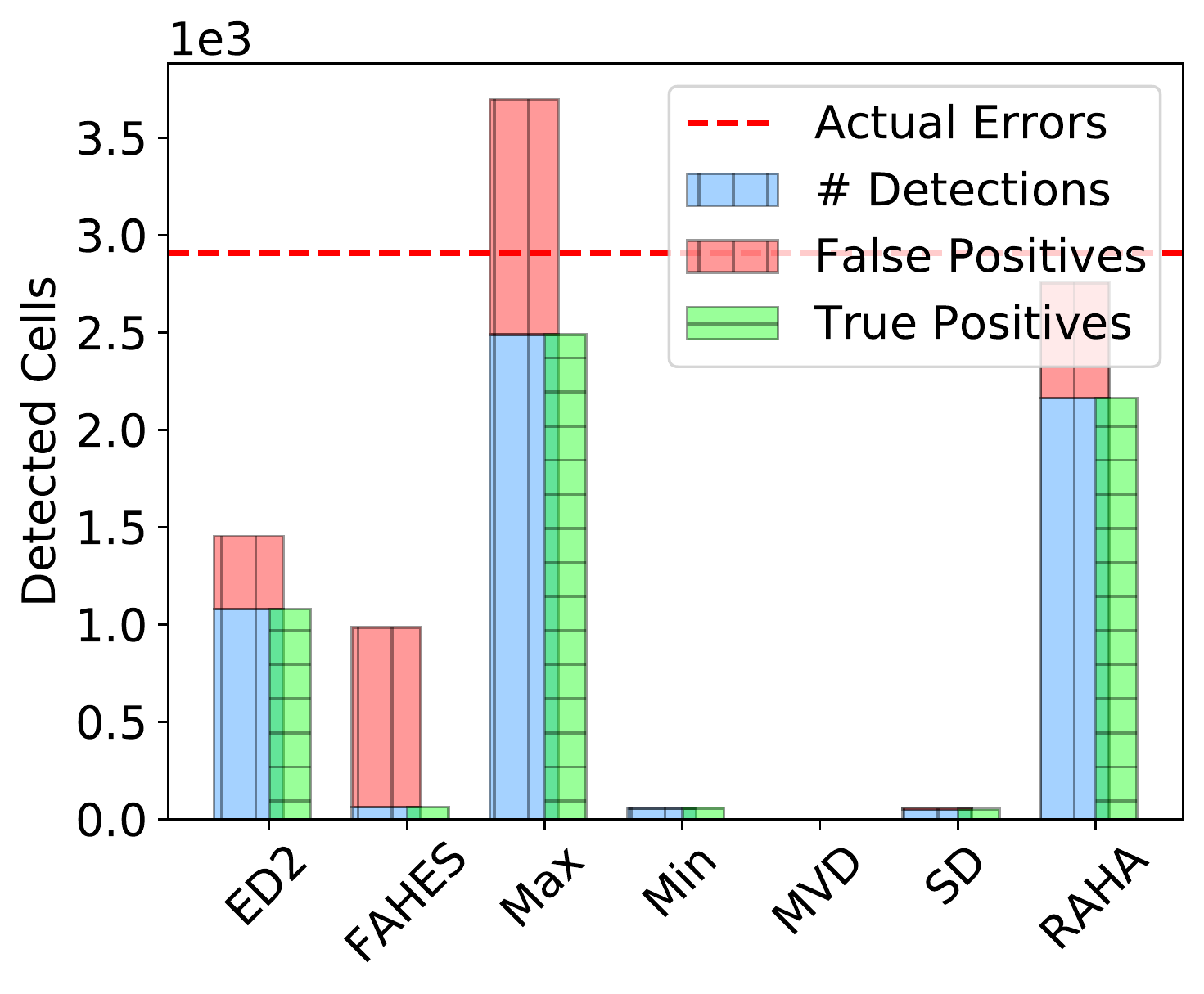}} 
%
%
%
	\subfloat[Power-Accuracy]{\label{fig:power_detect_acc}\includegraphics[width=0.21\textwidth]{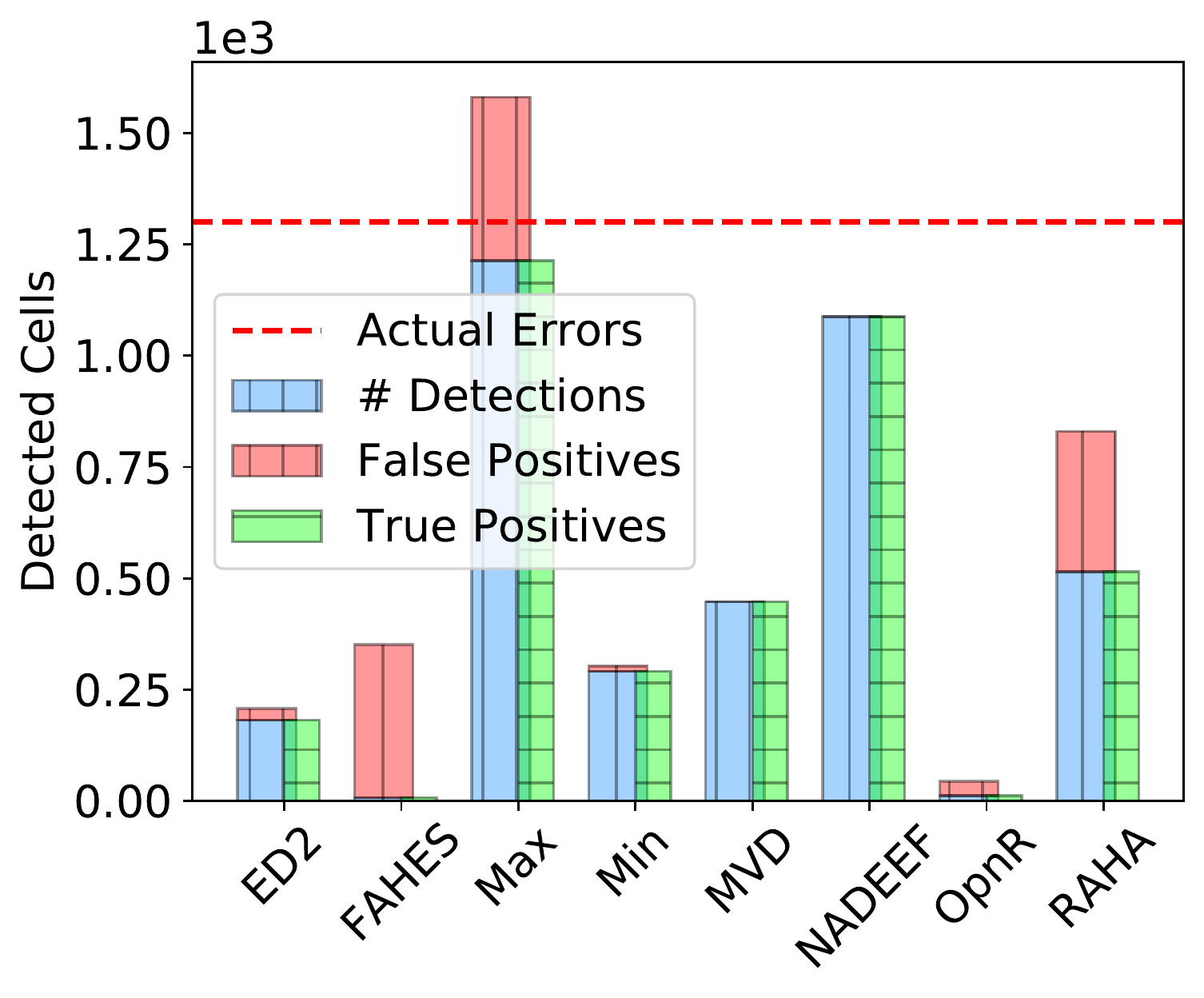}} 
%
%
%
	\subfloat[HAR-Accuracy]{\label{fig:har_detect_acc}\includegraphics[width=0.21\textwidth]{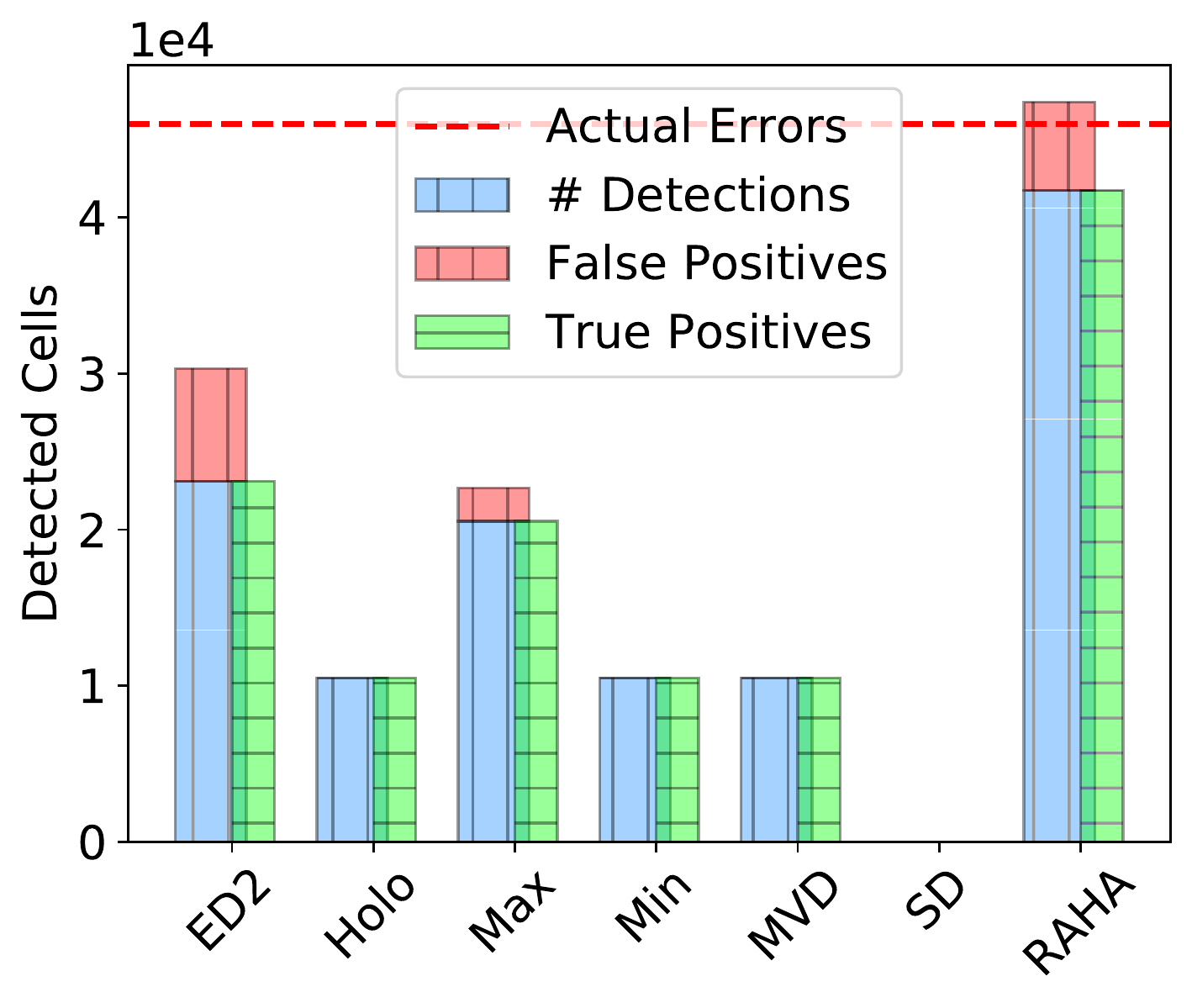}} 
	\subfloat[HAR-IoU]{\label{fig:har_iou}\includegraphics[width=0.18\textwidth]{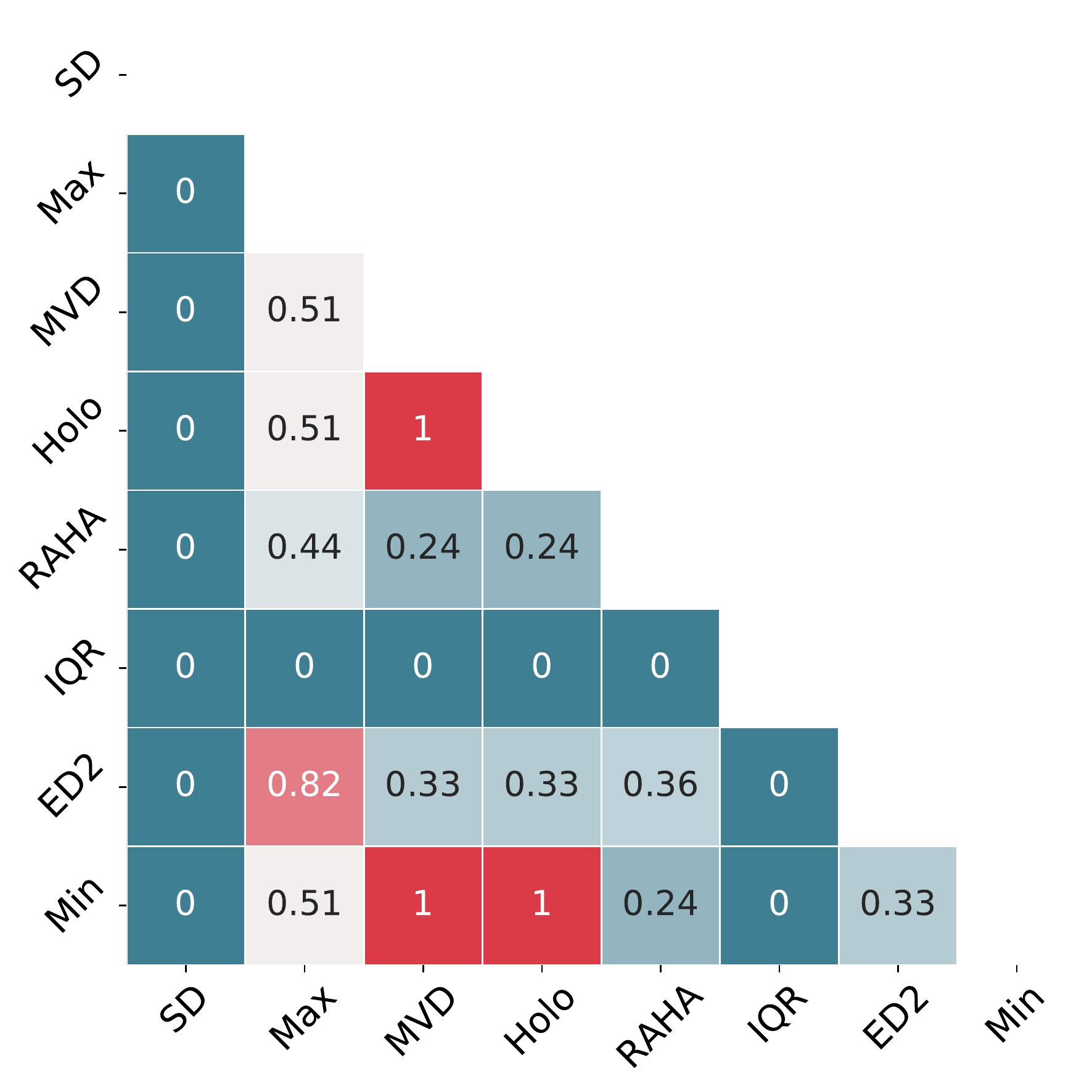}} 
	\subfloat[HAR-Runtime]{\label{fig:har_detect_time}\includegraphics[width=0.21\textwidth]{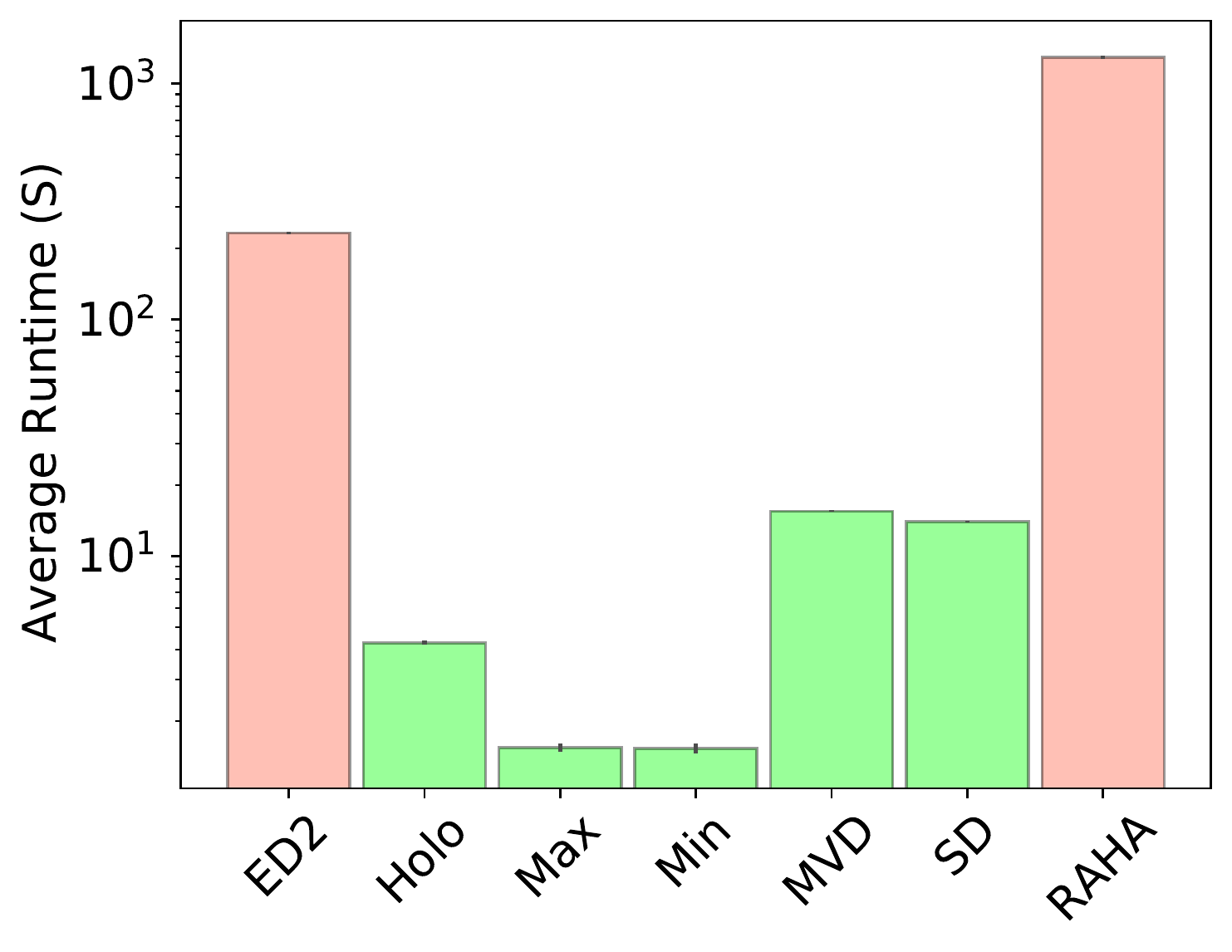}} 
	\caption{Detection results (In the accuracy plots, the blue bars are subdivided into red and green regions to show the false positives and true positives, respectively)}
\label{fig:det_results} \vspace{-3mm}
\end{figure*}

%
In this set of experiments, we assess the performance of several error detectors applied to different datasets. For each dataset, the number of examined detectors depends on the types of injected errors. Moreover, the detectors which fail to detect any cells are deliberately excluded from the figures. Figure~\ref{fig:beers_detect_acc} depicts the number of detected erroneous cells (blue bars) and the number of true positives (green bar) in the \textit{Beers} dataset using 14 error detection methods. The number of false positives is indicated by turning the color of the blue bars into red. The red dashed line represents the actual number of erroneous cells in the dataset. As depicted in the figure, most ML-based and ensemble methods, including ED2, RAHA, Min-k (Min), and Max-entropy (Max), outperform the other methods where their F1 score is between 0.92 and 0.99. As a result of converting the numerical attributes to categorical ones, several detectors, e.g., NADEEF and KATARA, erroneously flagged all clean numerical values in these converted attributes as noisy cells. The low precision of such methods (ranging from 0.08 to 0.16) typically has negative consequences on the repair phase (cf. Section~\ref{sec:data_repair_exp}). 

Figure~\ref{fig:beers_iou} demonstrates the IoU metric of detectors applied to the \textit{Beers} dataset. Obviously, the ML-based and ensemble methods have high similarity (at least IoU of 98\%). Furthermore, the figure shows a relatively high correlation (IoU of 87\%) between the detections of NADEEF (F1 of 0.74) and Metadata-driven (Meta, F1 of 0.48) methods. Accordingly, we can deduce that most detections of the Metadata-driven method, i.e., 2417 out of 2570 detected cells, are rule and pattern violations. Similarly, KATARA (F1 of 0.12) and FAHES (F1 of 0.35) have high similarity (IoU of 88\%) since both of them employ a syntactic pattern detection method. Figure~\ref{fig:beers_detect_time} depicts the average runtime (on the logarithmic scale) of the detectors, where the red bars indicate that the runtime exceeds one minute. As the figure depicts, the ML-based methods require long execution time due to searching for the optimal configurations, featurization, and training the classifiers. For instance, Max Entropy requires much less time (at least by 98\%) than ED2 while detecting the same erroneous cells (cf. Figure~\ref{fig:beers_iou}). 

Figure~\ref{fig:citation_detect_acc} depicts the number of detected cells in the \textit{Citation} dataset using seven detectors. Such a dataset contains duplicates and mislabeled samples. The figure shows that the key collision method (DuplD) outperforms all other methods, where it achieved an average F1 score of 0.86. Similarly, the ensemble methods (i.e., Min and Max) achieved better performance (F1 score between 0.74 and 0.78) than Picket (ML-based detection method, average F1 of 0.18) due to the low recall of Picket which relies on self-supervision to train its classifier. Moreover, CleanLab achieved a low F1 score of 0.19 where it captured only the mislabeled cells in the dataset while ignored the duplicates. Figure~\ref{fig:citation_iou} depicts a strong IoU relationship among the detections of key collision, ZeroER, Min-K, and Max Entropy. However, ZeroER requires much more time (by circa two orders of magnitude) to generate its detections.

For the \textit{Adult} dataset, Figure~\ref{fig:adult_detect_acc} depicts the number of detected cells using 11 detectors. Such a dataset suffers from rule violations and outliers, with a large error rate. In this case, both of RAHA and ED2 outperform all other methods (average F1 score of 0.8 and 0.78, respectively). According to their IoU values, the detections obtained by HoloClean, NADEEF, and Min-k exhibit high correlation where these methods captured most of the rule violations only. Conversely, dBoost captured most of the outliers while failed to identify the rule violations. Despite being effective while detecting erroneous cells in this dataset, ED2 and RAHA are less efficient where they required, on average, 35 minutes to find the erroneous cells compared to 2.3 and 0.73 minutes for dBoost and Min-k, respectively.
%
The \textit{Smart Factory} dataset represents an example of relatively large datasets suffering from explicit missing values and outliers with a moderate error rate. Figure~\ref{fig:sf_detect_acc} depicts the number of detected cells in the \textit{Smart Factory} dataset using eight detectors. In this case, Min-k outperforms (average F1 score of 0.75) other detectors while requiring much less time than other detectors (cf. Figure~\ref{fig:sf_detect_time}). RAHA and Meta have a relatively high correlation with Min-k, as depicted in Figure~\ref{fig:sf_iou}. Furthermore, Figure~\ref{fig:sf_detect_acc} shows that KATARA generated many false positives, which occurs since it failed to correctly interpret the data semantics. 

For the datasets with regression tasks, Figures~\ref{fig:nasa_detect_acc}-\ref{fig:bike_detect_time} show the detection accuracy and runtime of various detectors. For instance, Figure~\ref{fig:nasa_detect_acc} depicts the number of detected cells in the \textit{Nasa} dataset using 12 detectors. Such a dataset represent an example of small datasets suffering from explicit missing values and outliers with a small error rate. As the figure depicts, Max Entropy and dBoost outperform (average F1 score of 0.85) all other methods. Both detectors nearly generated the same detections where their IoU metric is 0.99, as illustrated in Figure~\ref{fig:nasa_iou}. Despite detecting mostly all erroneous cells, the ML-based methods have F1 score between 0.27 and 0.43 due to the large number of false positives. As the dataset is small, most detectors generated their detections in less than a minute, as depicted in Figure~\ref{fig:nasa_detect_time}. 
For the \textit{Bikes} dataset, it has rule violations and outliers with a small error rate. Figure~\ref{fig:bike_detect_acc} depicts the number of detected cells in the \textit{Bikes} dataset using 11 detectors. RAHA and Min-k outperform other detectors with average F1 scores of 0.72 and 0.75, respectively. The figure shows that KATARA and NADEEF (average F1 score of 0.25 and 0.4, respectively) have poor performance due to generating many false positives.  Similar to the \textit{Nasa} dataset, dBoost and Max Entropy have a high correlation. Figure~\ref{fig:bike_detect_time} shows that Min-k is more efficient than RAHA, where it required, on average, 9 seconds to generate the detections compared to 6.6 minutes for RAHA. 
%

Figures~\ref{fig:water_detect_acc}-\ref{fig:har_detect_time} depict the performance of various detectors using the datasets associated with clustering tasks. For the \textit{Water} dataset, it suffers from implicit missing values and outliers with a small error rate. Figure~\ref{fig:water_detect_acc} shows that Max Entropy and RAHA achieved the highest accuracy with average F1 scores of 0.74 and 0.76, respectively. The detections obtained by both detectors are highly correlated. However, Max Entropy required much less time (average runtime of 0.09 seconds) to generate its detections compared to RAHA (average runtime of 15.8 seconds with a standard deviation of 10.4) and ED2 (average runtime of 17.9 minutes). RAHA has typically high variance because it consumes a relatively long time in the first iteration to create the cleaning strategies utilized to generate the training features.    
For the \textit{Power} dataset, NADEEF and Max Entropy outperform other detectors with average F1 scores of 0.9 and 0.84, respectively, as shown in Figure~\ref{fig:power_detect_acc}. Clearly, both of NADEEF and MVD have high precision. However, each detector captured only the relevant errors. In other words, NADEEF detected 1088 pattern violations (corresponding to the typos and implicit missing values), while MVD found only the explicit missing values. For the efficiency, Max Entropy and NADEEF consumed circa the same time (average runtime of 0.05 seconds), while ED2 required, on average, 680 seconds to generate the detections. For the \textit{HAR} dataset, Figure~\ref{fig:har_detect_acc} shows that RAHA achieved the highest accuracy, with an average F1 score of 0.89, at the expense of consuming 20.5 minutes (standard deviation of 20 minutes) to generate its detections (cf. Figure~\ref{fig:har_detect_time}). Figure~\ref{fig:har_iou} demonstrates that MVD, HoloClean, and Min-K detected the same erroneous cells with missing values.
 
\vspace{-2mm}\subsubsection{Detection Robustness}
\begin{figure*}[tbph]
	\centering
	\subfloat[Adult]{\label{fig:adult_robust}\includegraphics[width=0.2\linewidth]{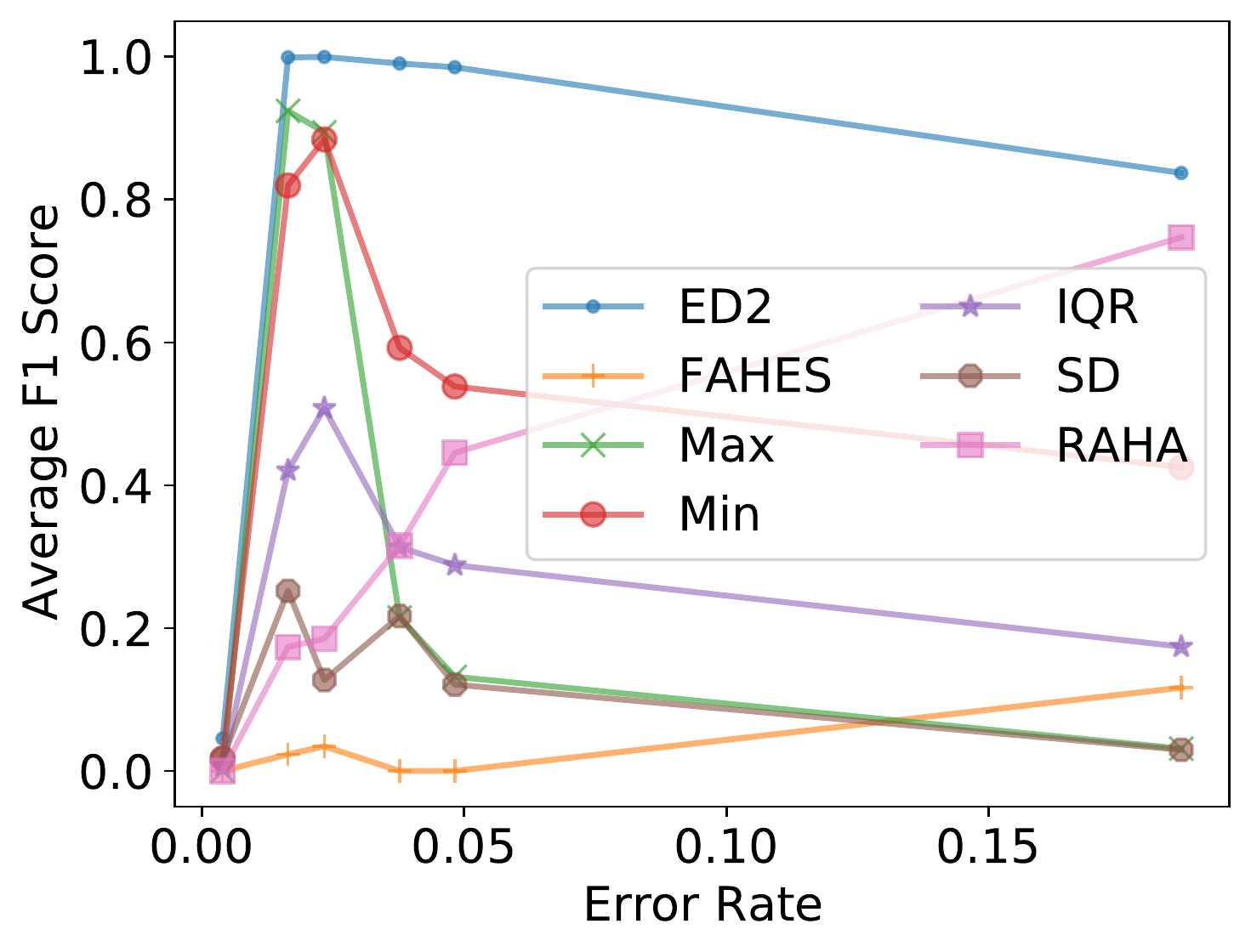}} \hfill
	\subfloat[Power]{\label{fig:power_robust}\includegraphics[width=0.2\linewidth]{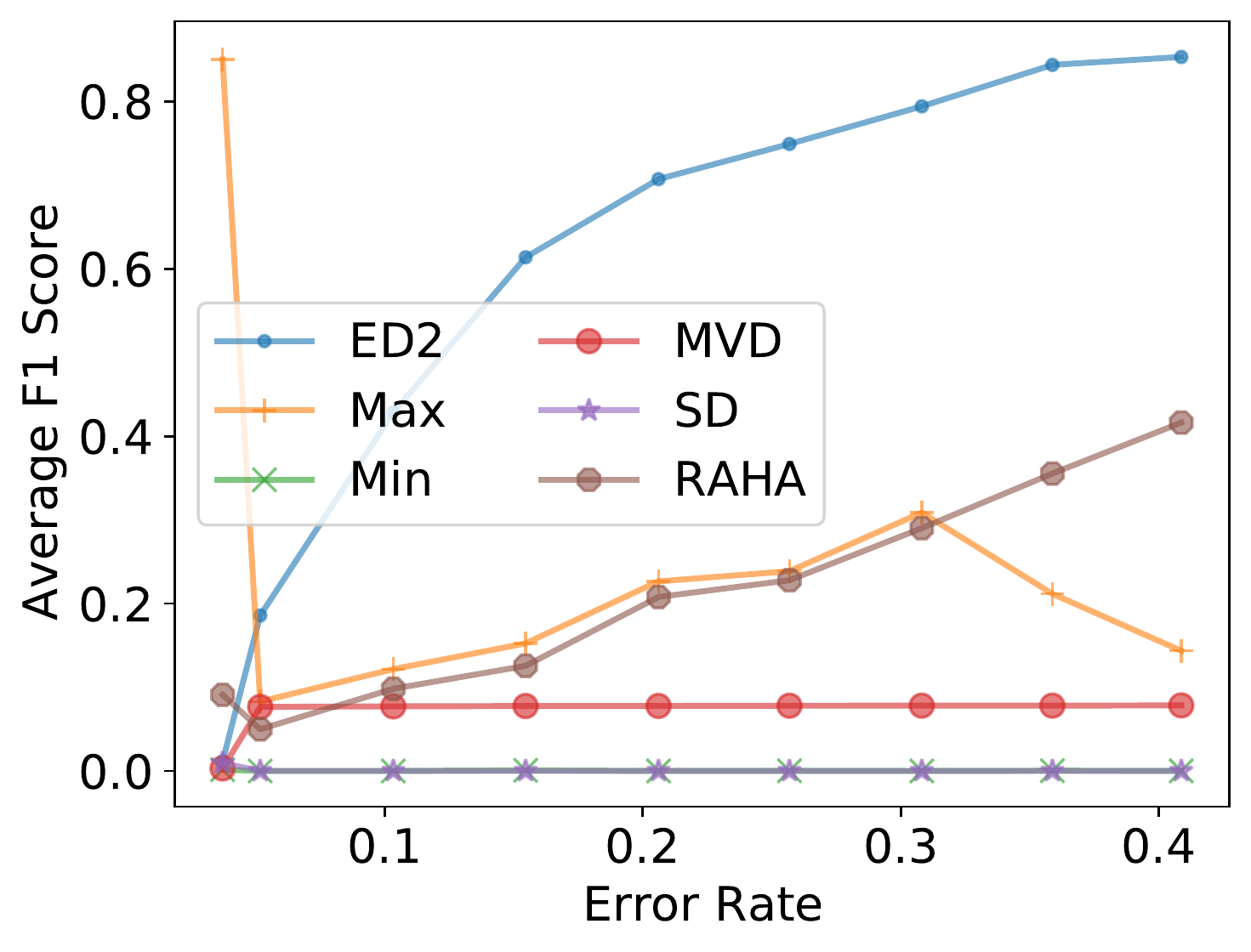}} \hfill
	%
	%
	\subfloat[SmartFactory-Outliers]{\label{fig:sf_outlier}\includegraphics[width=0.2\linewidth]{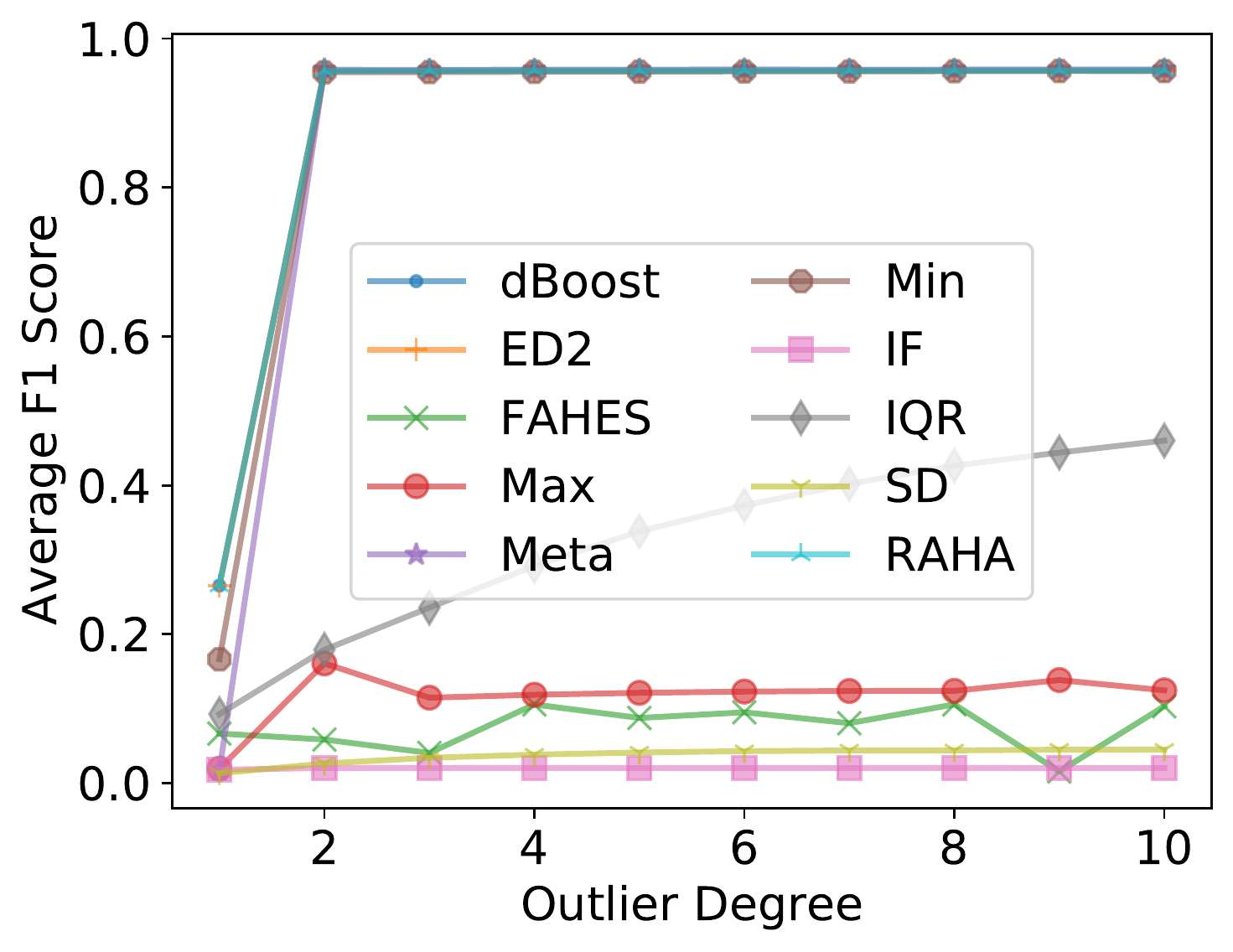}} \hfill
	\subfloat[Soccer-Accuracy]{\label{fig:soccer_accuracy}\includegraphics[width=0.2\linewidth]{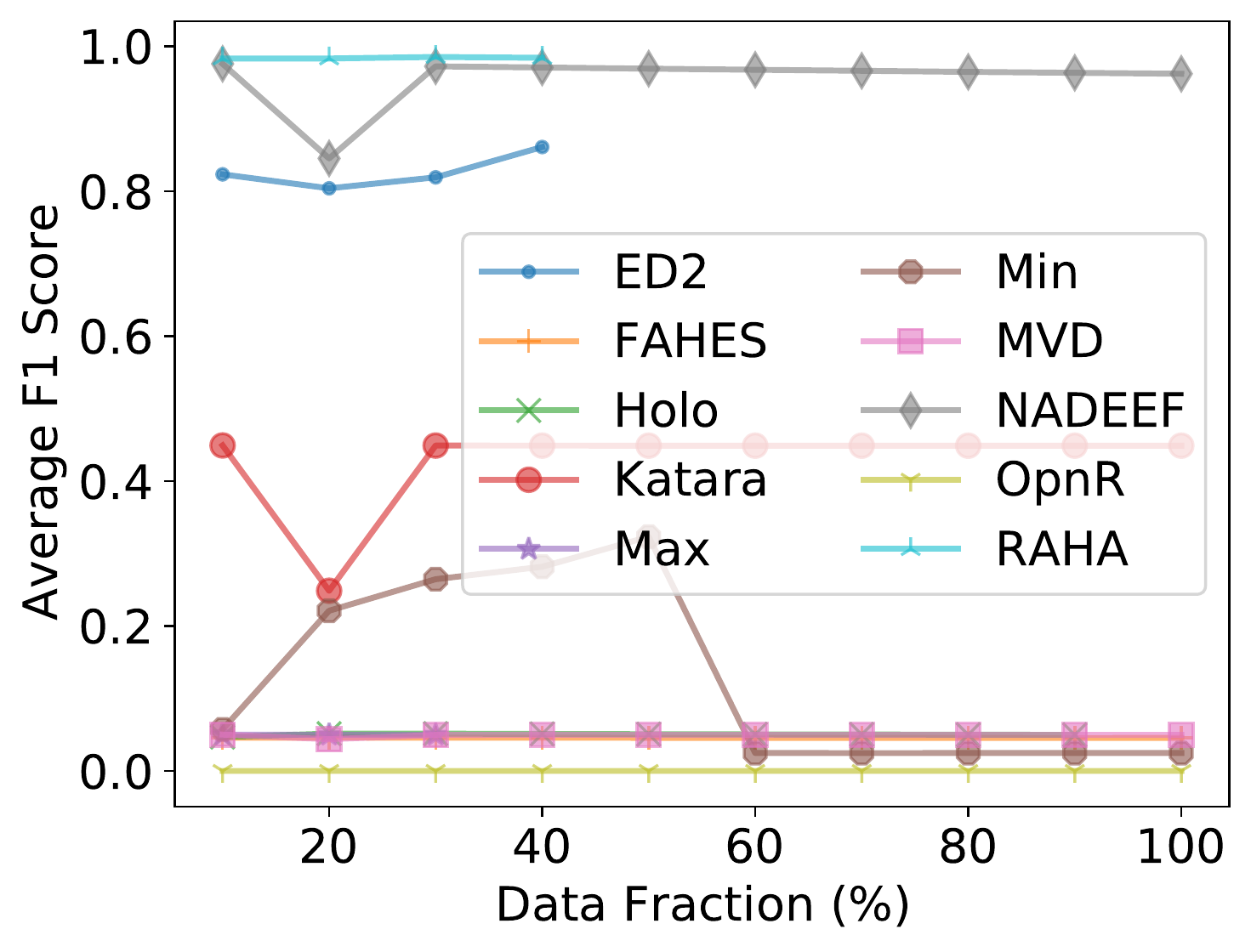}} \hfill
	\subfloat[Soccer-Runtime]{\label{fig:soccer_runtime}\includegraphics[width=0.2\linewidth]{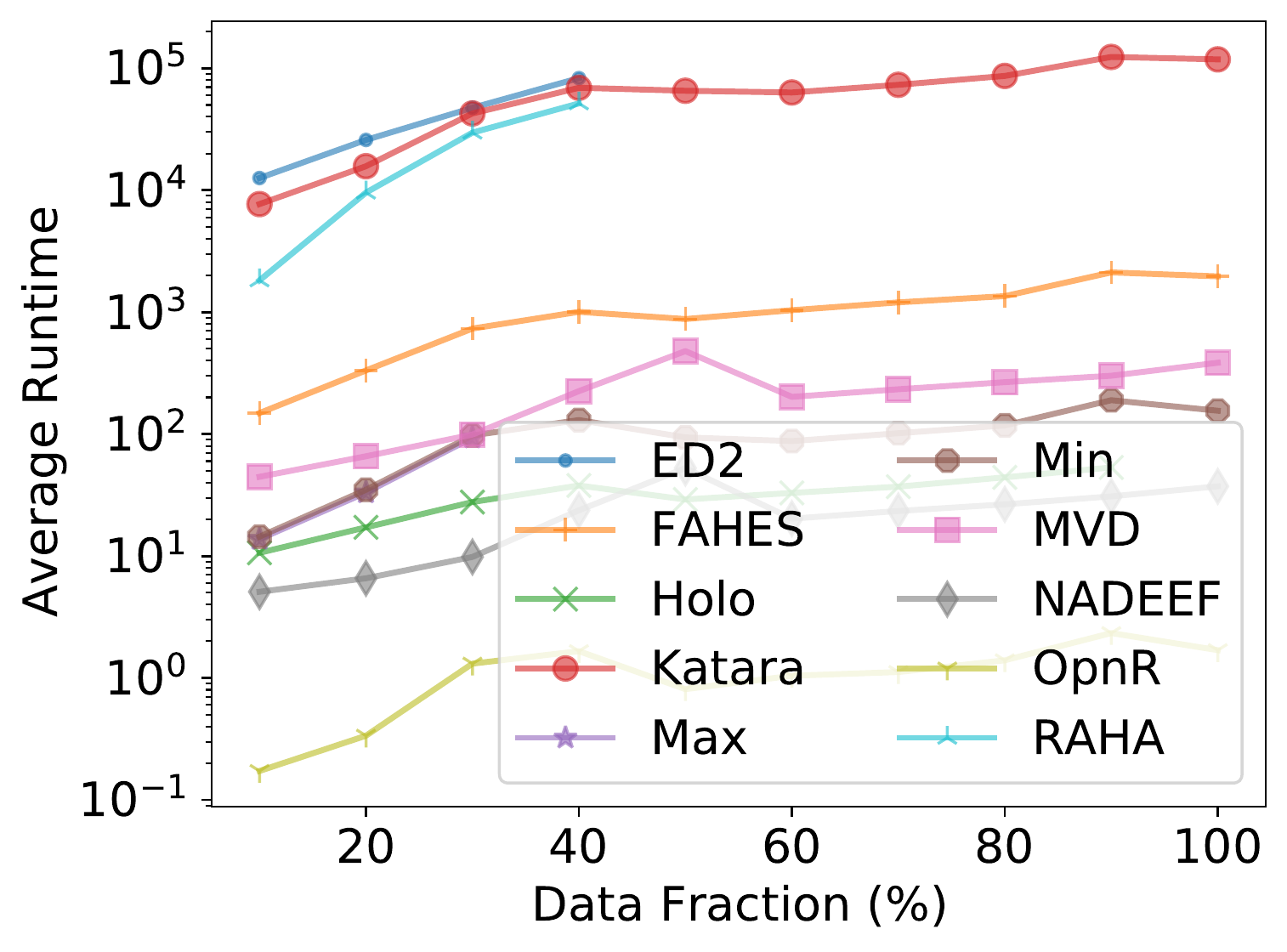}} \hfill
	
	\caption{Robustness and scalability results of the error detectors} \vspace{-4mm}
	\label{fig:robustness}
\end{figure*}

In this section, we examine the robustness of various error detectors in terms of their accuracy. To this end, we implemented two sets of experiments, including: (1) varying the \textit{error rate} of a dataset; and (2) varying the \textit{outlier degree}, defined as the number of standard deviations away from the mean. In the former set of experiments, we injected outliers and missing values where the outlier degree is set to 4. In the outlier degree experiment, we injected outliers with an error rate of 30\%. Figure~\ref{fig:adult_robust} compares the robustness of seven detectors while cleaning the \textit{Adult} dataset at different error rates. Clearly, the F1 score of all detectors increases linearly at low error rates (i.e., up to 0.02). In this range, several detectors (e.g., ED2, Max Entropy, and Min-k) have a large slope, which implies a high detection accuracy. When the error rate is further increased, the accuracy of most detectors, except RAHA, is gradually reduced. Figure~\ref{fig:power_robust} shows a similar experiment on the \textit{Power} dataset. As the figure depicts, ED2 achieved a higher accuracy, at low error rates, than all other models. For RAHA, its performance has been improved, when the error rate is increased. Figure~\ref{fig:sf_outlier} compares the performance of ten detectors when increasing the outlier degree injected into the \textit{Smart Factory} dataset. The figure shows that all detectors behave approximately the same when the outlier degree is relatively small (i.e., below two). However, the performance of RAHA, ED2, Min-k, dBoost, and Meta is broadly improved when the value of the outlier degree goes beyond two.   


\subsubsection{Scalability Analysis}
In this section, we evaluate the efficiency of several error detectors when dealing with large datasets. To this end, we ran several experiments to detect errors in different data fractions. Figures~\ref{fig:soccer_accuracy} and \ref{fig:soccer_runtime} compares the accuracy and efficiency of ten detectors for different factions of the \textit{Soccer} dataset. For this dataset, Figure~\ref{fig:soccer_accuracy} shows that ED2, NADEEF, and RAHA achieved the highest F1 score (i.e., 0.83, 0.93, and 0.98, respectively). Furthermore, the figure illustrates that some detectors work only with small data fractions. For instance, RAHA, ED2 stopped working at a  data fraction of 50\%, while HoloClean is terminated with 90\% of the data. Figure~\ref{fig:soccer_runtime} shows the comparison in terms of the average runtime (in logarithmic scale). Obviously, RAHA, ED2 and KATARA required much more time (average runtime of 3.5, 10.1, 13.8 hours, respectively) than other detectors. In contrast to ED2 and RAHA, KATARA managed to detect errors for all data fractions.


\vspace{-2mm}\subsection{Data Repair}\label{sec:data_repair_exp}
%
In this section, we introduce the results of the repair methods while being used to generate repair candidates based on the detections obtained from various error detectors. We divide the experiments according to the data type in each dataset. Moreover, we introduce the results of the ML-oriented repair methods, whose outputs are ML models rather than repaired datasets.
%
\subsubsection{Categorical Attributes}
\begin{figure*}[htbp] 
	\centering
	\subfloat[Beers-Accuracy]{\label{fig:beers_repair_acc}\includegraphics[width=0.26\textwidth]{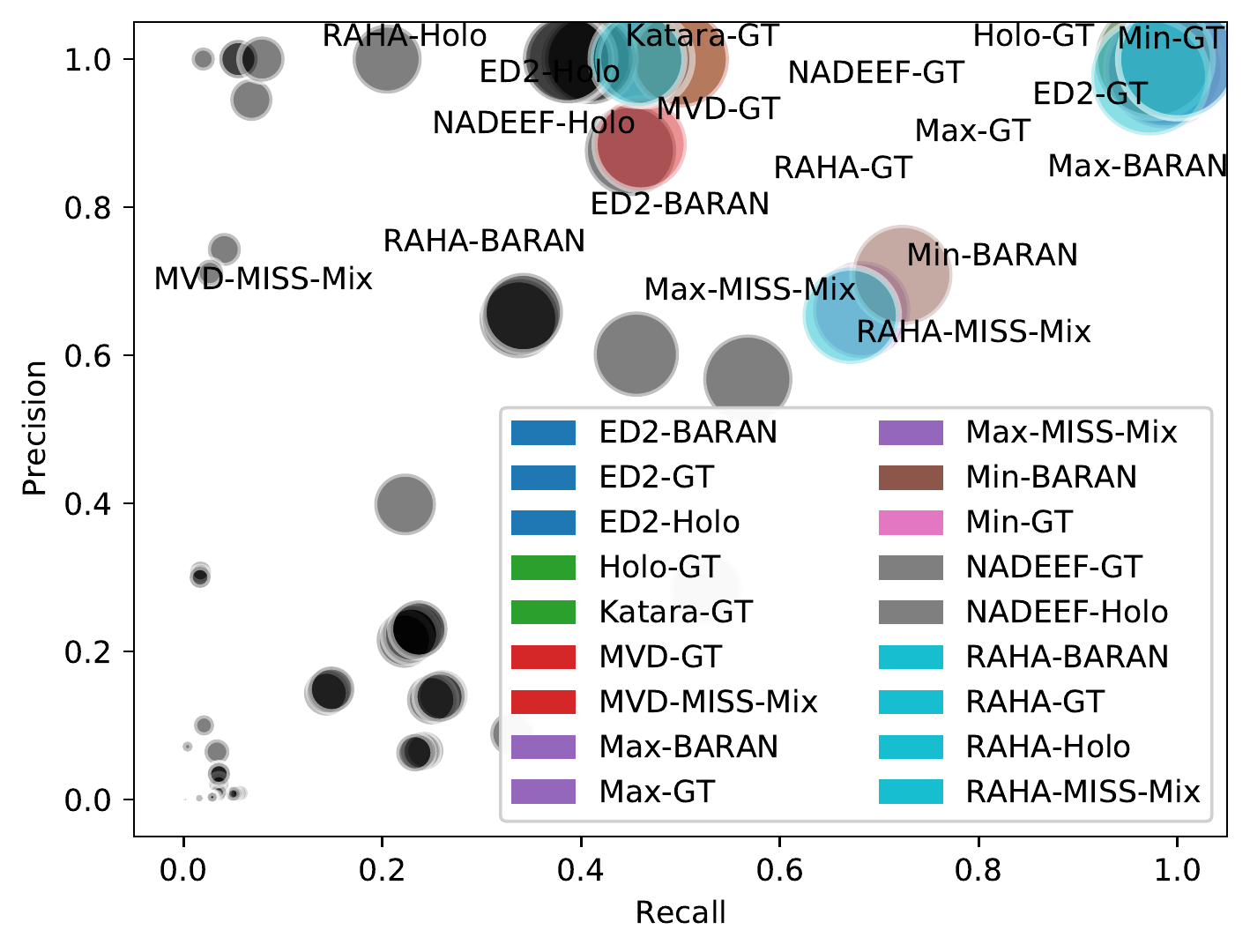}} \hfill
	%
	%
	\subfloat[Beers-Runtime]{\label{fig:beers_repair_runtime}\includegraphics[width=0.24\textwidth]{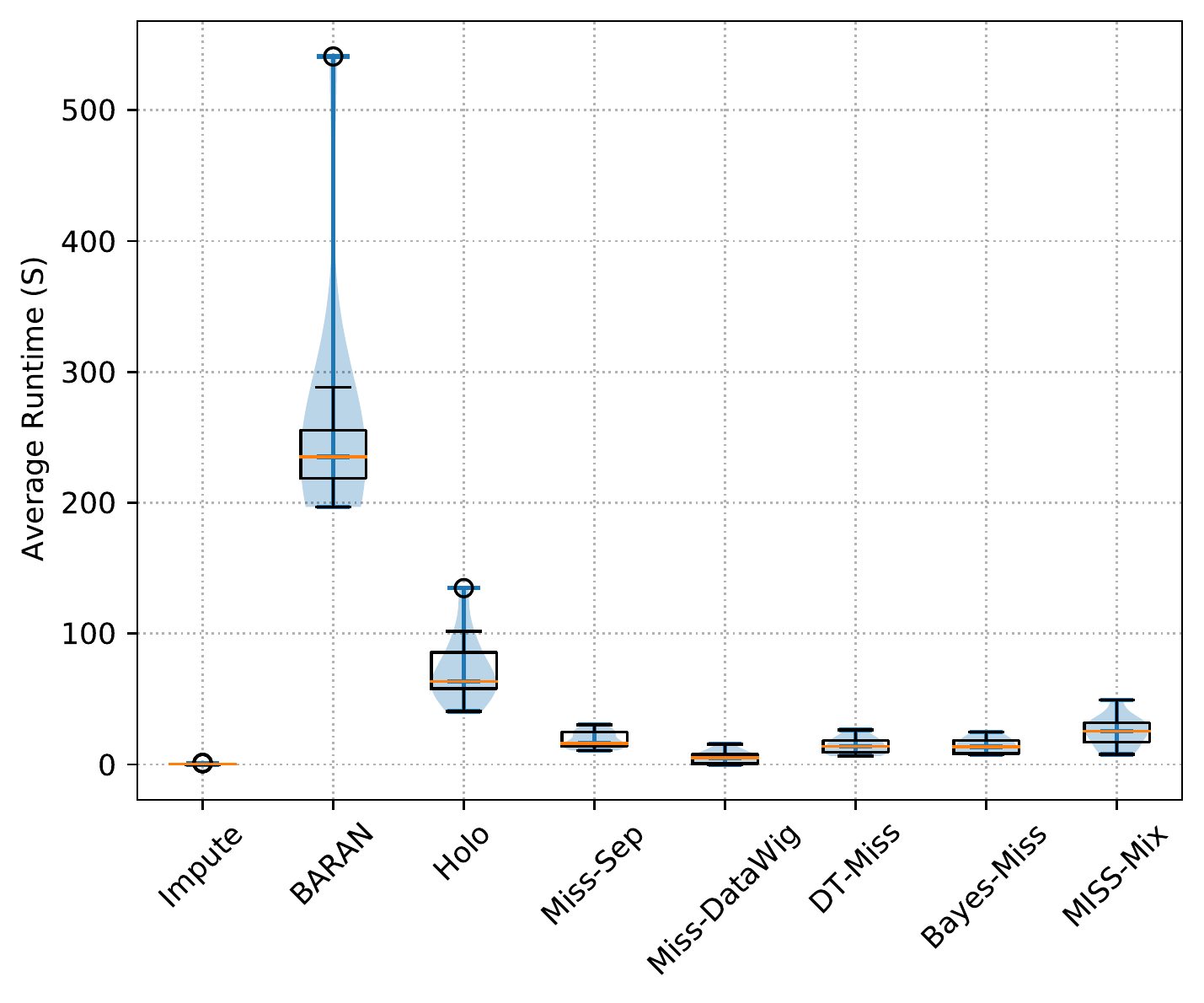}} \hfill
	%
	%
	%
	%
	\subfloat[BreastCancer-Accuracy]{\label{fig:bc_repair_acc}\includegraphics[width=0.26\textwidth]{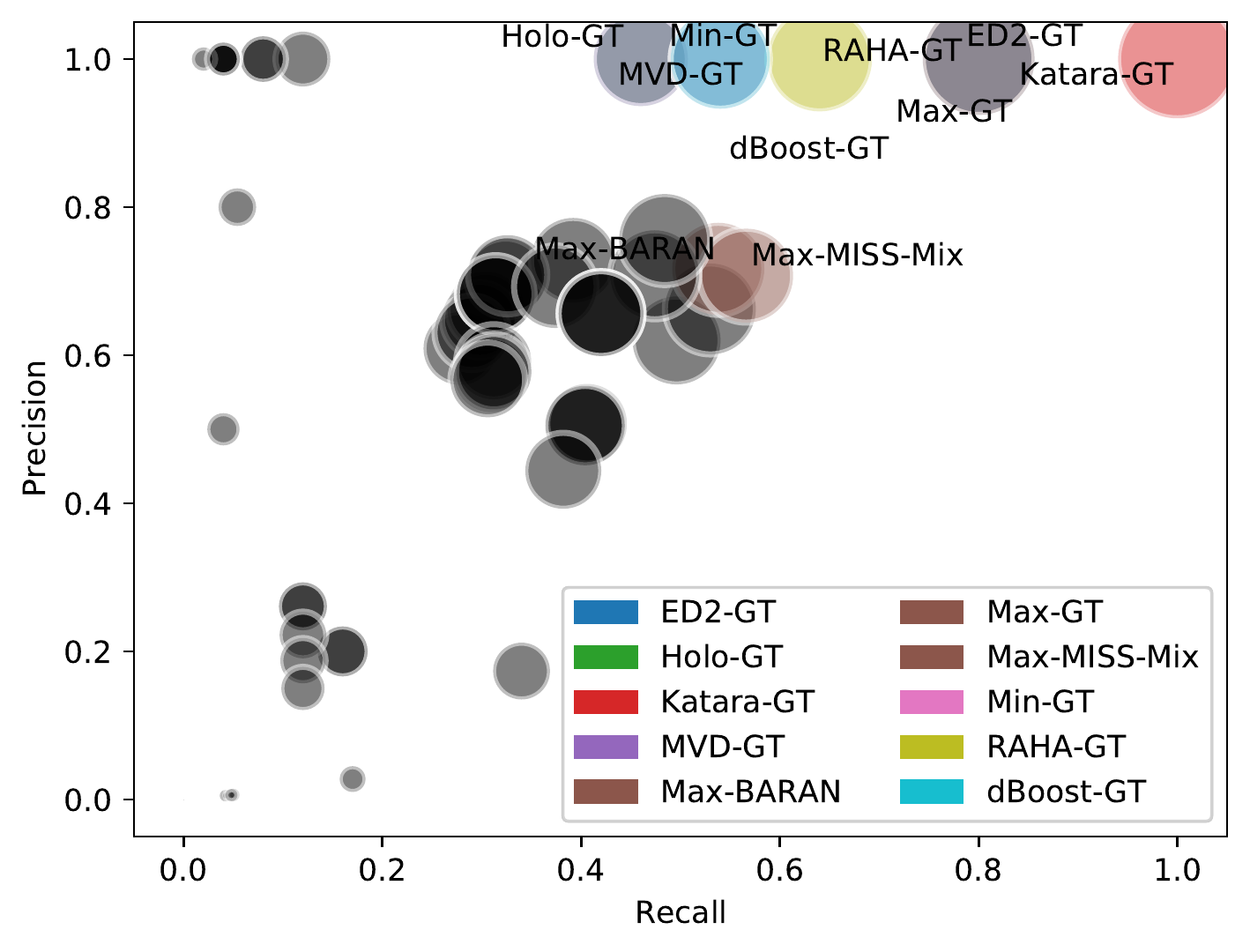}} \hfill
	\subfloat[BreastCancer-Runtime]{\label{fig:bc_repair_time}\includegraphics[width=0.24\textwidth]{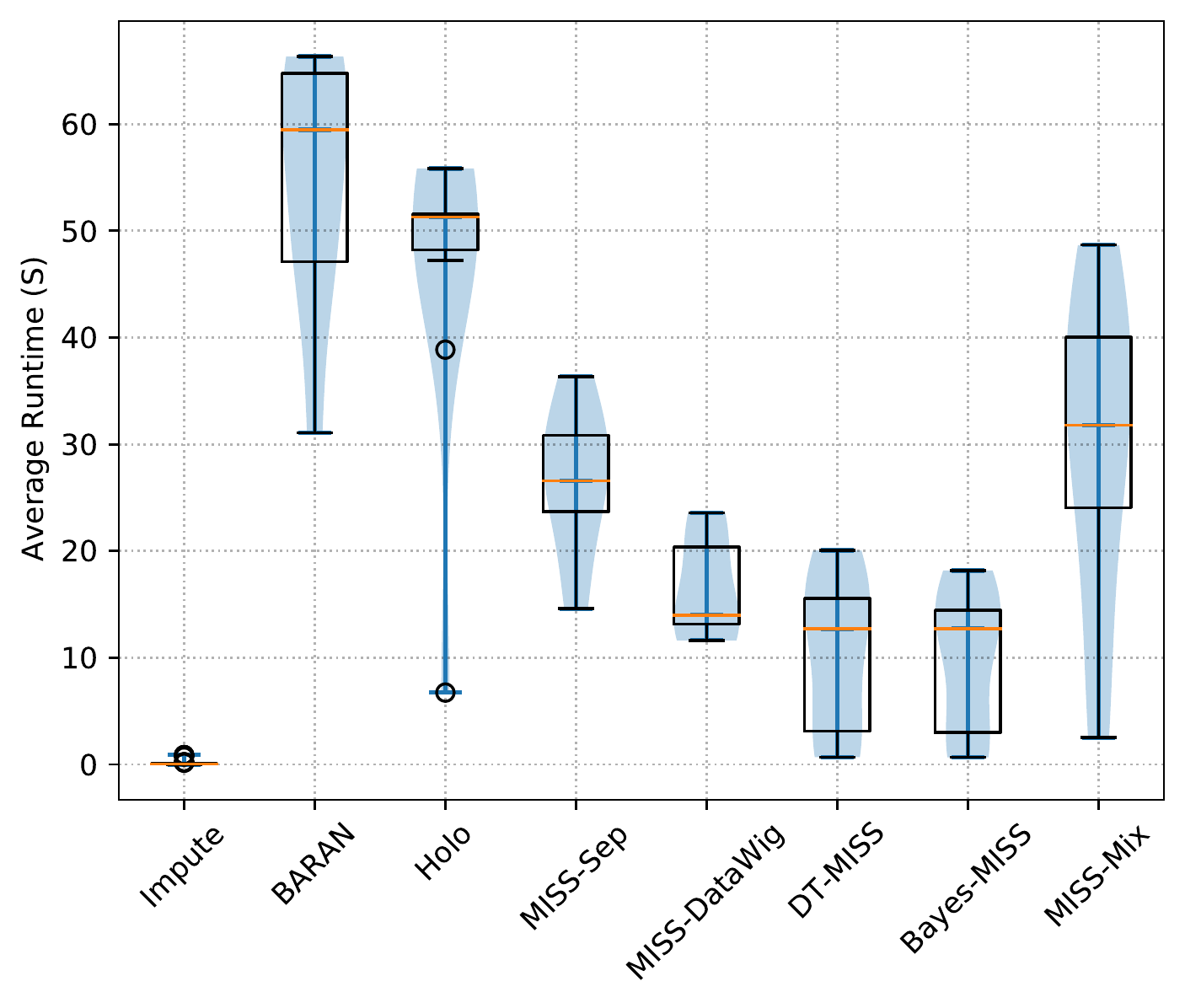}} \hfill
	%
	%
	\caption{Repair results considering only the categorical attributes (In the accuracy figures, each bubble represents a different cleaning strategy and the size of the bubbles denotes the F1 score. To highlight the most effective cleaning strategies, we colored only the bubbles whose F1 score is above 0.6)}
	\label{fig:acc_onlycat} \vspace{-6mm}
\end{figure*}

Figure~\ref{fig:acc_onlycat} shows the repair results in terms of the repair accuracy and runtime for two datasets which include categorical attributes. For instance, Figure~\ref{fig:beers_repair_acc} delineates the repair accuracy, in terms of the precision and recall, when cleaning the \textit{Beers} dataset. The figure shows that the detections obtained by several detectors, including RAHA, ED2, Min-k, Max Entropy, HoloClean, and NADEEF, can result in a high repair accuracy (average F1 score of 0.99) if being repaired by an optimal repair method (simulated by GT). The high performance of HoloClean-GT is achieved, despite the low recall of HoloClean as shown in Figure~\ref{fig:beers_detect_acc}, since HoloClean detected 248 out of 254 actual erroneous categorical cells. For this dataset, BARAN achieved the highest accuracy (average repair F1 score of 0.98) when generating repair candidates for the detections obtained by RAHA, ED2, and Max Entropy. Due to the large number of false negatives (127 cells out of 254 erroneous categorical cells) obtained by KATARA (cf. Figure~\ref{fig:beers_detect_acc}), the maximum repair F1 score, when repaired using the ground truth, is limited to only 0.66. Figure~\ref{fig:beers_repair_runtime} compares the runtime of eight repair methods. The blue band enveloping the boxes represents the standard deviation of the  runtime at a given point. Clearly, BARAN consumed much more time (an average runtime of 4.4 minutes with a standard deviation of 1.5 minutes) than all other detectors. 

Figure~\ref{fig:bc_repair_acc} shows the repair accuracy of various detector/repair combinations adopted to clean the \textit{Breast Cancer} dataset. As the figure illustrates, the detections obtained by Max Entropy led to a moderate accuracy, when MissForest (F1 score of 0.63) and BARAN (F1 score of 0.6) are utilized. Furthermore, the figure shows that KATARA achieved a repair F1 score of one when the detections are repaired using the ground truth. In fact, KATARA generated many false positives (6,843 cells) and few false negatives (86 cells, all numerical values). Accordingly, we can deduce that in the presence of highly-effective repair methods, the detection false negatives are more harmful to the repair accuracy than the detection false positives. Figure~\ref{fig:bc_repair_time} depicts that HoloClean and BARAN are the most time-consuming repair methods (average runtime of 45.7 and 53.8 and seconds, respectively). 

\subsubsection{Numerical Attributes}
\begin{figure*}[htbp] 
	\centering
	%
	%
	%
	\subfloat[SmartFactory-Accuracy]{\label{fig:sf_repair_acc}\includegraphics[width=0.39\textwidth]{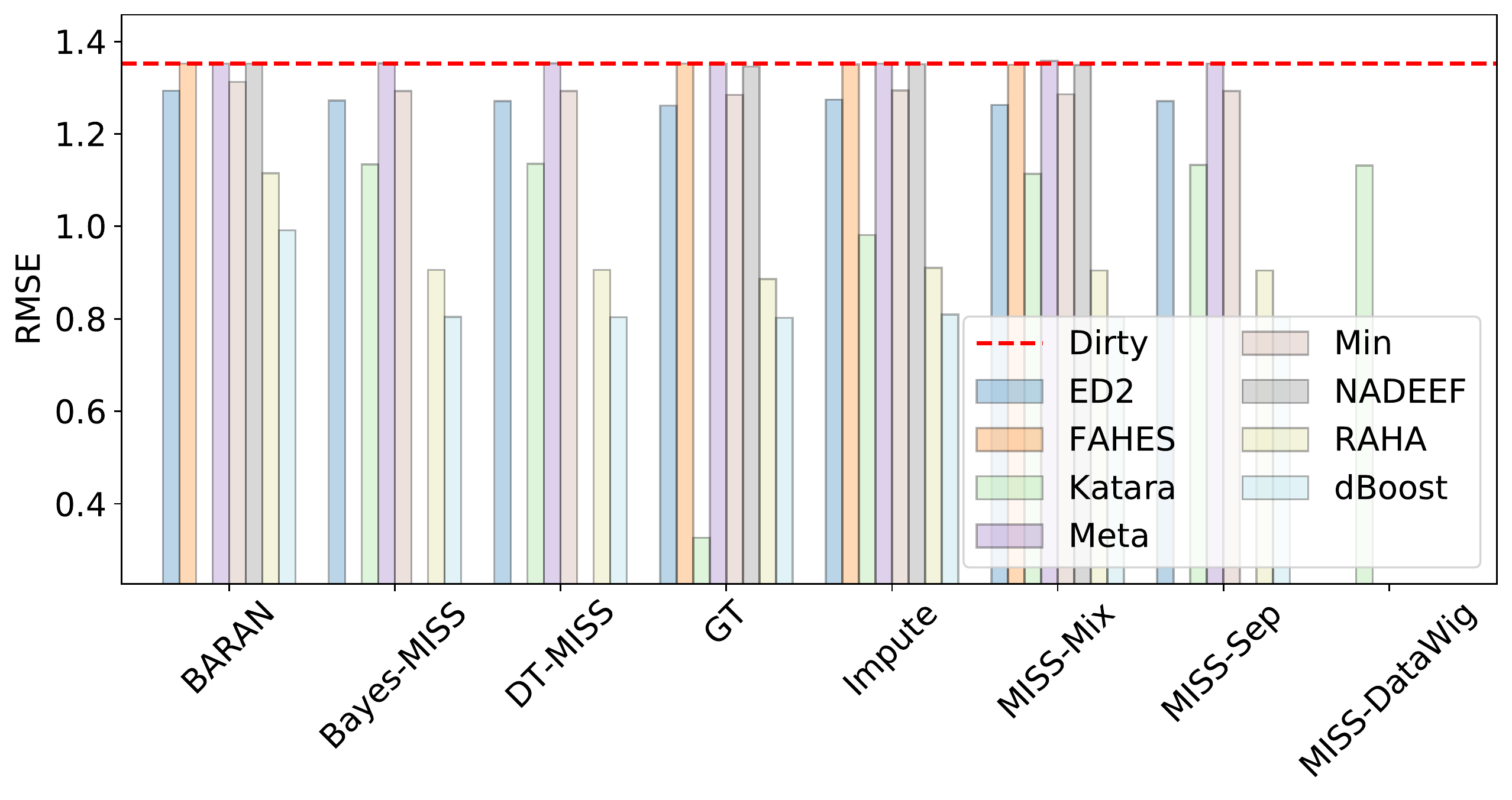}} \hfill
	\subfloat[SmartFactory-Runtime]{\label{fig:sf_repair_time}\includegraphics[width=0.22\textwidth]{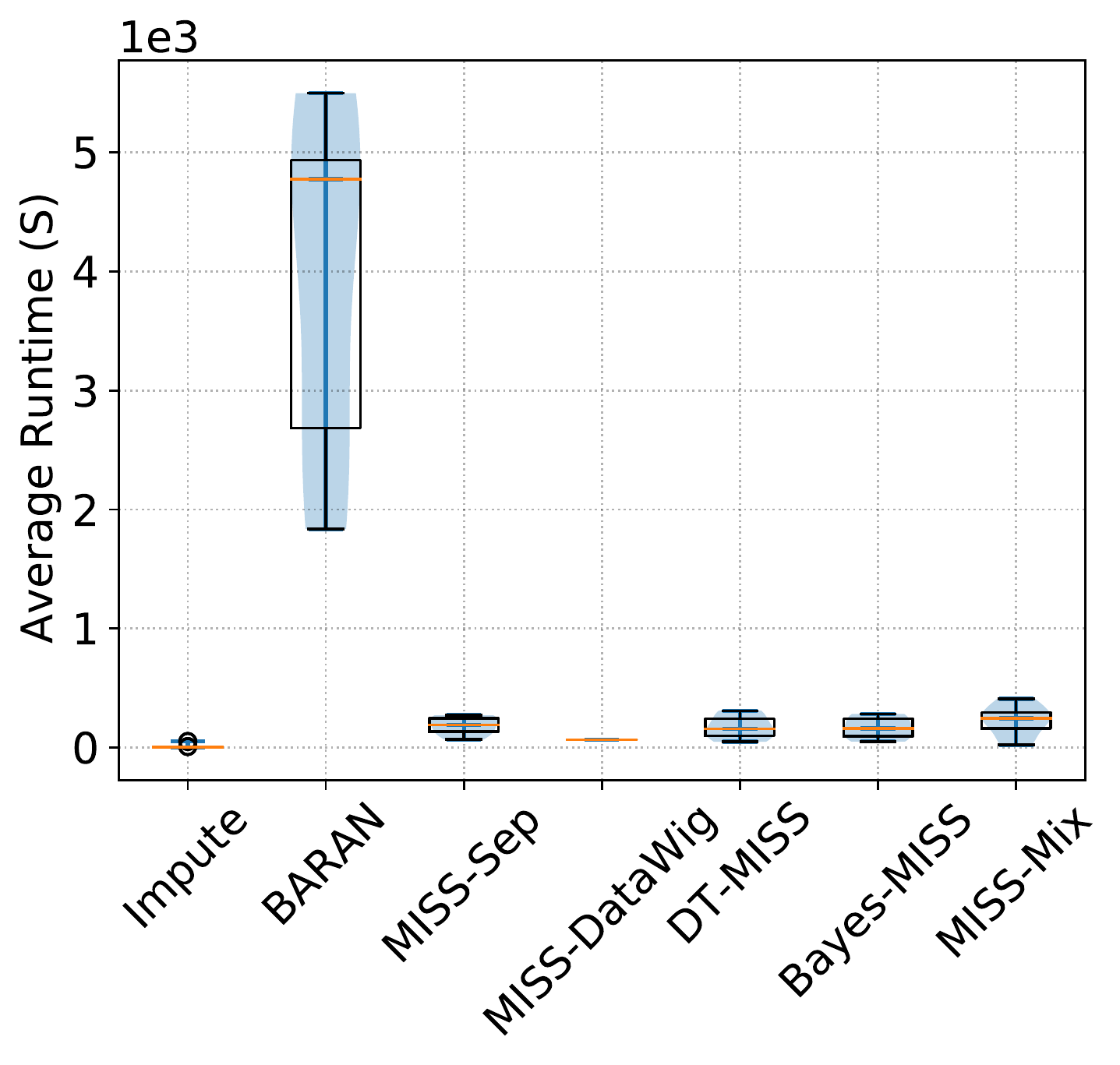}} \hfill
	%
	%
	%
	\subfloat[Breast Cancer-Accuracy]{\label{fig:bc_onlynum}\includegraphics[width=0.39\textwidth]{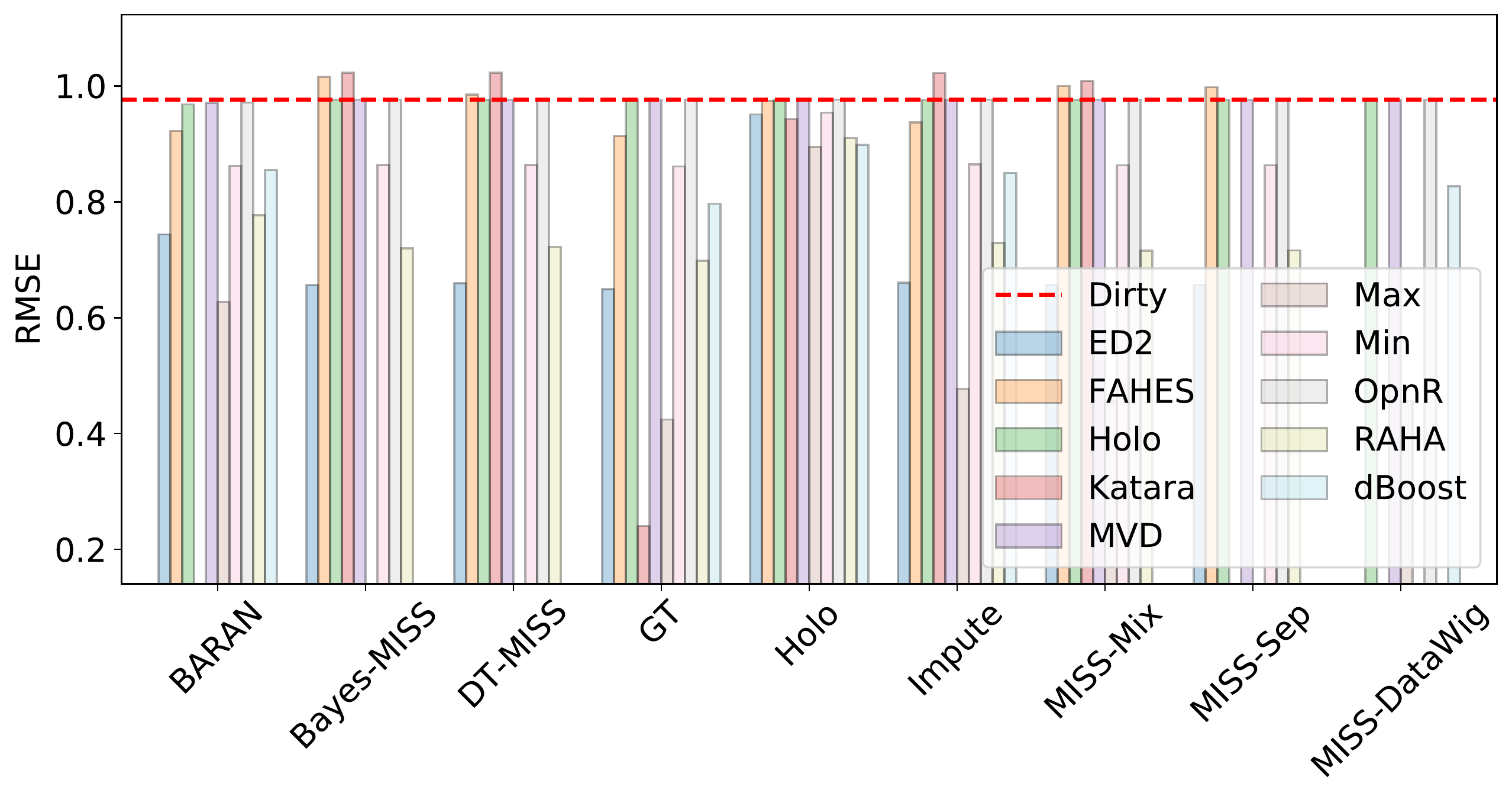}} \hfill
	\subfloat[Bikes-Accuracy]{\label{fig:bike_repair_acc}\includegraphics[width=0.385\textwidth]{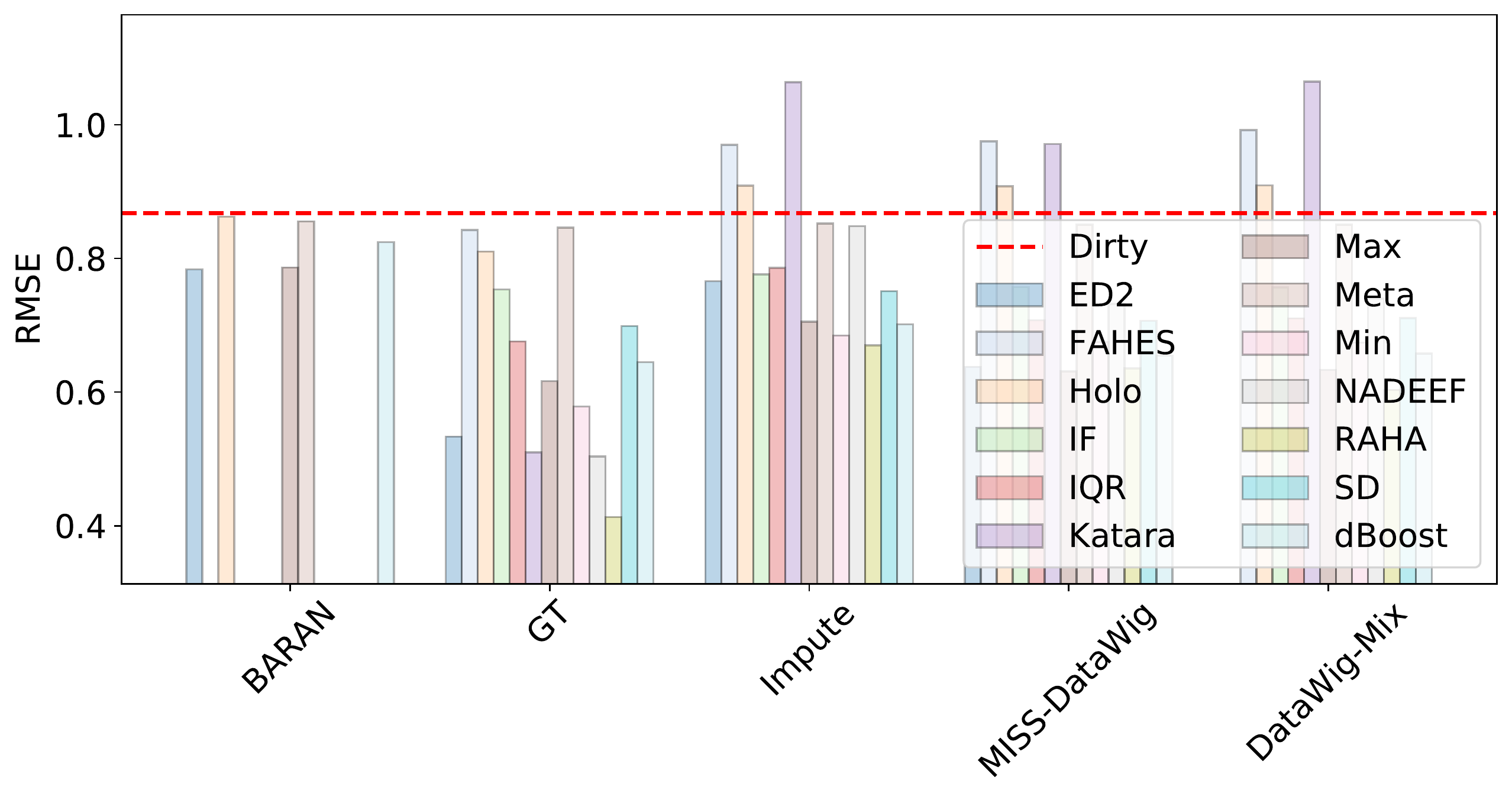}} \hfill
	%
	%
	%
	\subfloat[Water-Accuracy]{\label{fig:water_repair_acc}\includegraphics[width=0.395\textwidth]{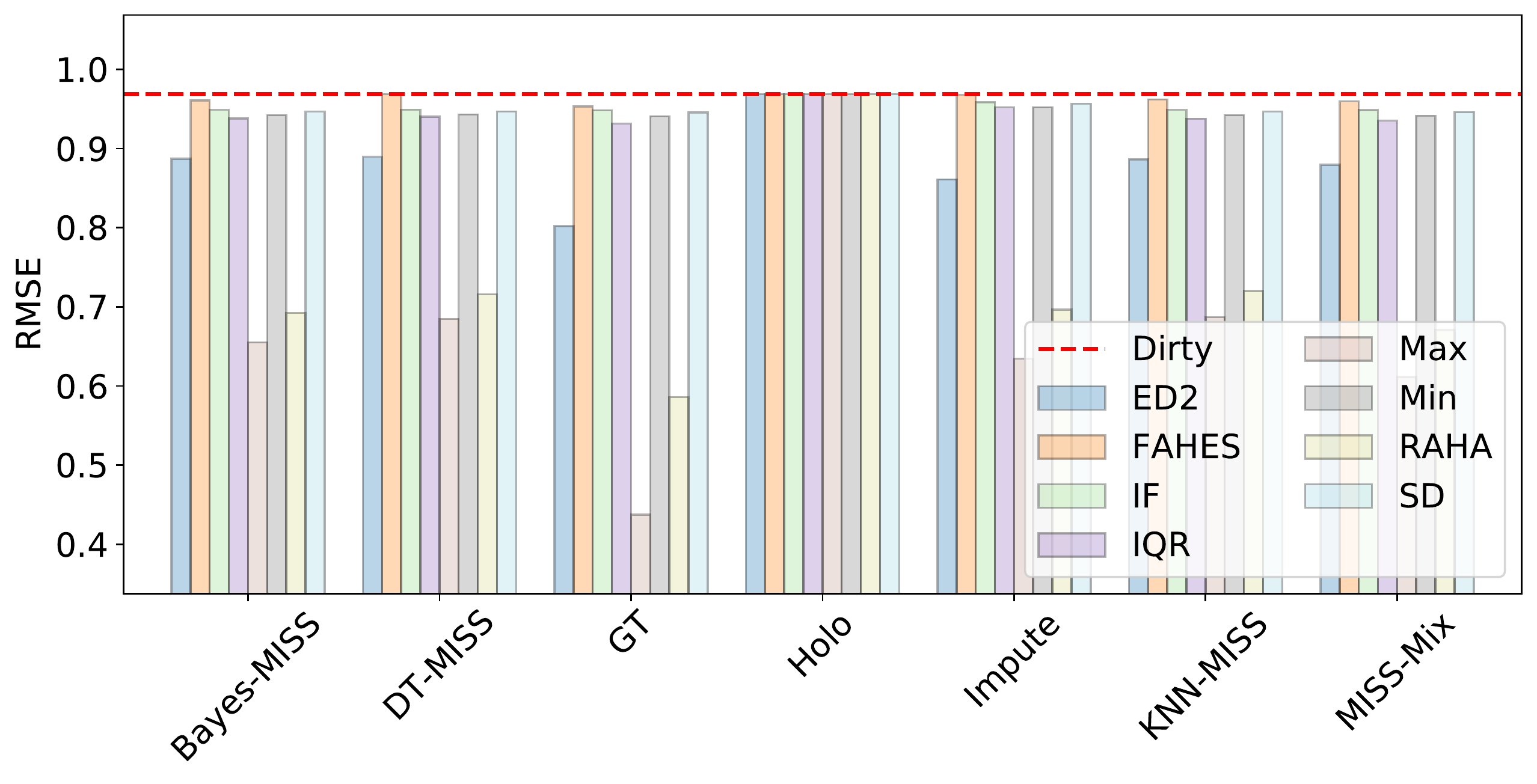}} \hfill
	\subfloat[Water-Runtime]{\label{fig:water_repair_time}\includegraphics[width=0.22\textwidth]{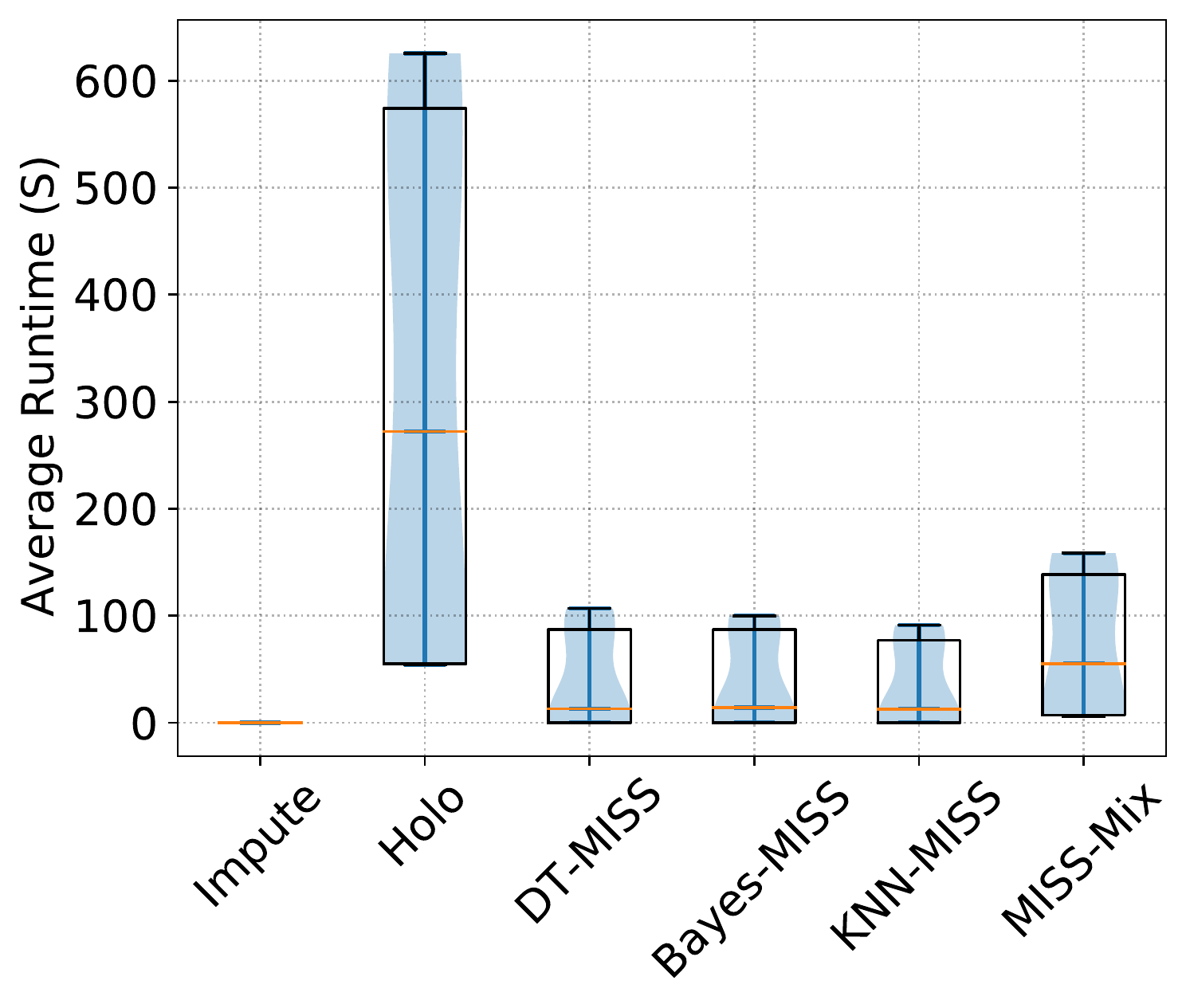}} \hfill
	\caption{Repair results considering only the numerical attributes}\vspace{-3mm}
	\label{fig:acc_onlynum} 
\end{figure*}

%
Figure~\ref{fig:acc_onlynum} depicts the repair results of the numerical attributes in terms of the RMSE values and the runtime. For instance, Figure~\ref{fig:sf_repair_acc} compares the performance of eight repair methods while cleaning the \textit{Smart Factory} dataset. Each repair method comprises a group of bars representing the different detection methods. The red dashed line denotes the RMSE value of the dirty version of the dataset. The Figure depicts that the detections of RAHA and dBoost achieved the highest performance (average RMSE of 0.93 and 0.82 for RAHA and dBoost, respectively) for different repair methods. Furthermore, the figure depicts that GT may generate repaired versions with RMSE comparable to the dirty version (cf. the bars of FAHES, Meta, and NADEEF in the GT group). Such a repair performance typically occurs due to the low accuracy of these detections. Accordingly, we can conclude that without an accurate error detection process, the highly-effective repair methods can achieve poor results. Figure~\ref{fig:bc_onlynum} shows that the detections of ED2 and RAHA, in the \textit{Breast Cancer} dataset, achieved the highest repair accuracy over mostly all repaired methods.

For the \textit{Bikes} dataset, Figure~\ref{fig:bike_repair_acc} shows that most cleaning strategies generate repaired versions relatively better than the dirty data. However, the repaired versions, resulted from the detections of FAHES, HoloClean, and KATARA, have higher RMSE values than the dirty version (cf. the bars above the dashed line for standard and ML-based imputation methods). For this dataset, BARAN required much more time (an average runtime of 58.4 $\pm$ 40.2 minutes) than all other methods. Figure~\ref{fig:water_repair_acc} compares the accuracy of ten repair methods while cleaning the \textit{Water} dataset. The figure shows that all repaired versions have either similar or better performance than the dirty version. Obviously, RAHA and Max Entropy achieved the highest accuracy over all repair methods (an average RMSE of 0.7 and 0.65, respectively). In terms of runtime, Figure~\ref{fig:water_repair_time} shows that HoloClean is the most time-consuming method with an average runtime of 5.2 $\pm$ 4 minutes.  
%
%
%
\vspace{-6mm}\subsubsection{ML-Oriented Repair Methods}
%
%
\begin{figure}[htbp] 
	\centering
	\subfloat[Adult]{\label{fig:adult_repair_models}\includegraphics[width=0.25\textwidth]{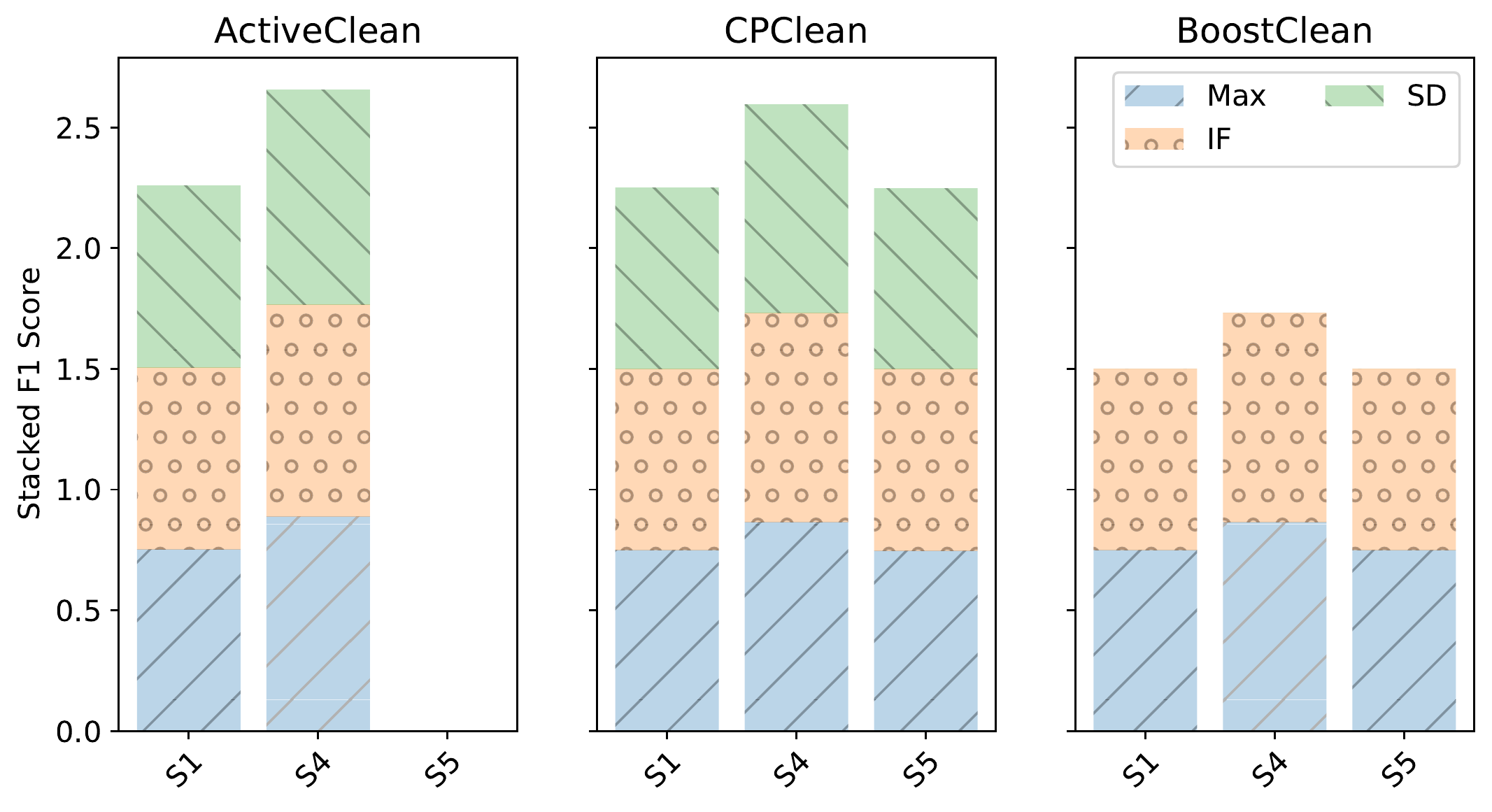}} 
	\subfloat[BreastCancer]{\label{fig:bc_repair_models}\includegraphics[width=0.25\textwidth]{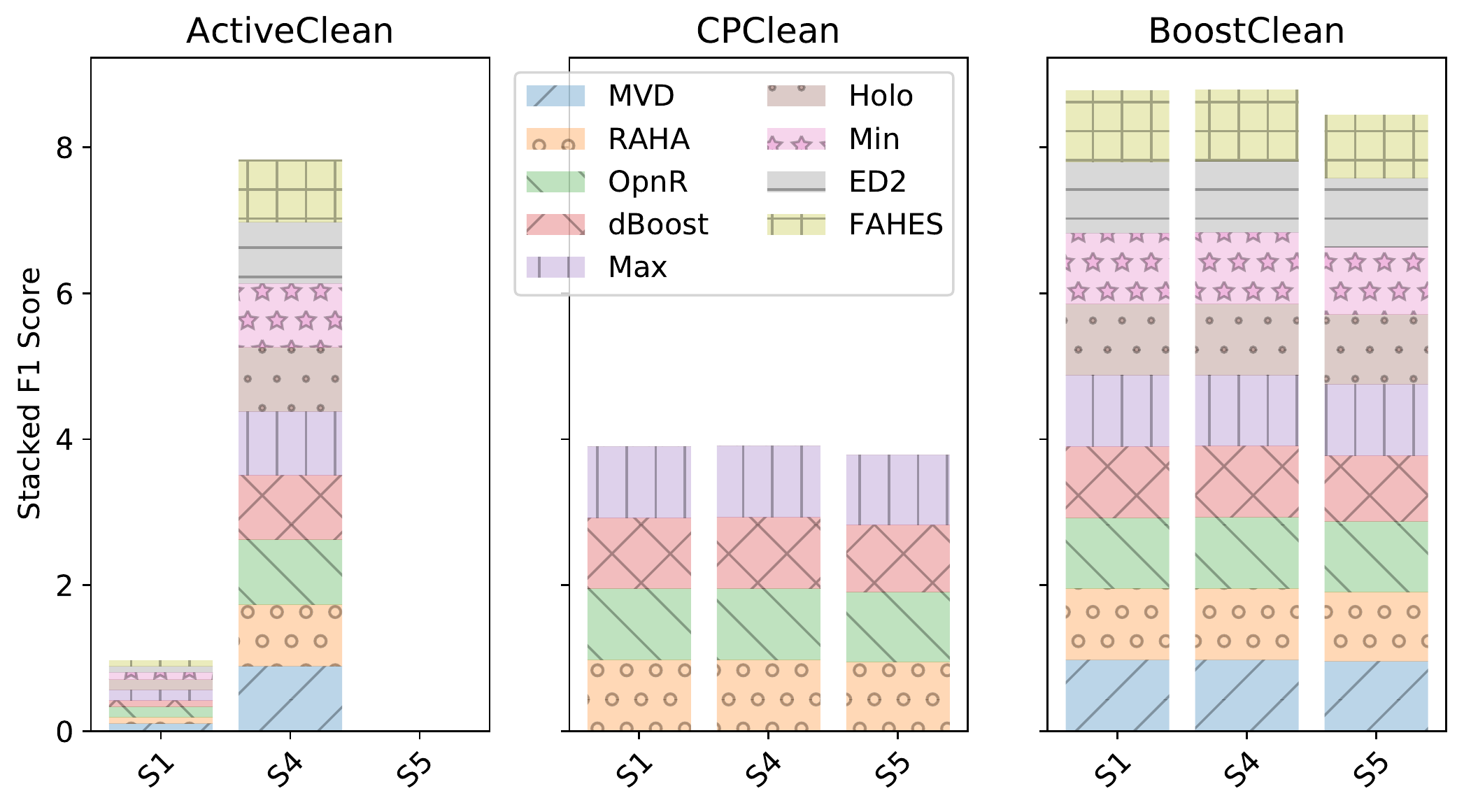}} \hfill
	%
	%
	\caption{Accuracy of ML-oriented repair methods}
	\label{fig:repair_models} \vspace{-3mm}
\end{figure}
In this section, we present the results of the ML-oriented methods, including ActiveClean, CPClean, and BoostClean. Figure~\ref{fig:repair_models} compares the performance of these methods in terms of the modeling accuracy. In particular, Figure~\ref{fig:adult_repair_models} shows the F1 score of the generated models in scenarios S1, S4, and S5 for the \textit{Adult} dataset. The figure shows that the datasets, repaired using the three cleaning methods, slightly lag behind the ground truth versions (on average by 15\%, 0.13\%, and 0.13\%, for ActiveClean, CPClean, and BoostClean, respectively). Furthermore, the results of CPClean and BoostClean in S1 are approximately the same as in S5. The reason behind such a result lies in the relatively comparable accuracy of the dirty and the repaired versions, as shown in Figure~\ref{fig:acc_onlynum}. For the \textit{Breast Cancer} dataset, Figure~\ref{fig:bc_repair_models} depicts that the models generated by ActiveClean in S1 broadly suffer from low accuracy, where the average F1 score in S4 is higher than in S1 by circa 88\%. This result mostly occurred due to the small size of the dataset and the relatively low detection accuracy of all detectors (i.e., the highest F1 score of 0.75 by Max Entropy). For CPClean and BoostClean, the results are close to each other in the three scenarios.\vspace{-4mm}
%
\subsection{Modeling Accuracy}
\begin{figure*}[htbp] 
	\centering
	\subfloat[Beers]{\label{fig:beers_all_models}\includegraphics[width=0.22\textwidth]{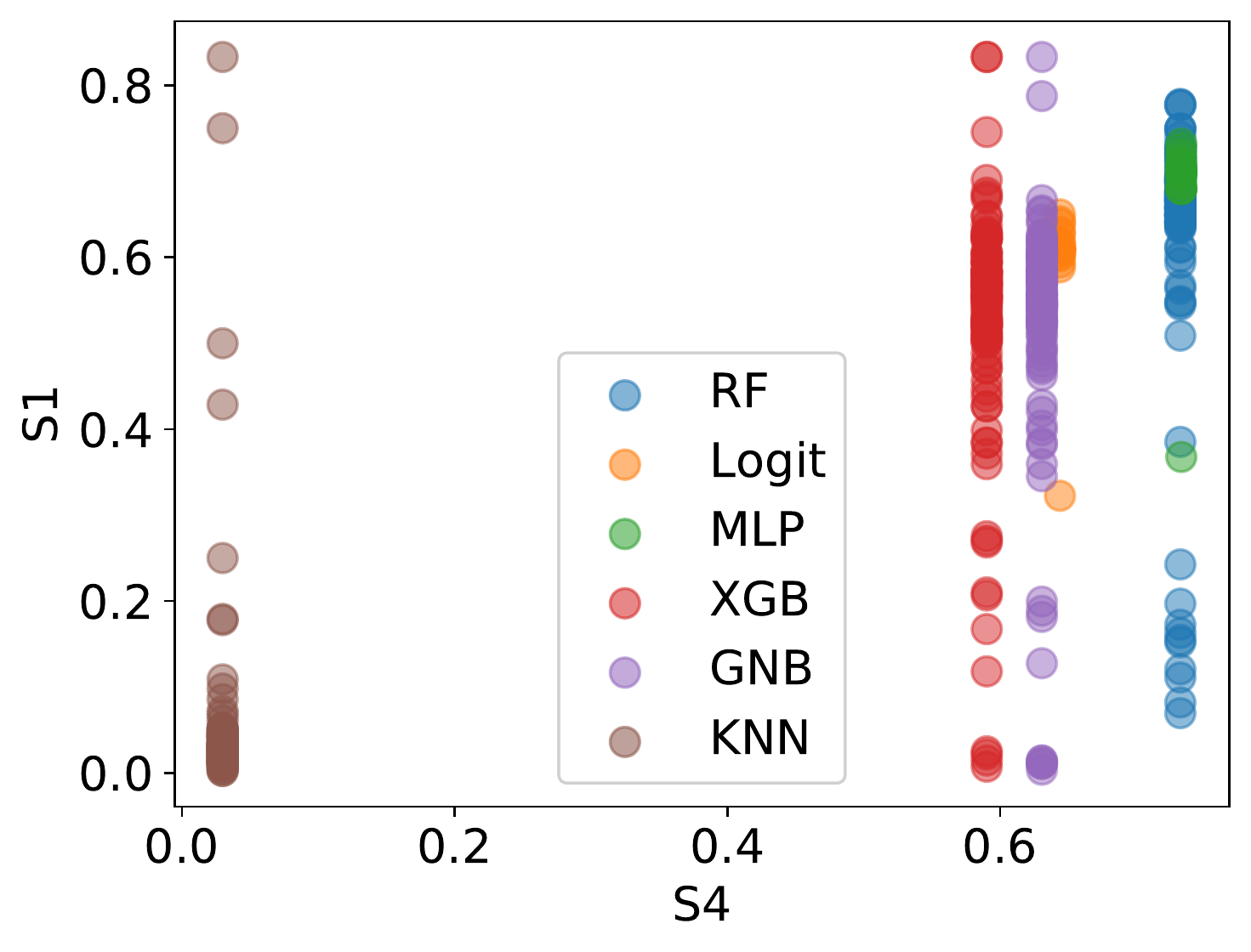}} 
	\subfloat[Beers-MLP]{\label{fig:beers_mlp}\includegraphics[width=0.17\textwidth]{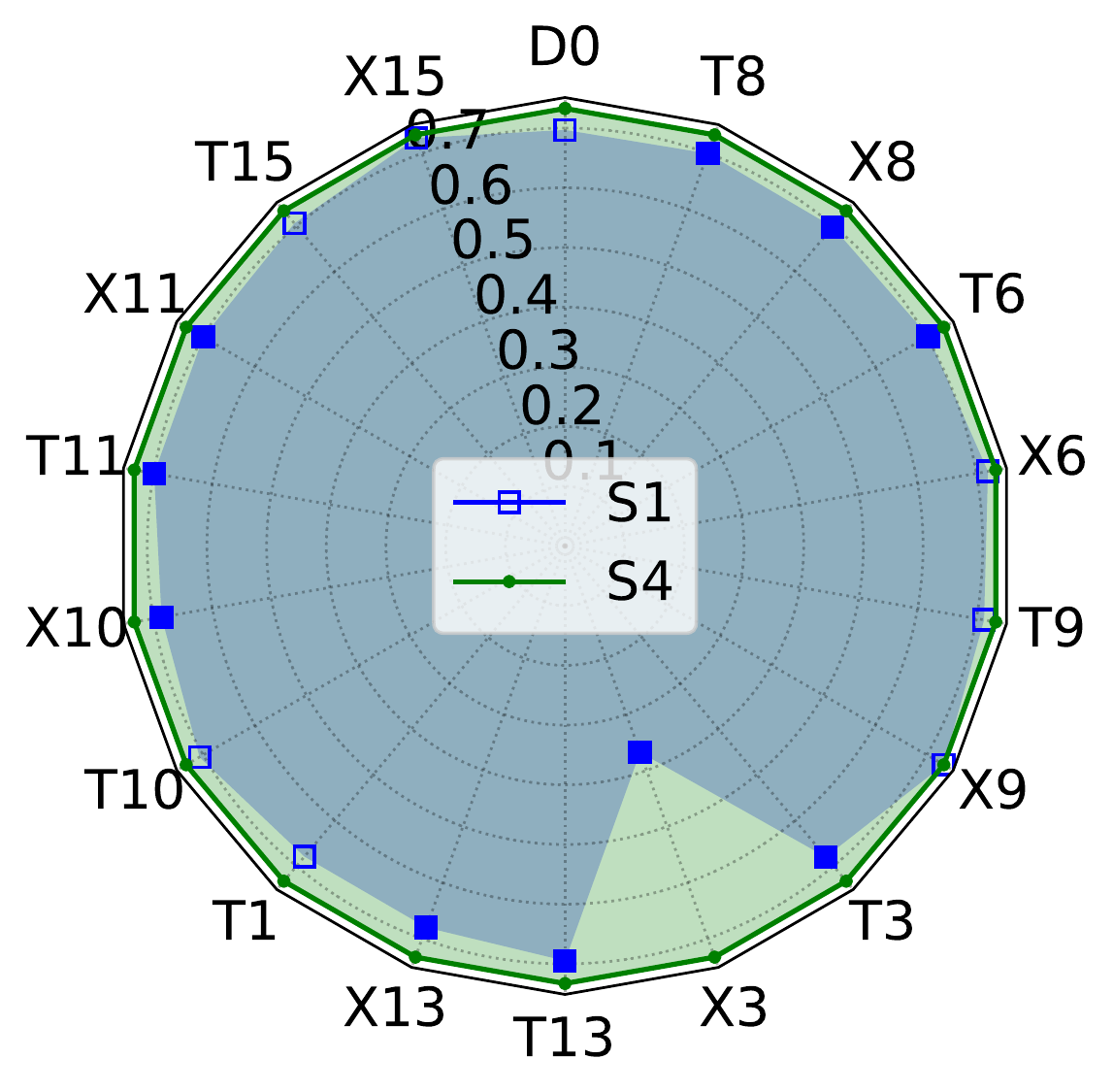}} 
	\subfloat[Adult]{\label{fig:adult__all_models}\includegraphics[width=0.22\textwidth]{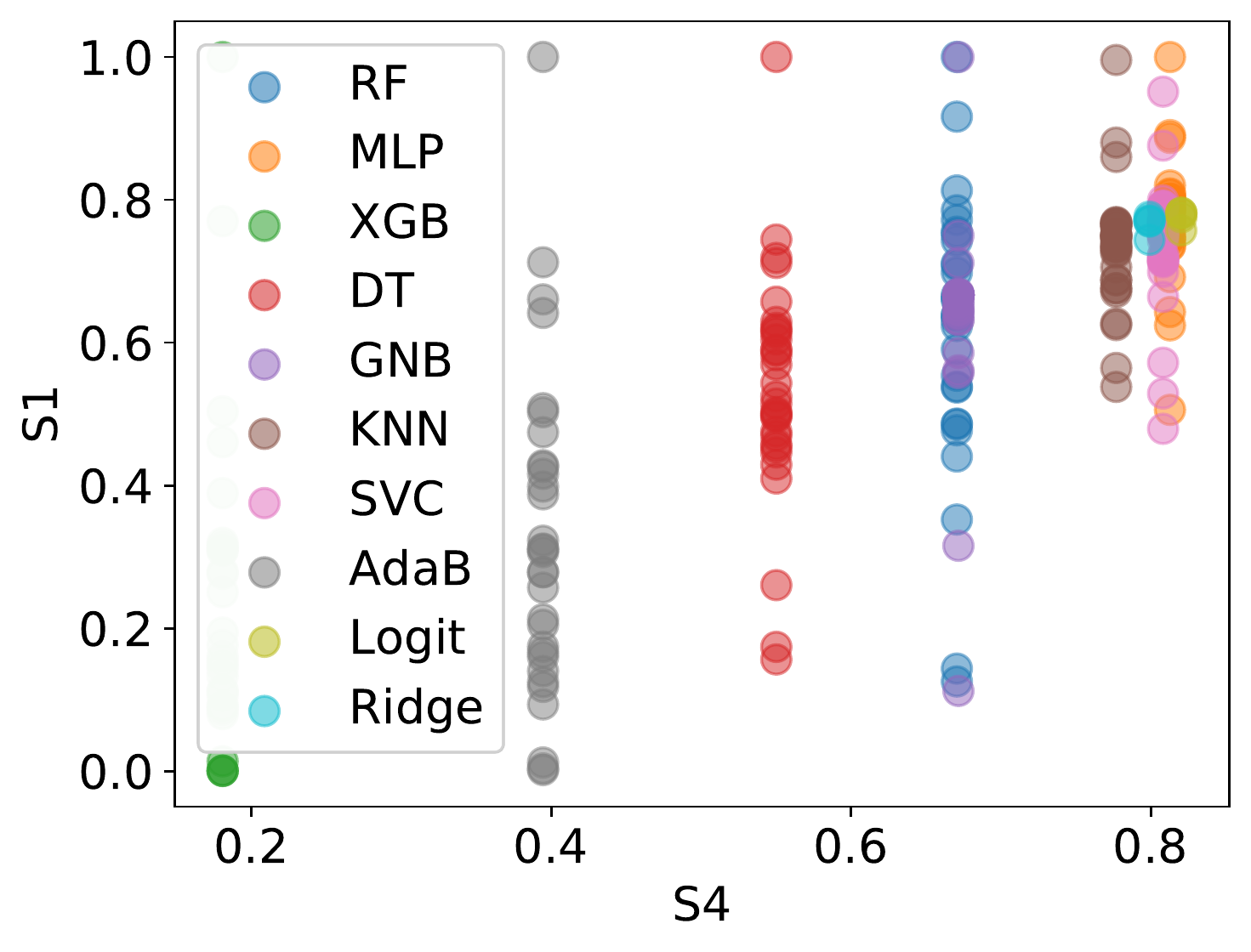}} 
	\subfloat[Adult-SVC]{\label{fig:adult_svc}\includegraphics[width=0.17\textwidth]{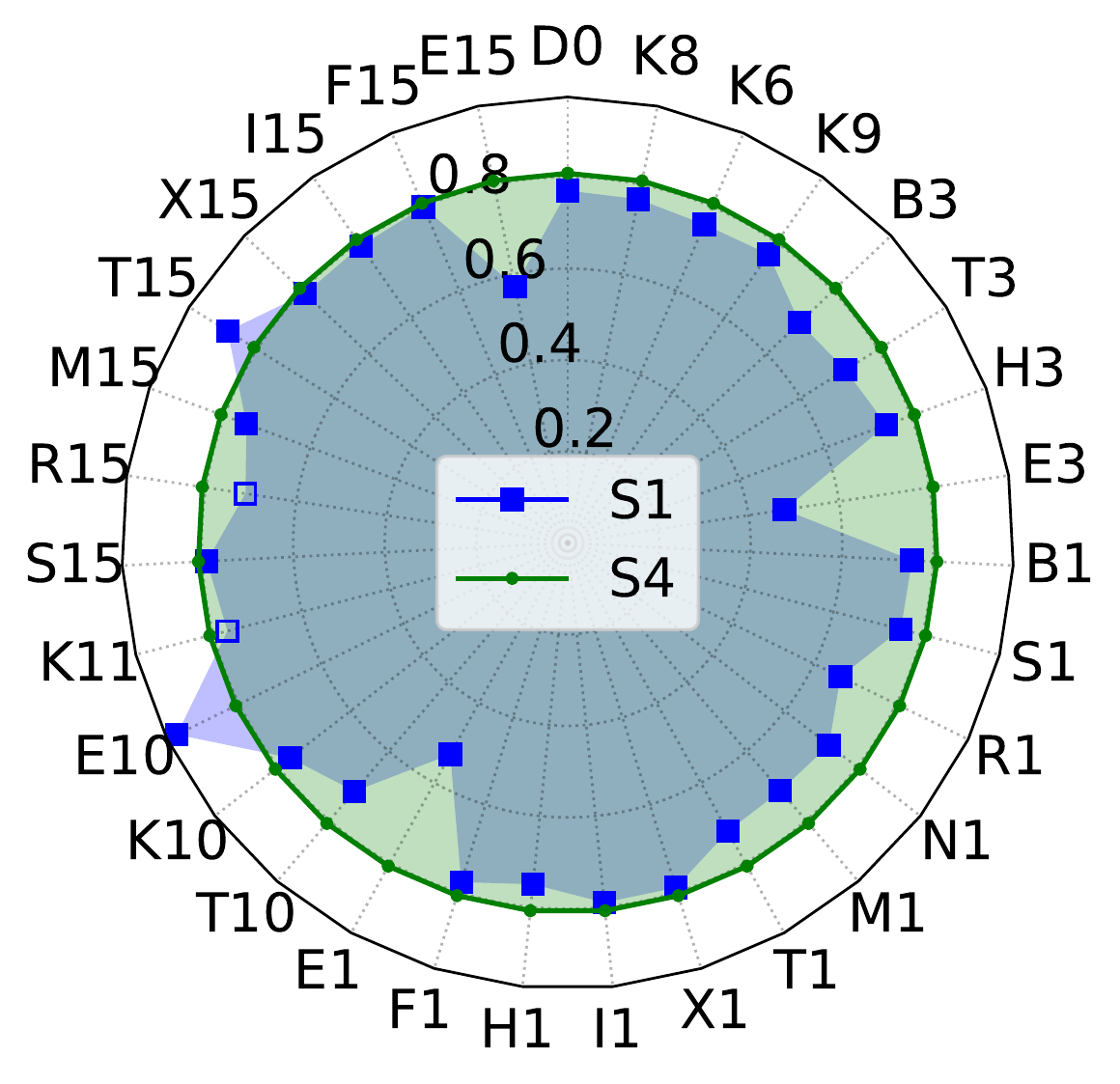}} 
	\subfloat[BreastCancer]{\label{fig:bc_all_models}\includegraphics[width=0.22\textwidth]{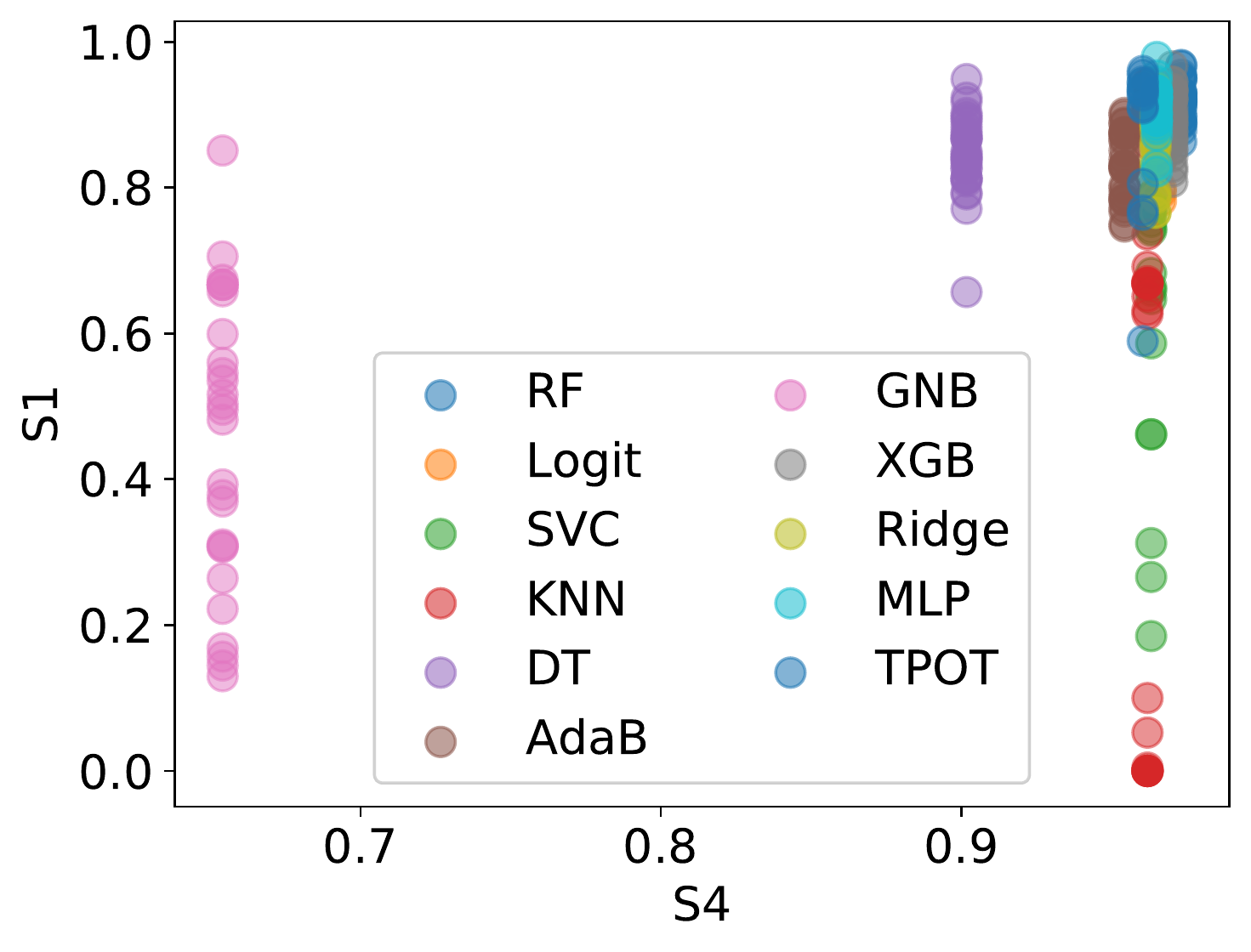}} \hfill
	\subfloat[BreastCancer-XGB]{\label{fig:bc_adab}\includegraphics[width=0.18\textwidth]{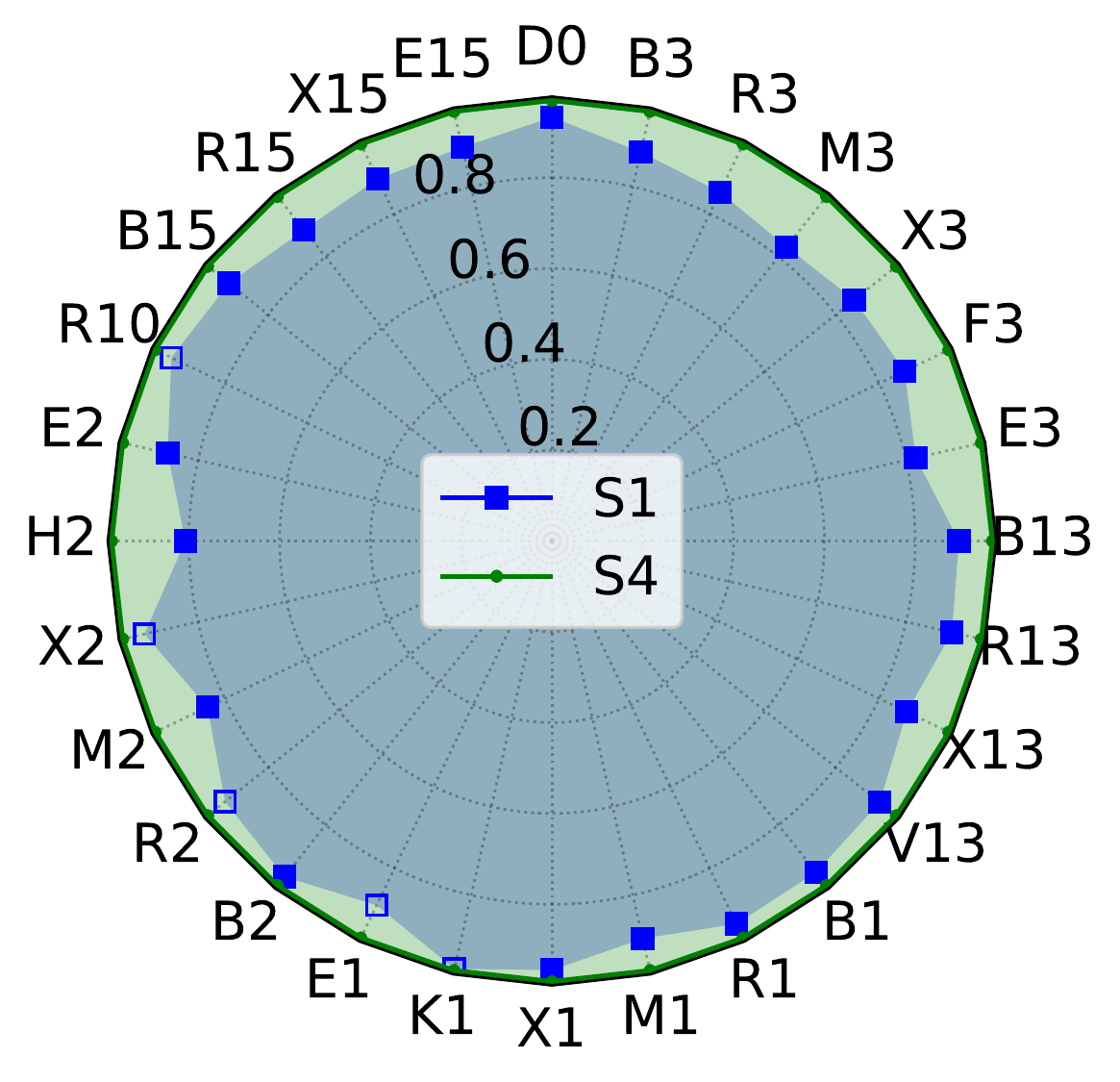}} \hfill
	\subfloat[Citation]{\label{fig:bc_tpot}\includegraphics[width=0.22\textwidth]{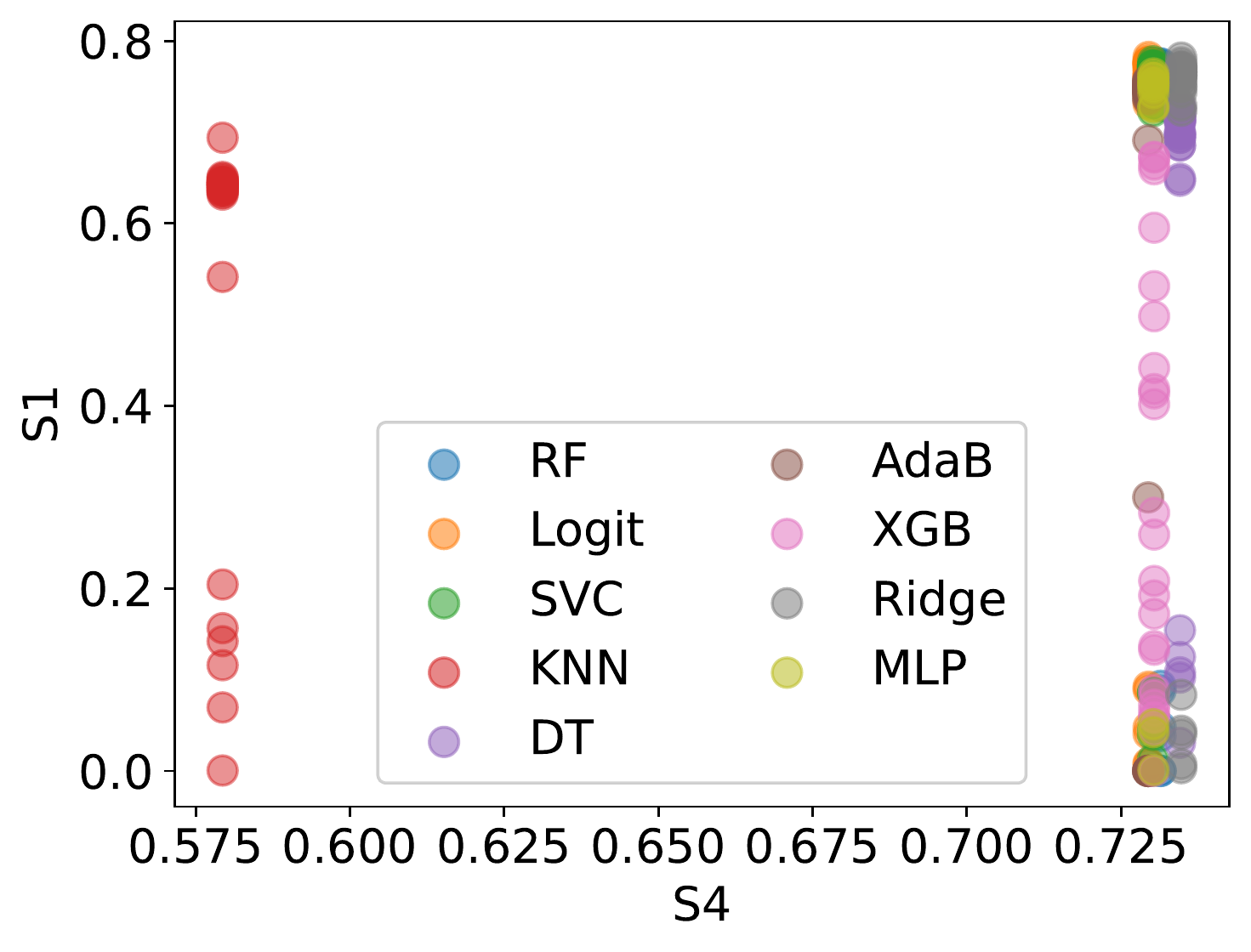}} \hfill
	\subfloat[Citation-Ridge]{\label{fig:citation_ridge}\includegraphics[width=0.18\textwidth]{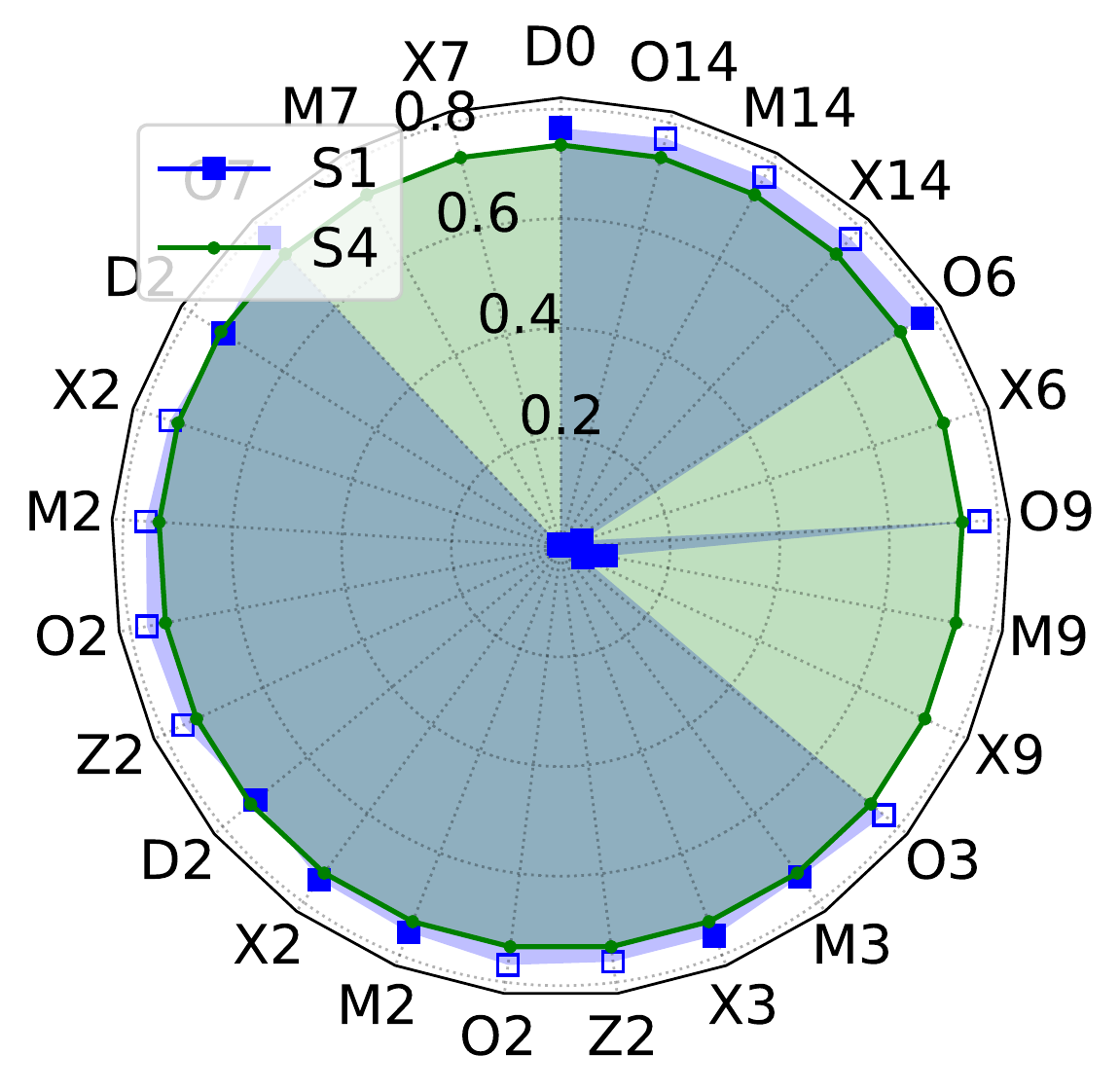}} \hfill
	\subfloat[SmartFactory-RF]{\label{fig:sf_rf}\includegraphics[width=0.18\textwidth]{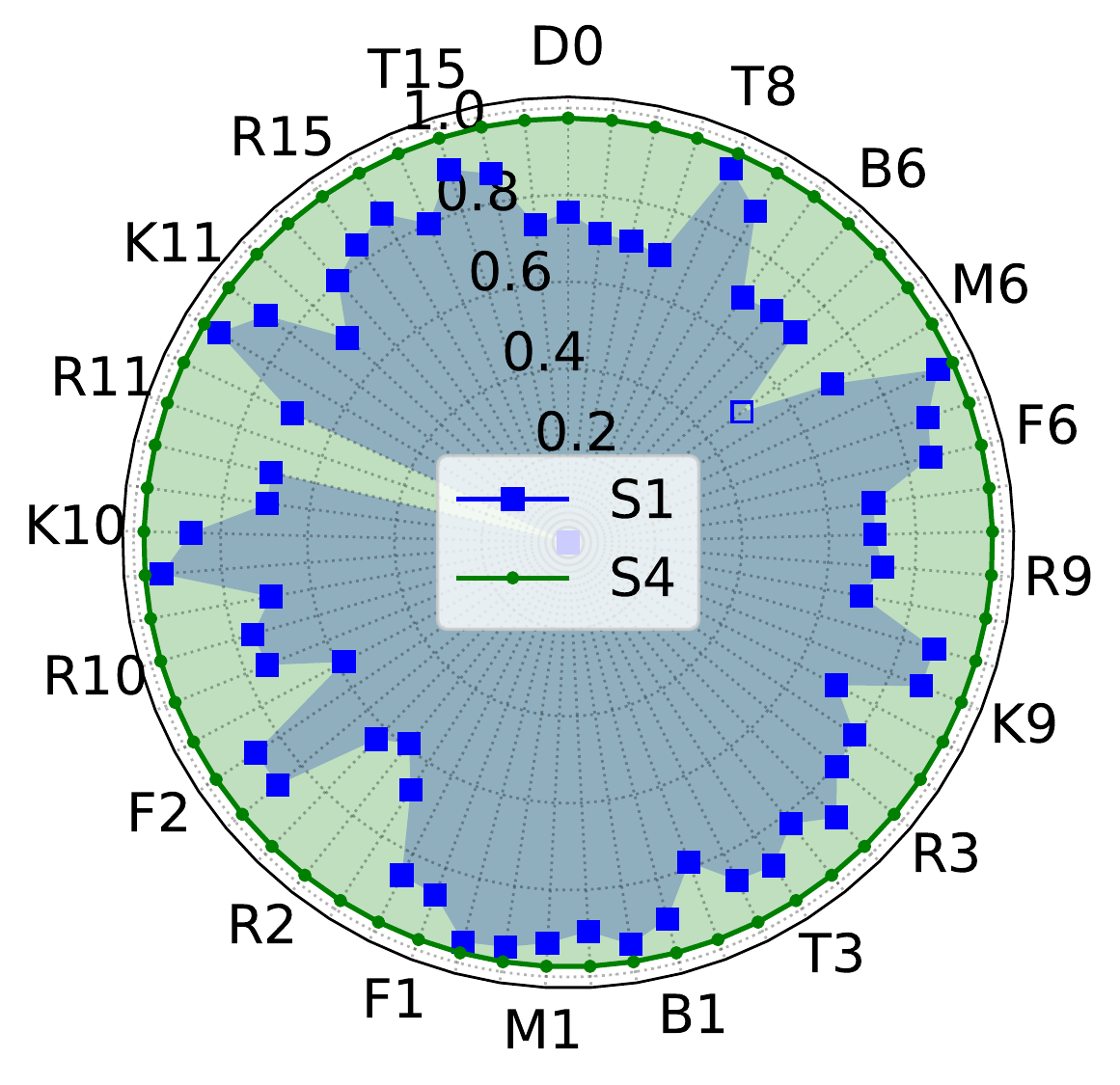}} 
	\subfloat[Nasa]{\label{fig:ns_all_models}\includegraphics[width=0.22\textwidth]{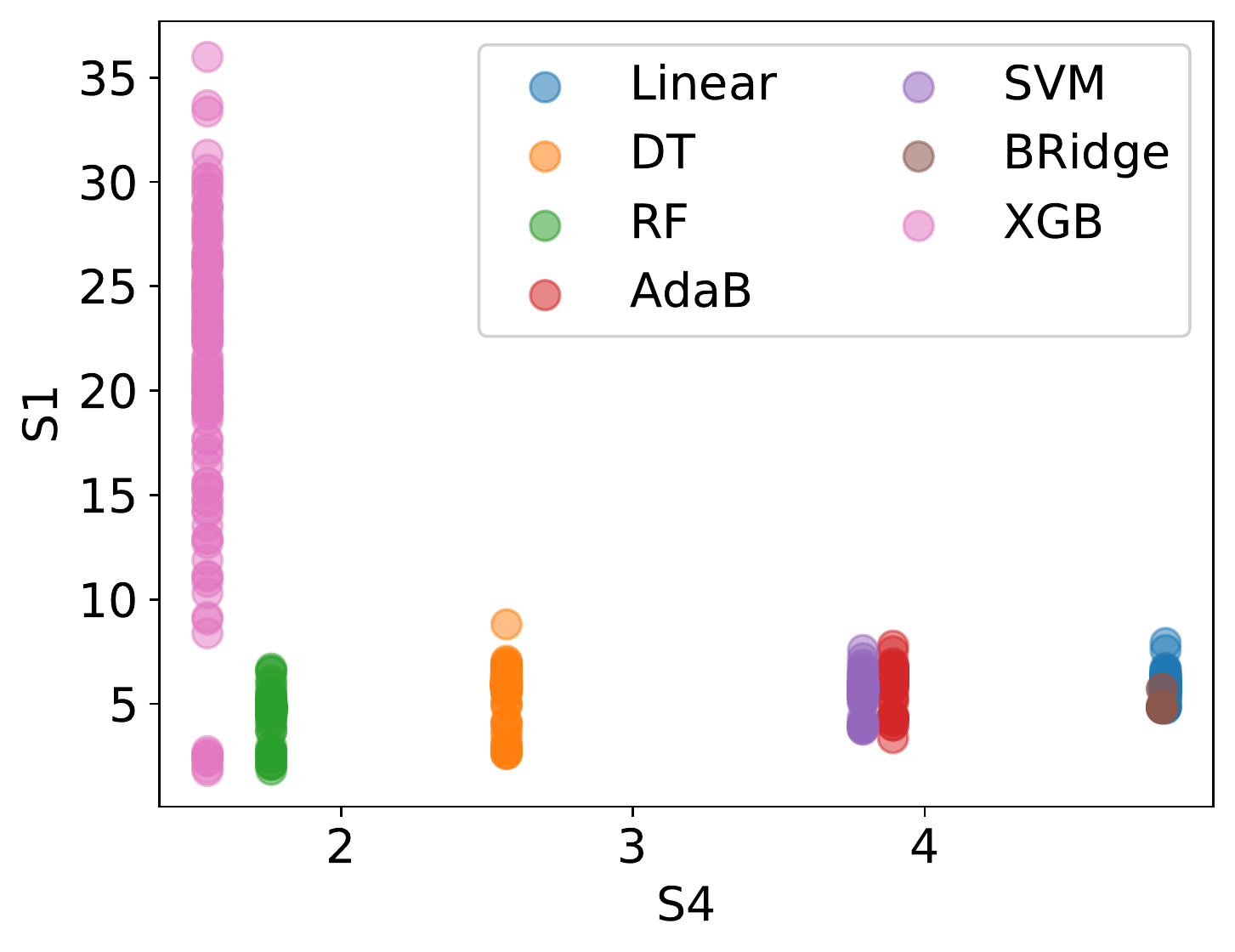}} \hfill
	\subfloat[Nasa-DT]{\label{fig:ns_dt}\includegraphics[width=0.19\textwidth]{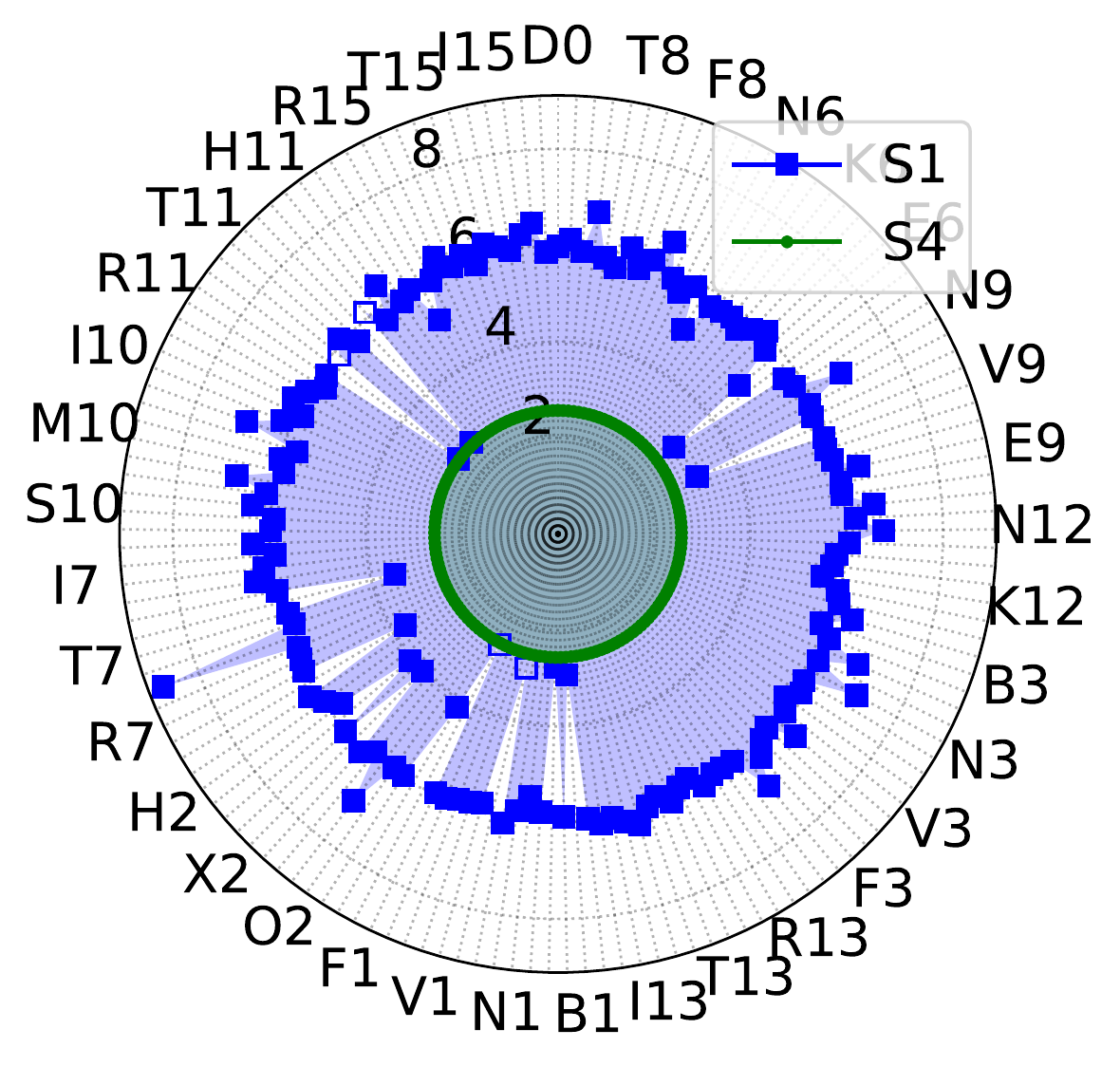}} 
	\subfloat[SoilMoisture]{\label{fig:sm_all_models}\includegraphics[width=0.22\textwidth]{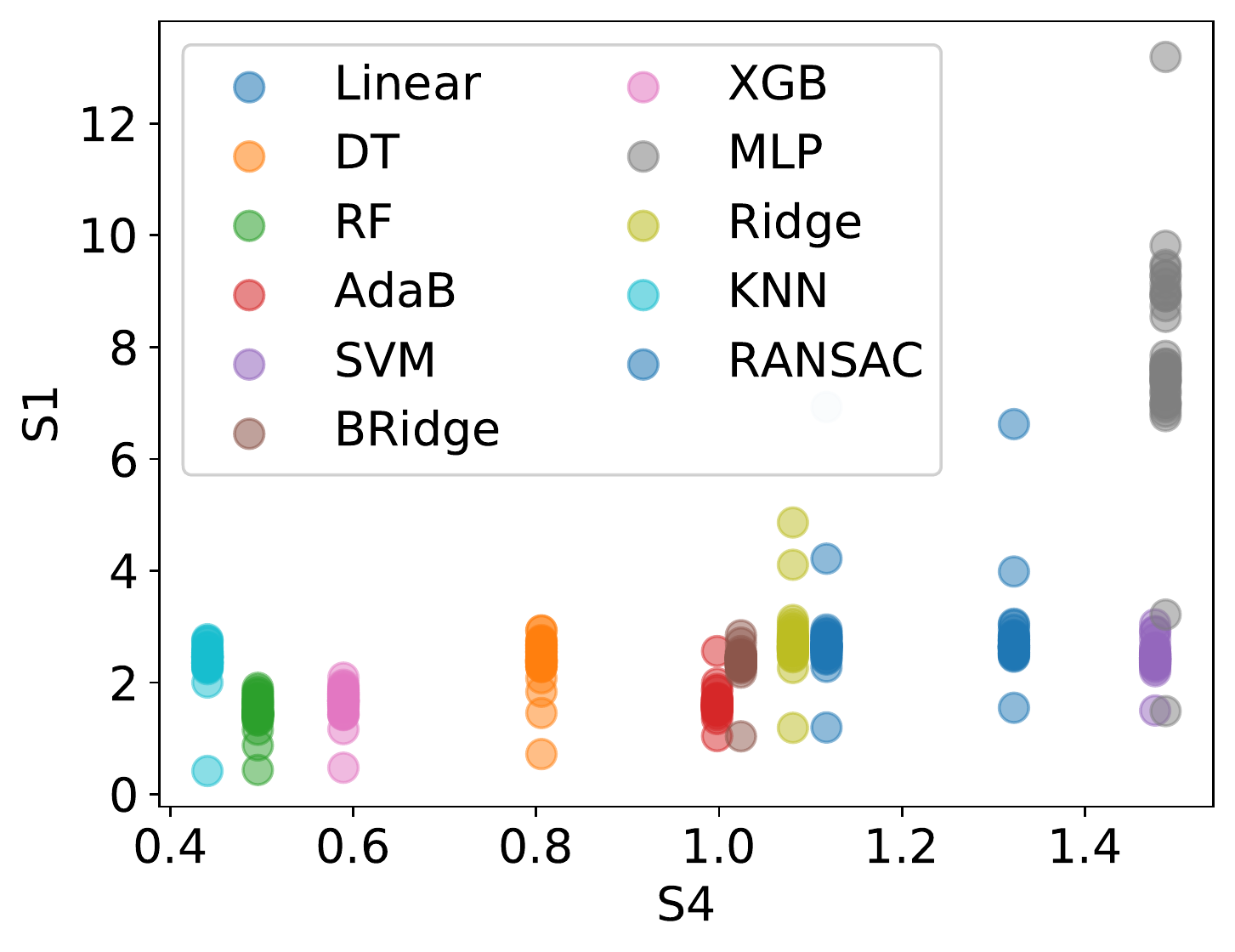}} 
	\subfloat[SoilMoisture-KNN]{\label{fig:sm_knn}\includegraphics[width=0.19\textwidth]{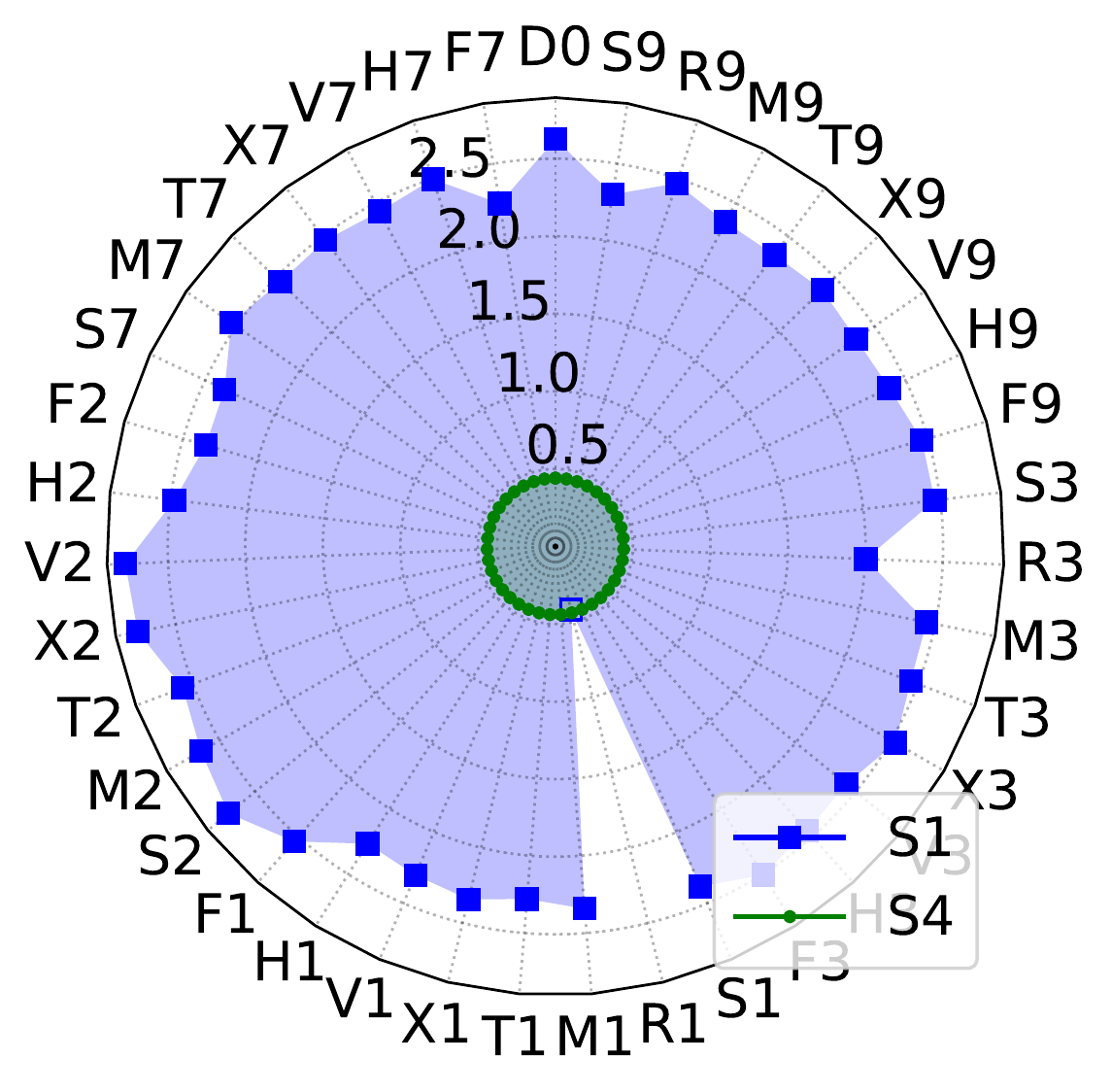}} 
	\subfloat[SoilMoisture-RANSAC]{\label{fig:sm_ransac_s2s3}\includegraphics[width=0.19\textwidth]{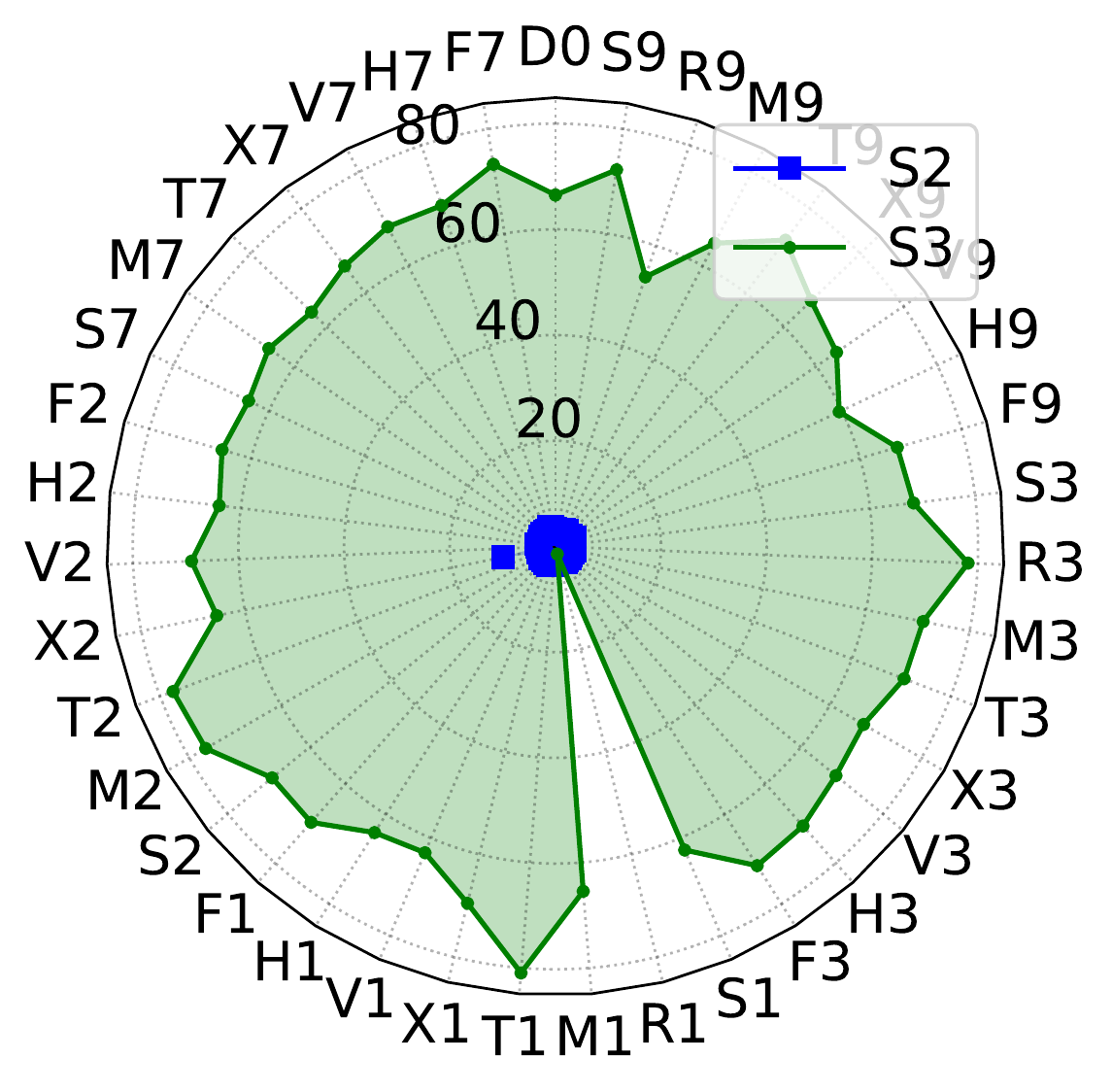}} 
	\subfloat[SoilMoisture-BayesRidge]{\label{fig:sm_bridge}\includegraphics[width=0.19\textwidth]{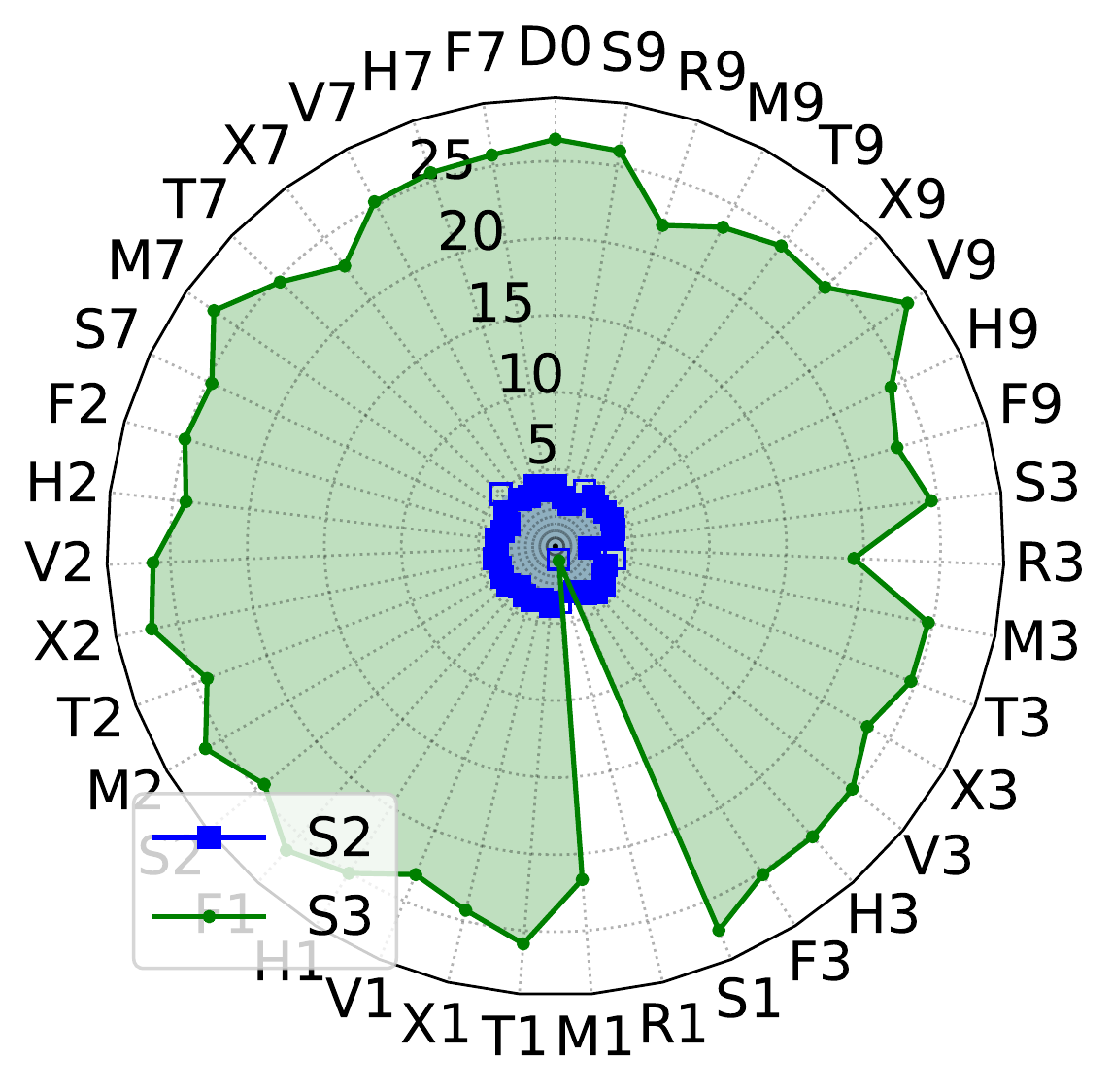}} \hfill
	%
	\subfloat[Water]{\label{fig:water_all_models}\includegraphics[width=0.22\textwidth]{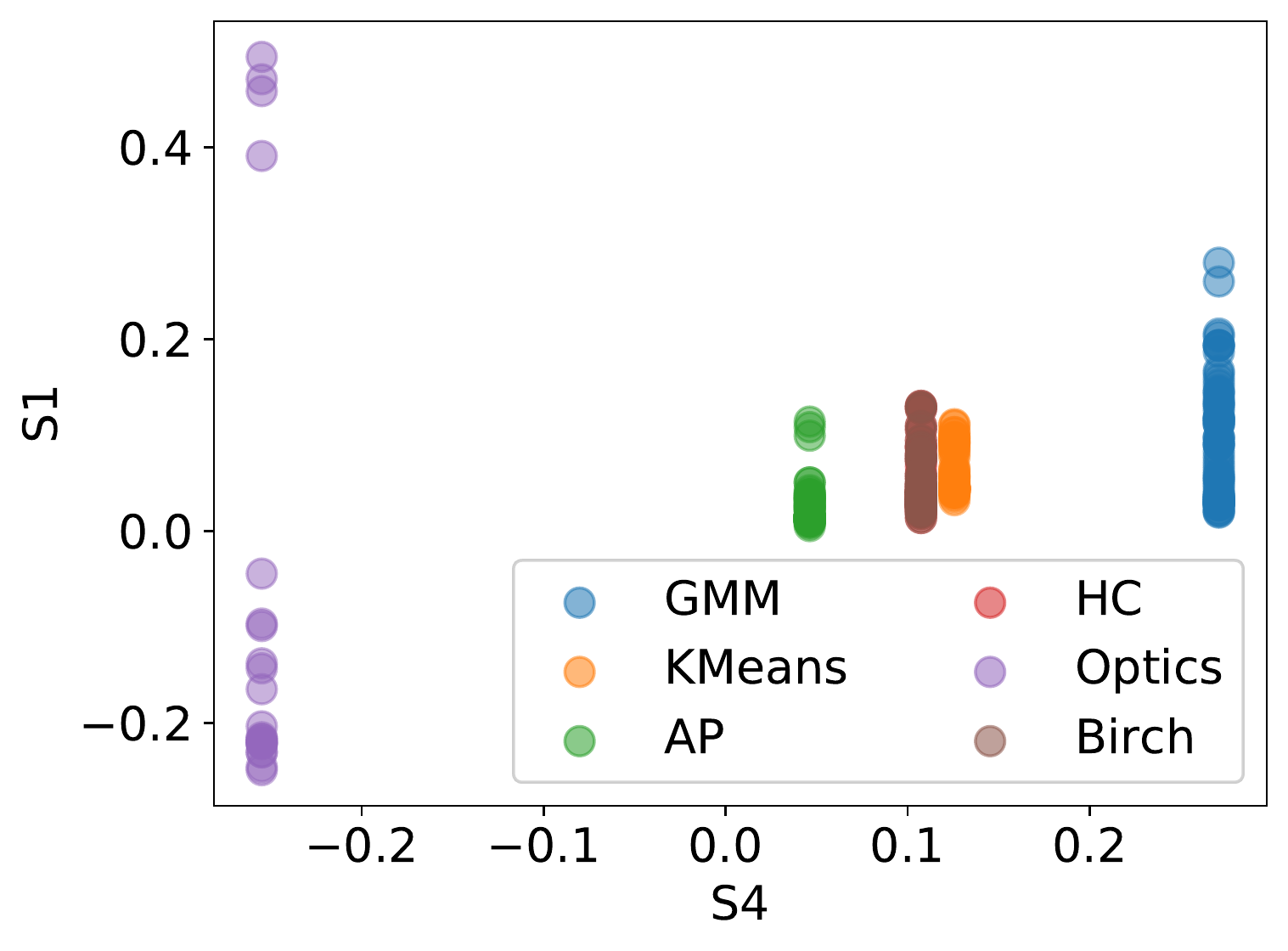}} 
	\subfloat[Water-Birch]{\label{fig:water_birch}\includegraphics[width=0.17\textwidth]{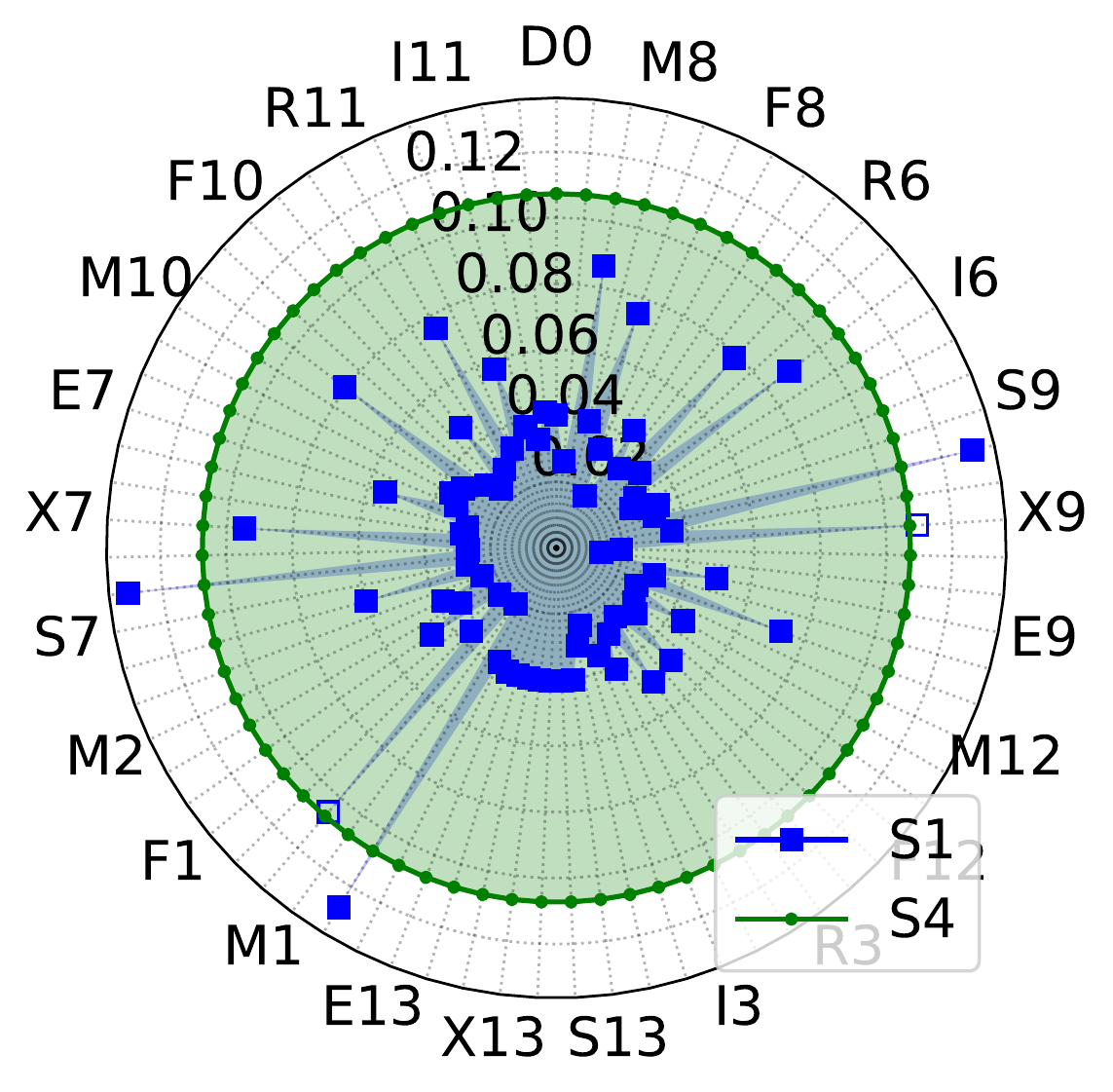}} 
	\subfloat[Power]{\label{fig:power_all_models}\includegraphics[width=0.22\textwidth]{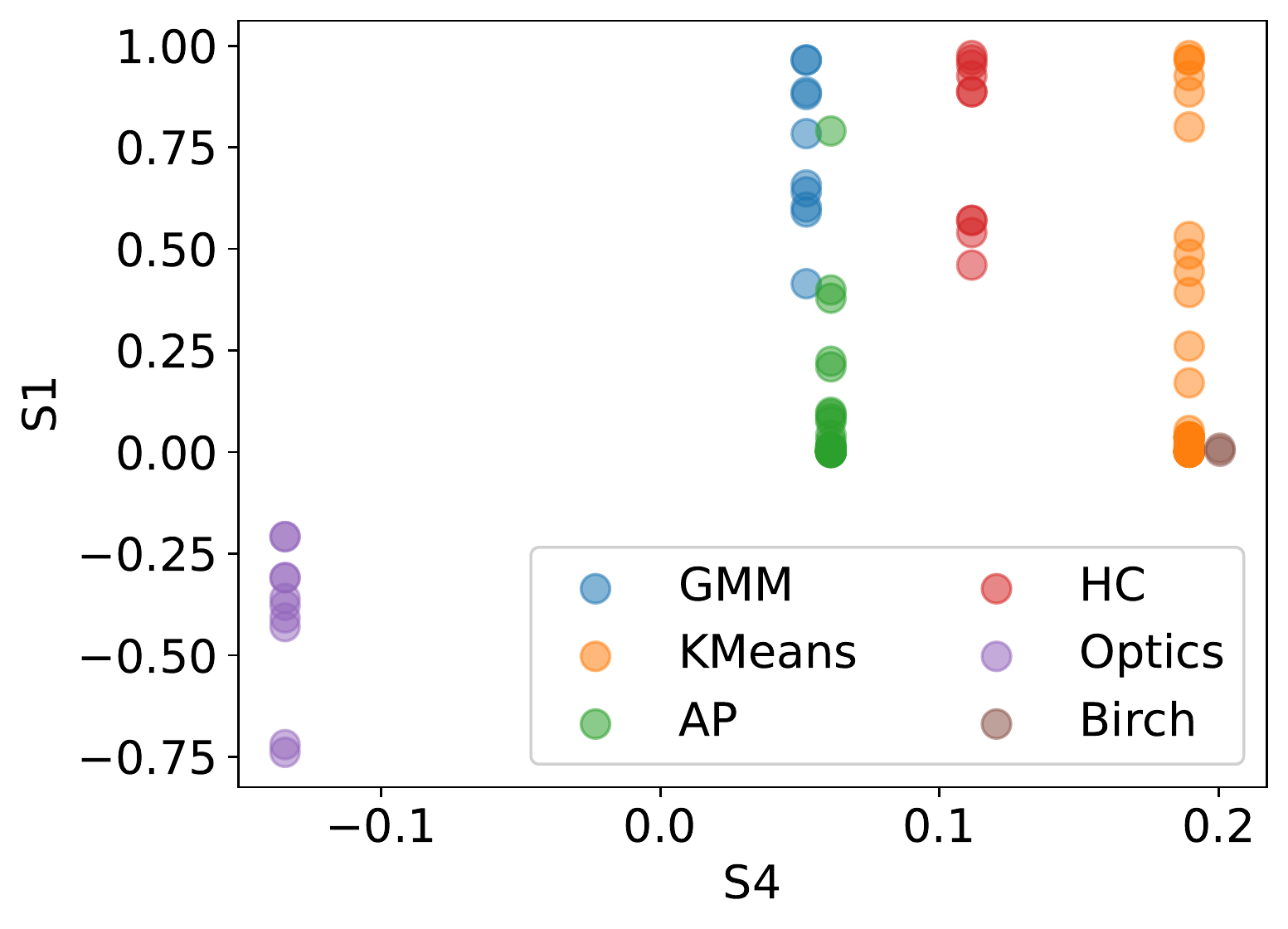}} 
	\subfloat[Power-KMeans]{\label{fig:power_kmeans}\includegraphics[width=0.17\textwidth]{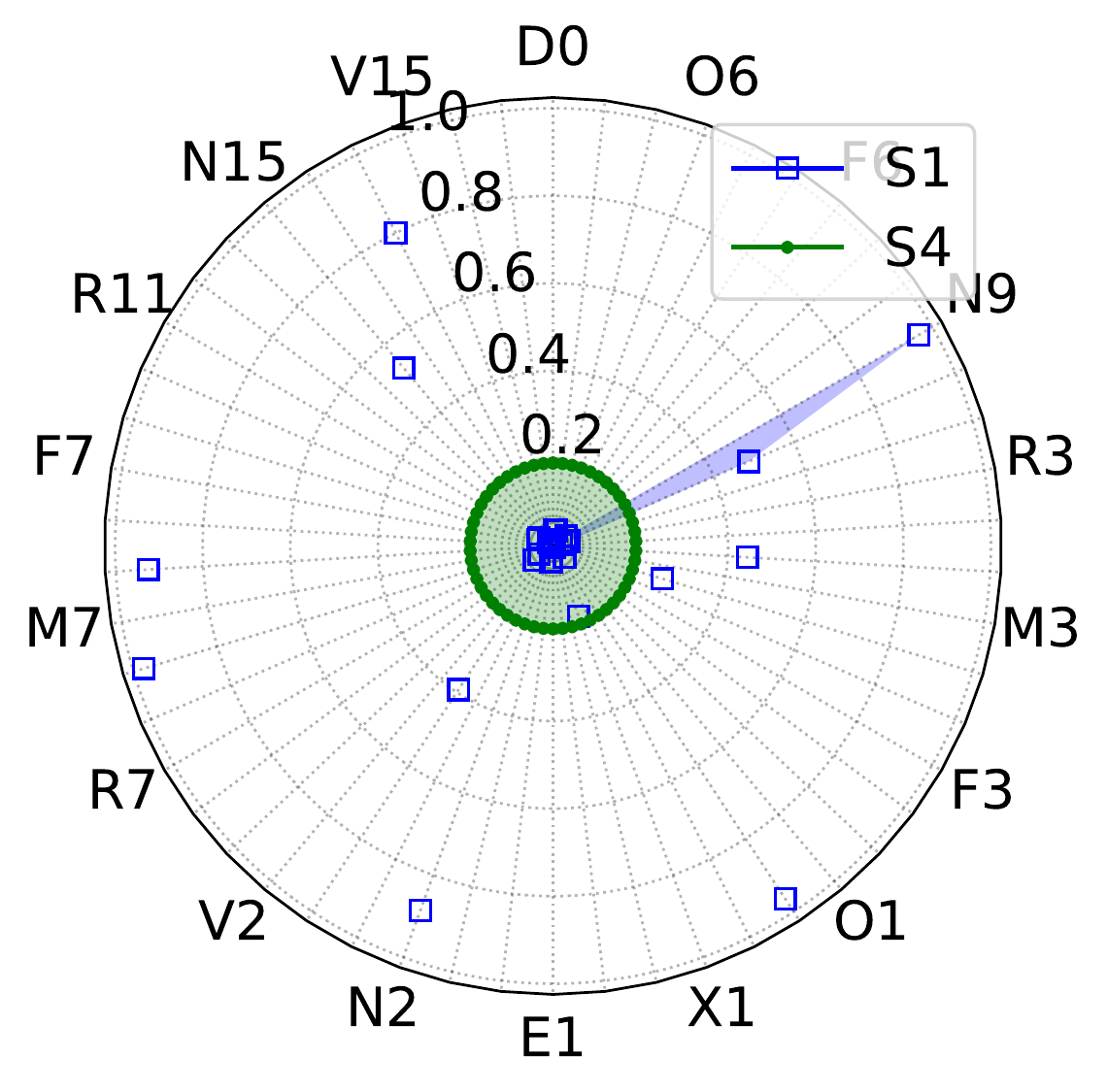}} 
	\subfloat[HAR]{\label{fig:har_all_models}\includegraphics[width=0.22\textwidth]{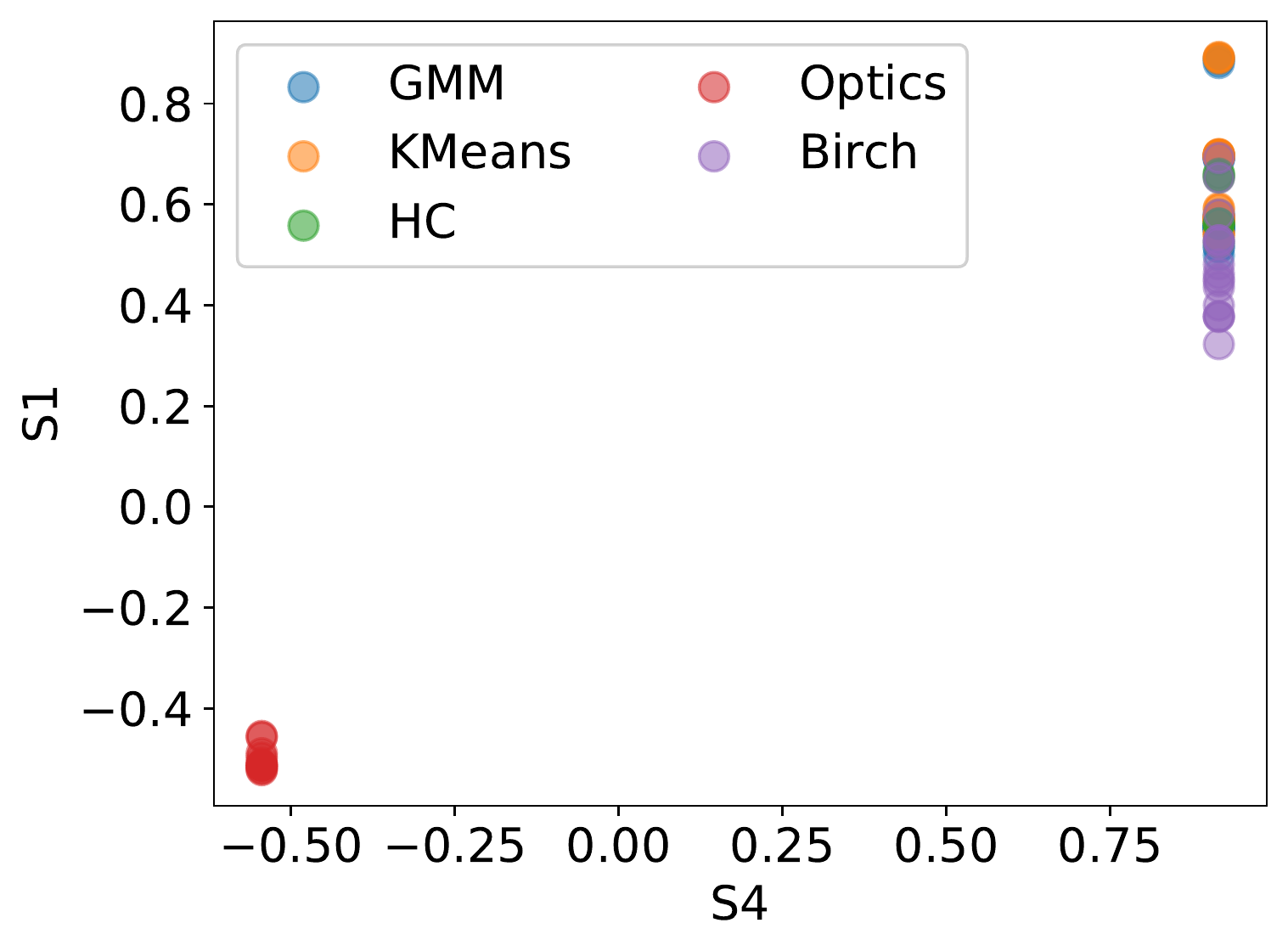}} 
\caption{Accuracy of ML Models trained on different data versions in different scenarios (F1 score, RMSE, and Silhouette metric for datasets with associated classification, regression, and clustering tasks, respectively)}
\label{fig:model_accuracy} 
\vspace{-3mm}
\end{figure*}

%
In this section, we present the results of modeling the various datasets in different scenarios. Figure~\ref{fig:model_accuracy} demonstrates the accuracy of different classification, regression, and clustering models trained on different data versions. For the \textit{Beers} dataset, Figure~\ref{fig:beers_all_models} shows the average F1 score of six classifiers in scenarios S1 and S4. As the figure depicts, the performance of all classifiers changes according to the quality of the repaired data. For example, the MLP classifier achieved an average F1 score of 0.732 in S4, while the accuracy in S1 ranges from 0.368 to 0.727. Figure~\ref{fig:beers_mlp} clarifies these results via comparing the performance of MLP models trained on different versions, i.e., dirty (D0), ground truth, and repaired, of the \textit{Beers} dataset in S1 (in blue) and S4 (in green). Obviously, the blue and green regions mostly overlap with each other. The only exception occurs with the combination X3, representing Max Entropy and standard imputation. Such a low accuracy, repeated with several classifiers, is usually caused by the low-quality repairs generated by different standard imputation methods. In this figure, the results of the A/B statistical test are delineated in the form of blue filled/empty square markers. In this context, a filled marker denotes that the null hypothesis $H_0$ can be rejected (i.e., the two MLP models in S1 and S4 are different), whereas an empty marker means failing to reject $H_0$. Thus, we can confirm that the performance difference of the models in S1 and S4 will remain, if we run the experiments for more than ten times.
%
Figure~\ref{fig:adult__all_models} compares the results of ten classifiers trained on different versions of the \textit{adult} dataset. In this figure, the distribution of the results in S1 enables us to identify the ML models robust to data quality problems. For instance, the results of DT in S1 range from 0.17 to 0.99, whereas the results of Ridge range from 0.74 to 0.78. Figure~\ref{fig:adult_svc} shows the performance of SVC when trained on different versions of the \textit{Adult} dataset. For most data versions, the accuracy of SVC is comparable in both scenarios. Despite achieving high detection accuracy, the detections of ED2 led to quality problems in most of its repaired versions, e.g., E1, E3, E10, E15. This behavior occurs due to the large number of false positives (118,741 cells) generated by ED2 (cf. Figure~\ref{fig:adult_detect_acc}). Similarly, Figures~\ref{fig:bc_all_models} and \ref{fig:bc_adab} depict the modeling accuracy for different versions of the \textit{Breast Cancer} dataset. For this dataset, DT performed well with a relatively tight range of F1 scores from 0.65 to 0.94, compared to GNB whose F1 scores range from 0.15 to 0.85. Figure~\ref{fig:bc_adab} depicts that the performance of XGBoost is slightly better in S4 than in S1 for most repaired versions of the \textit{Breast Cancer} dataset. 

For the \textit{Citation} dataset, which includes duplicates and mislabels, Figure~\ref{fig:bc_tpot} demonstrates the F1 score of several classification models in the scenarios S1 and S4. As it can be seen in the top right corner of the figure, most classifiers yield similar performance as the ground truth when applying the \quotes{Delete} strategy. Other cleaning strategies which rely on ML-based imputation, e.g., M6, M7, M9, X7, X6, and X9, cause the predictive performance to be substantially deteriorated (cf. Figure~\ref{fig:citation_ridge}). Unlike other classifiers, XGBoost exhibits poor performance over the dirty and the most repaired data versions (F1 score ra   nges from 0.05 to 0.8 and has a high density under the value of 0.26, as depicted in Figure~\ref{fig:bc_tpot}). To further understand the impact of mislabels, we carried out experiments on the \textit{Adult} and \textit{Breast Cancer} datasets after adding noise to the labels (i.e., flipping some binary labels). The results of such experiments show that several ML models, e.g., MLP, RF, DT, trained on dirty versions have slightly worse performance than the same models trained on the ground truth (for RF, an average F1 score of 0.9 for the dirty dataset and 0.93 for the ground truth).  

Figures~\ref{fig:ns_all_models}-\ref{fig:sm_bridge} illustrate the performance of various regression models trained on different datasets. As depicted in Figures~\ref{fig:ns_all_models}, XGB achieved the highest accuracy in S4 (RMSE of 1.54). However, its performance broadly depends on the quality of the repairs (cf. the RMSE values in S1 which range from 1.78 to 35.9). Conversely, DT and RF have tighter distribution of RMSE values in S1. Figure~\ref{fig:ns_dt} demonstrates that DT has approximately the same predictive performance over the most repaired data versions. The figures also some cleaning strategies, e.g., X2, X7, X8, N11, and K11, which achieve similar performance as the ground truth. For the \textit{Soil Moisture} dataset, Figure~\ref{fig:sm_all_models} depicts that KNN outperforms other models with a relatively tight distribution of the RMSE values in S1. For this dataset, the detections of RAHA repaired using the ground truth led to a comparable RMSE as that obtained in S4, as depicted in Figure~\ref{fig:sm_knn}. In Figures~\ref{fig:sm_ransac_s2s3} and \ref{fig:sm_bridge}, we demonstrate the performance of RANSAC and Bayesian Ridge in scenario S2 and S3 (cf. Table~\ref{tab:scenarios}). Obviously, RANSAC and Bayesian Ridge perform in S2 much better than in S3. Since this result appeared in all other datasets, we can deduce that models trained on dirty or relatively low-quality repaired data may perform well whenever they are tested/served using high-quality data. 

Aside from regression, the accuracy of several clustering methods also have been measured in terms of the silhouette index, as illustrated in Figures~\ref{fig:water_all_models}-\ref{fig:har_all_models}. The results showed that some clustering methods, e.g., Optics, GMM, and HC, yielded a comparable performance in S1 and in S4, or even better in S1 for several repaired versions, as depicted in Figures~\ref{fig:water_all_models} and \ref{fig:power_all_models}. For instance, Figure~\ref{fig:water_birch} compares the performance of Birch when clustering different versions of the \textit{Water} dataset. In general, Birch performed in S4 better than in S1. However, there exist several repaired methods which exhibit better clustering performance (on average by 16\%, 18\%, and 17\% for R1, R7, and R9, respectively) than the ground truth. Figure~\ref{fig:power_kmeans} shows similar results for K-Means while clustering the \textit{power} dataset. Finally, Figure~\ref{fig:har_all_models} compares the performance of five clustering methods trained on the \textit{HAR} dataset. The figure shows that all models have a relatively tight distribution in S1, which implies non-sensitivity to the quality of the repaired versions. Several repaired versions, generated using the detections of RAHA (e.g., R1, R2, and R6), led to similar performance as the ground truth.\vspace{-4.3mm}
%
\subsection{Lessons Learned}
%
\paragraph{Main Findings:} In this section, we highlight the main findings and lessons learned throughout this study. Through extensive experiments, \PaperAcronym proved that evaluating the error detection and repair methods in isolation from the downstream applications, e.g., predictive tasks, can be broadly misleading. For instance, Figures~\ref{fig:beers_detect_acc},~\ref{fig:sf_detect_acc}, and \ref{fig:bike_detect_acc} show that KATARA suffers from many false positives. Moreover, the quality of repairs generated for the detections of KATARA is sometimes worse than the dirty versions of the datasets (cf. Figure~\ref{fig:bike_repair_acc}). Nevertheless, Figures~\ref{fig:adult_svc},~\ref{fig:bc_tpot}, \ref{fig:sf_rf}, and \ref{fig:ns_dt} clearly depict that the ML models trained on the KATARA-based repaired data versions have a comparable predictive performance to the other models. Similar conclusions can also be drawn for other detectors, such as FAHES, NADEEF, and HoloClean. In fact, most error detection and repair methods are typically evaluated using their performance relative to the ground truth \cite{raha19,baran20,ed219,holoclean17,holodetect19}. Accordingly, the finding above represents a major result which guides researchers and developers on how they can effectively evaluate their data cleaning methods.

Another interesting finding is that classification models are more robust to attribute errors than regression models and clustering methods. Through comparing the performance of different models in Figure~\ref{fig:model_accuracy}, it is clear that the differences between S1 (blue regions) and S4 (green regions) for almost all classifiers are relatively small. Conversely, regression models and clustering methods remarkably perform in S4 better than in S1. Accordingly, data cleaning is a necessary component in the pipelines of regression and clustering applications. Furthermore, classification applications may not need to implement a sophisticated data cleaning method. Simple cleaning methods can supply the classification models with the necessary quality level that is needed to train the models. At the same time, simple error detection and repair methods do not require excessive time, hence we can broadly accelerate the data preparation process. In the presence of class errors, some classifiers exhibited relatively poor performance. Hence, automated mislabels detection methods are necessary to produce accurate predictions. For the examined AutoML algorithms, i.e., TPOT and Auto-Sklearn, the results showed that they do not \textit{always} produce the most accurate models. For example, in case of the \textit{Breast Cancer} dataset, the models generated by TPOT with X13 and X15 have F1 scores of 0.75 and 0.6, respectively. Compared to TPOT with B15 and X2, which have F1 scores of circa 0.98 and 0.99, respectively. Thus, these algorithms may fail to generate accurate models in case of improper data cleaning.

\vspace{-2mm}\paragraph{Error Detectors:} Regarding the error detection methods, it is obvious that ML-based and ensemble methods, in most cases, have a higher detection accuracy than the other non-learning methods, as illustrated in Figure~\ref{fig:det_results}. However, the results also showed that most detectors lack consistency over different datasets, i.e., their performance varies from one dataset to another. For instance, Figure~\ref{fig:beers_detect_acc} shows that ED2 detected all errors with high precision in the \textit{Beers} dataset. Nevertheless, it suffered from false positives and false negatives in other datasets, such as \textit{Adult}, \textit{Nasa}, and \textit{HAR} (cf. Figures~\ref{fig:adult_detect_acc},~\ref{fig:nasa_detect_acc}, and \ref{fig:har_detect_acc}). Similarly, NADEEF performed poorly (an average F1 score of 0.12) in case of the \textit{Nasa} dataset, whereas it achieved a reasonable performance (an average F1 score of 0.91) in case of the \textit{Power} dataset. Other shortcomings of ML-based detectors are as follows: (1) They are not able to recognize the error type, i.e., they only provide a binary decision for each cell of whether it is erroneous. This behavior may make it complex to select a well-suited data repair tool. (2) They suffer from poor scalability (cf. results in Figures~\ref{fig:soccer_accuracy}-\ref{fig:soccer_runtime}). (3) They require users intervention to label data. Accordingly, it is necessary to exert more efforts to advance the ML-based detectors for the sake of resolving the above shortcomings.

The results illustrated that the performance of rule-based error detectors broadly relies on the number and the quality of the user-provided rules/constraints. For instance, the F1 score of HoloClean, in case of the \textit{Adult} dataset, is dropped from 0.51 to 0.12, when the number of provided rules is reduced from 17 to seven. Accordingly, it is crucial to integrate an automated rules/constraints generator with such detectors to improve their performance. In this context, we highlight that configuration-free methods are generally simple and easy to be employed, but they usually need long times to find the most suitable configurations, e.g., dBoost and RAHA (cf. Figures~\ref{fig:beers_detect_time},~\ref{fig:sf_detect_time}, and \ref{fig:har_detect_time}). It is worthwhile mentioning that the current implementation of RAHA, ED2, and Meta do not work in the presence of duplicates in the dirty datasets. This problem mainly occurs since the dirty and ground truth versions of the dataset become of different lengths. In this case, these detectors are not able to use the ground truth to simulate a human annotator, i.e., for labeling the dirty cells. Picket represents an exception to this fact since it relies on self-supervision. Therefore, it does not mandate user-provided labels. However, the results showed that Picket is only suitable for small datasets, where it does not scale well due to the complexity of self-supervision. For larger datasets, e.g., \textit{Adult} and \textit{Smart Factory}, Picket was terminated since it caused memory faults.


\vspace{-3mm}\paragraph{Repair Methods:} For a better repair experience, it is found that the detection precision has a relatively higher impact on the repair quality than the detection recall (cf. Figures~\ref{fig:beers_detect_acc} and \ref{fig:bike_detect_acc}). The reason behind such a superiority is to avoid false positives which may drive the adopted repair method to either introduce new erroneous cells or remove all the detected cells, causing the repaired dataset to be entirely out of sync with the ground truth. However, an effective repair method can even avoid the negative impacts of false positives in the detection phase. For instance, NADEEF, in the case of the \textit{Beers} dataset, generated many false positives. Nevertheless, these false positives have circa no impact when the detections are repaired using GT (simulates a highly-effective repair method). In this case, false negatives in the detection phase become more harmful than false positives, in the presence of highly-effective repair methods.


For ML-oriented repair methods, we noticed that CPClean and BoostClean are hardly applicable to datasets associated with multi-class classification tasks. The underlying reason is that the methods divide each dirty dataset into batches, and each batch has to include samples from all classes. However, obtaining samples from each class is not always possible when there are several minority classes. For the datasets which have a binary classification problem, if the labels comprise erroneous cells, CPClean and BoostClean may not work due to introducing new values in the labels, turning the problem into a multi-class classification. For ActiveClean, it starts with partitioning the dirty dataset to obtain a clean fraction (i.e., data fraction without any errors) for warming up. Such a partition needs to represent all possible classes in the dataset. Therefore, ActiveClean searches for a partition that meets this condition. If it does not find such a partition, it returns an exception. Such a problem may happen in the following situations: (1) a dataset has too many classes with multiple minor classes (e.g., \textit{Beers}) and (2) there exist no sufficient clean cells in the dataset. 

\vspace{-3mm}\paragraph{Actionable Suggestions:} Based on the results obtained in \PaperAcronym, we provide the following suggestions while designing or selecting data cleaning tools: (1) tailor the design and evaluation of data cleaning methods to the planned downstream applications to properly select a well-suited cleaning method; (2) adopt simple cleaning strategies (non-learning detectors and generic repair methods) with classification tasks to combat attribute errors and more advanced cleaners (ML-based) with regression and clustering tasks; (3) exploit advanced techniques to combat class errors, e.g., CleanLab, data valuation, label smoothing, and noise-aware learning \cite{picado2020learning,song2022learning}; (4) employ automated tools, e.g., FDX profiler and Metanome \cite{metanome2015}, to extract integrity constraints and functional dependency rules to properly use cleaning tools, such as NADEEF and HoloClean, with minimal user involvement; (5) adopt duplicates detection tools, e.g., ZeroER, record linkage and data hashing, as early as possible, in the ML pipeline, to prevent data leakage between the training, the validation, and test sets; and (6) avoid ML-based error detectors, e.g., ED2, RAHA, and Picket, while preparing large volumes of data (i.e., over 50k rows, as shown in Figure~\ref{fig:soccer_accuracy}) due to their poor scalability.

\vspace{-3mm}\section{Related Work}\label{sec:rw}
%
%
In fact, there exist few studies which survey or compare the already-existing data cleaning methods \cite{lee2021survey,ridzuan2019review,max_min16,cleanml19}. Lee et al. \cite{lee2021survey} introduce a survey of five data cleaning methods and propose several research directions, such as integrating data cleaning methods with visual interface and the usage of high-performance memory management hardware solutions. Similarly, Ridzuan et al. \cite{ridzuan2019review} presents a review of data cleaning methods together with their challenges for dealing with big data. CleanML \cite{cleanml19} introduces a relational database schema designed to organize the experimental results of investigating the impact of data cleaning on ML classification tasks. Since it does not consider the ground truth of each dataset, CleanML overlooks comparing the performance of ML models when trained using ground truth and repaired datasets. Moreover, CleanML limits the evaluations to simple classification tasks, while ignoring other ML tasks such as regression, clustering, and AutoML algorithms. In addition, CleanML does not consider the holistic, semi-supervised, or ML-oriented error detection and repair methods. In \PaperAcronym, we tackle these shortcomings to generalize our findings to properly guide practitioners and data scientists while dealing with data cleaning problems in tabular data.

\vspace{-2mm}\section{Conclusion}\label{sec:conclusion}
In this study, we introduced a benchmark framework, called \PaperAcronym, to properly evaluate the error detection and repair methods. \PaperAcronym enables ML engineers and practitioners to select the most well-suited data cleaning methods in ML pipelines. We carried out an extensive experimental study which involves 19 detectors, 19 repair methods, 33 ML models, and 14 datasets. The obtained results revealed that evaluating the data cleaning method in isolation from the downstream applications can be broadly misleading.

%
%
%

\balance
\bibliographystyle{ACM-Reference-Format}
\bibliography{main}

\end{document}